\documentclass[twocolumn,appendixfloats,numberedappendix,twocolappendix]{openjournal}

\usepackage{amsmath}
\usepackage{lineno}
\usepackage{color}
\usepackage{url}
\usepackage{hyperref}
\hypersetup{colorlinks=true,linkcolor=blue,citecolor=blue,filecolor=blue,urlcolor=blue}
\usepackage{cleveref}
\usepackage{flafter} 
\usepackage{float}
\usepackage{textcase}
\usepackage{graphicx}
\usepackage{xcolor}
\usepackage{appendix}
\usepackage{bm}

\newcommand{\msun}{\ensuremath{M_\odot/h_{70} }}
\newcommand{\mass}{\mbox{$M_{\mbox{\scriptsize 500c}}$}}
\newcommand{\mmass}{\mbox{$M^{\mbox{\scriptsize med}}_{\mbox{\scriptsize 500c}}$}}

\renewcommand{\vec}[1]{\mbox{\boldmath$#1$}}

\newcommand{\spitzer}{{\sl Spitzer}}

\newcommand{\Tcmb}{\mbox{$T_{\mbox{\tiny CMB}}$}}
\newcommand{\muk}{\ensuremath{\mu {\rm K_{CMB}}}}

\newcommand{\sqdeg}{deg$^2$}

\newcommand{\fcont}{{$f_\mathrm{cont}$}}
\newcommand{\bigysz}{\ensuremath{Y_\mathrm{\mbox{\tiny{SZ}}}}}

\newcommand{\webaddress}{\url{https://pole.uchicago.edu/public/data/spt3g-clusters}}

\newcommand{\skyarea}{1,604.8}
\newcommand{\maskfrac}{$3.1\%$}

\newcommand{\fsz}{\ensuremath{f_\mathrm{\mbox{\tiny{SZ}}}}}
\newcommand{\ysz}{\ensuremath{y_\mathrm{\mbox{\tiny{SZ}}}}}

\newcommand{\beq}{\begin{equation}}
\newcommand{\eeq}{\end{equation}}
\newcommand{\beqar}{\begin{eqnarray}}
\newcommand{\eeqar}{\end{eqnarray}}

\newcommand{\ncand}{8,892} 
\newcommand{\nconfirm}{7,190} 
\newcommand{\medianmass}{$1.65 \times 10^{14}$ \mbox{$M_\odot/h_{70}$}} 
\newcommand{\mlow}{$7.9 \times 10^{13}$ \mbox{$M_\odot/h_{70}$}} 
\newcommand{\mhigh}{$1.6 \times 10^{15}$ \mbox{$M_\odot/h_{70}$}}
\newcommand{\medianredshift}{0.726}
\newcommand{\nzgtone}{1,780}
\newcommand{\ngtonepfive}{271}
\newcommand{\fourstarconfirmed}{36}
\newcommand{\nwenhan}{24}
\newcommand{\nmadcows}{7}
\newcommand{\nspecz}{187}

\newcommand{\nstronglensing}{195}
\newcommand{\nnewstronglens}{$>60$}

\graphicspath{{figures/}}

\begin{document}

\begin{nolinenumbers}
\vspace*{-\headsep}\vspace*{\headheight}
\footnotesize \hfill FERMILAB-PUB-26-0364-V\\
\vspace*{-\headsep}\vspace*{\headheight}
\footnotesize \hfill DES-2026-0968
\end{nolinenumbers}

\title{Galaxy Clusters Selected via the Sunyaev-Zel'dovich Effect in 5 year data from the SPT-3G Main Survey\vspace{-1.25cm}}

\def\ANLHEP{1}
\def\KICPChicago{2}
\def\AAUChicago{3}
\def\LMU{4}
\def\Villanova{5}
\def\FNAL{6}
\def\Linea{7}
\def\INAF{8}
\def\UCDavisStats{9}
\def\Zurich{10}
\def\Melbourne{11}
\def\IAP{12}
\def\UNM{13}
\def\Cardiff{14}
\def\Cincinnati{15}
\def\KIPAC{16}
\def\Stanford{17}
\def\SLAC{18}
\def\UCLondon{19}
\def\CfA{20}
\def\Queensland{21}
\def\EFIChicago{22}
\def\PhysicsUChicago{23}
\def\Canarias{24}
\def\Laguna{25}
\def\IFAE{26}
\def\IRSOL{27}
\def\Geneva{28}
\def\NTU{29}
\def\Berkeley{30}
\def\Trieste{31}
\def\IFPU{32}
\def\Saclay{33}
\def\ILAst{34}
\def\KEK{35}
\def\CIEMAT{36}
\def\IITHyderabad{37}
\def\McGill{38}
\def\CIFAR{39}
\def\PhysicsPrinceton{40}
\def\Caltech{41}
\def\Michigan{42}
\def\MichiganAstro{43}
\def\MichiganTheory{44}
\def\USTCAst{45}
\def\USTCPhys{46}
\def\ILPhys{47}
\def\UCLA{48}
\def\MSU{49}
\def\Portsmouth{50}
\def\MIT{51}
\def\UAM{52}
\def\UCDavis{53}
\def\Innsbruck{54}
\def\CASA{55}
\def\ColoradoPhys{56}
\def\SantaCruz{57}
\def\Washington{58}
\def\eScience{59}
\def\AAO{60}
\def\Lowell{61}
\def\TTU{62}
\def\STAR{63}
\def\TexasAM{64}
\def\Aix{65}
\def\Grenoble{66}
\def\ILNCSA{67}
\def\Catalana{68}
\def\AstroPrinceton{69}
\def\Surrey{70}
\def\ANLMSD{71}
\def\Tohoku{72}
\def\Rio{73}
\def\SKAI{74}
\def\Ruhr{75}
\def\Copenhagen{76}
\def\Sussex{77}
\def\CaseWestern{78}
\def\Chulalongkorn{79}
\def\OakRidge{80}
\def\Cerca{81}
\def\Kerala{82}
\def\BCCP{83}
\def\LBNL{84}
\def\MPE{85}

\author{
  L.~E.~Bleem\altaffilmark{\ANLHEP,\KICPChicago,\AAUChicago},
  M.~Klein\altaffilmark{\LMU},
  K.~Kornoelje\altaffilmark{\AAUChicago,\KICPChicago,\ANLHEP},
  S.~Bocquet\altaffilmark{\LMU},
  J.~A.~Sobrin\altaffilmark{\Villanova,\FNAL},
  M.~Aguena\altaffilmark{\Linea,\INAF},
  E.~Anderes\altaffilmark{\UCDavisStats},
  A.~J.~Anderson\altaffilmark{\FNAL,\KICPChicago,\AAUChicago},
  F.~Andrade-Oliveira\altaffilmark{\Zurich},
  B.~Ansarinejad\altaffilmark{\Melbourne}
% ...
}

\altaffiltext{\ANLHEP}{High-Energy Physics Division, Argonne National Laboratory, 9700 South Cass Avenue, Lemont, IL, 60439, USA}
\altaffiltext{\KICPChicago}{Kavli Institute for Cosmological Physics, University of Chicago, 5640 South Ellis Avenue, Chicago, IL, 60637, USA}
\altaffiltext{\AAUChicago}{Department of Astronomy and Astrophysics, University of Chicago, 5640 South Ellis Avenue, Chicago, IL, 60637, USA}
\altaffiltext{\LMU}{University Observatory, Faculty of Physics, LMU Munich, Scheinerstr.~1, 81679 Munich, Germany}
\altaffiltext{\Villanova}{Department of Physics, Villanova University, 800 E Lancaster Ave, Villanova, PA 19085, USA}
\altaffiltext{\FNAL}{Fermi National Accelerator Laboratory, MS209, P.O. Box 500, Batavia, IL, 60510, USA}
\altaffiltext{\Linea}{Laborat\'orio Interinstitucional de e-Astronomia - LIneA, Rua Gal. Jos\'e Cristino 77, Rio de Janeiro, RJ - 20921-400, Brazil}
\altaffiltext{\INAF}{INAF-Osservatorio Astronomico di Trieste, via G. B. Tiepolo 11, I-34143 Trieste, Italy}
\altaffiltext{\UCDavisStats}{Department of Statistics, University of California, One Shields Avenue, Davis, CA 95616, USA}
\altaffiltext{\Zurich}{Physik-Institut, University of Z\"urich, Winterthurerstrasse 190, CH-8057 Z\"urich, Switzerland}
\altaffiltext{\Melbourne}{School of Physics, University of Melbourne, Parkville, VIC 3010, Australia}
\altaffiltext{\IAP}{Sorbonne Universit\'e, CNRS, UMR 7095, Institut d'Astrophysique de Paris, 98 bis bd Arago, 75014 Paris, France}
\altaffiltext{\UNM}{Department of Physics and Astronomy, University of New Mexico, Albuquerque, NM, 87131, USA}
\altaffiltext{\Cardiff}{School of Physics and Astronomy, Cardiff University, Cardiff CF24 3YB, United Kingdom}
\altaffiltext{\Cincinnati}{Department of Physics,University of Cincinnati, Cincinnati,OH 45221, USA}
\altaffiltext{\KIPAC}{Kavli Institute for Particle Astrophysics and Cosmology, Stanford University, 452 Lomita Mall, Stanford, CA, 94305, USA}
\altaffiltext{\Stanford}{Department of Physics, Stanford University, 382 Via Pueblo Mall, Stanford, CA, 94305, USA}
\altaffiltext{\SLAC}{SLAC National Accelerator Laboratory, 2575 Sand Hill Road, Menlo Park, CA, 94025, USA}
\altaffiltext{\UCLondon}{Department of Physics \& Astronomy, University College London, Gower Street, London, WC1E 6BT, UK}
\altaffiltext{\CfA}{Center for Astrophysics \textbar{} Harvard \& Smithsonian, 60 Garden Street, Cambridge, MA, 02138, USA}
\altaffiltext{\Queensland}{School of Mathematics and Physics, University of Queensland,  Brisbane, QLD 4072, Australia}
\altaffiltext{\EFIChicago}{Enrico Fermi Institute, University of Chicago, 5640 South Ellis Avenue, Chicago, IL, 60637, USA}
\altaffiltext{\PhysicsUChicago}{Department of Physics, University of Chicago, 5640 South Ellis Avenue, Chicago, IL, 60637, USA}
\altaffiltext{\Canarias}{Instituto de Astrofisica de Canarias, E-38205 La Laguna, Tenerife, Spain}
\altaffiltext{\Laguna}{Universidad de La Laguna, Dpto. Astrofísica, E-38206 La Laguna, Tenerife, Spain}
\altaffiltext{\IFAE}{Institut de F\'{\i}sica d'Altes Energies (IFAE), The Barcelona Institute of Science and Technology, Campus UAB, 08193 Bellaterra (Barcelona) Spain}
\altaffiltext{\IRSOL}{Istituto ricerche solari Aldo e Cele Dacc\`o (IRSOL), Faculty of Informatics, Universit\`a della Svizzera italiana, 6605 Locarno, Switzerland}
\altaffiltext{\Geneva}{Universit\'e de Gen\`eve, D\'epartement de Physique Th\'eorique, 24 Quai Ansermet, CH-1211 Gen\`eve 4, Switzerland}
\altaffiltext{\NTU}{National Taiwan University, No. 1, Sec. 4, Roosevelt Road, Taipei 106319, Taiwan}
\altaffiltext{\Berkeley}{Department of Physics, University of California, Berkeley, CA, 94720, USA}
\altaffiltext{\Trieste}{Astronomy Unit, Department of Physics, University of Trieste, via Tiepolo 11, I-34131 Trieste, Italy}
\altaffiltext{\IFPU}{Institute for Fundamental Physics of the Universe, Via Beirut 2, 34014 Trieste, Italy}
\altaffiltext{\Saclay}{Universit\'e Paris-Saclay, Universit\'e Paris Cit\'e, CEA, CNRS, AIM, 91191, Gif-sur-Yvette, France}
\altaffiltext{\ILAst}{Department of Astronomy, University of Illinois Urbana-Champaign, 1002 West Green Street, Urbana, IL, 61801, USA}
\altaffiltext{\KEK}{High Energy Accelerator Research Organization (KEK), Tsukuba, Ibaraki 305-0801, Japan}
\altaffiltext{\CIEMAT}{Centro de Investigaciones Energ\'eticas, Medioambientales y Tecnol\'ogicas (CIEMAT), Madrid, Spain}
\altaffiltext{\IITHyderabad}{Department of Physics, IIT Hyderabad, Kandi, Telangana 502285, India}
\altaffiltext{\McGill}{Department of Physics and McGill Space Institute, McGill University, 3600 Rue University, Montreal, Quebec H3A 2T8, Canada}
\altaffiltext{\CIFAR}{Canadian Institute for Advanced Research, CIFAR Program in Gravity and the Extreme Universe, Toronto, ON, M5G 1Z8, Canada}
\altaffiltext{\PhysicsPrinceton}{Joseph Henry Laboratories of Physics, Jadwin Hall, Princeton University, Princeton, NJ 08544, USA}
\altaffiltext{\Caltech}{California Institute of Technology, 1200 East California Boulevard., Pasadena, CA, 91125, USA}
\altaffiltext{\Michigan}{Department of Physics, University of Michigan, 450 Church Street, Ann Arbor, MI, 48109, USA}
\altaffiltext{\MichiganAstro}{Department of Astronomy, University of Michigan, Ann Arbor, MI 48109, USA}
\altaffiltext{\MichiganTheory}{Leinweber Center for Theoretical Physics, University of Michigan, Ann Arbor, MI 48109, USA}
\altaffiltext{\USTCAst}{Department of Astronomy, University of Science and Technology of China, Hefei 230026, China}
\altaffiltext{\USTCPhys}{School of Astronomy and Space Science, University of Science and Technology of China, Hefei 230026}
\altaffiltext{\ILPhys}{Department of Physics, University of Illinois Urbana-Champaign, 1110 West Green Street, Urbana, IL, 61801, USA}
\altaffiltext{\UCLA}{Department of Physics and Astronomy, University of California, Los Angeles, CA, 90095, USA}
\altaffiltext{\MSU}{Department of Physics and Astronomy, Michigan State University, East Lansing, MI 48824, USA}
\altaffiltext{\Portsmouth}{Institute of Cosmology and Gravitation, University of Portsmouth, Portsmouth, PO1 3FX, UK}
\altaffiltext{\MIT}{Department of Physics, Massachusetts Institute of Technology, 77 Massachusetts Avenue, Cambridge, MA 02139, USA}
\altaffiltext{\UAM}{Instituto de Fisica Teorica UAM/CSIC, Universidad Autonoma de Madrid, 28049 Madrid, Spain}
\altaffiltext{\UCDavis}{Department of Physics \& Astronomy, University of California, One Shields Avenue, Davis, CA 95616, USA}
\altaffiltext{\Innsbruck}{Universit\"at Innsbruck, Institut f\"ur Astro- und Teilchenphysik, Technikerstr. 25/8, 6020 Innsbruck, Austria}
\altaffiltext{\CASA}{CASA, Department of Astrophysical and Planetary Sciences, University of Colorado, Boulder, CO, 80309, USA }
\altaffiltext{\ColoradoPhys}{Department of Physics, University of Colorado, Boulder, CO, 80309, USA}
\altaffiltext{\SantaCruz}{Santa Cruz Institute for Particle Physics, Santa Cruz, CA 95064, USA}
\altaffiltext{\Washington}{Department of Astronomy \& the DiRAC Institute, University of Washington, Physics-Astronomy Building, Box 351580, Seattle, WA 98195-1700, USA}
\altaffiltext{\eScience}{eScience Institute, University of Washington, Physics-Astronomy Building, Box 351580, Seattle, WA 98195-1700, USA}
\altaffiltext{\AAO}{Australian Astronomical Optics, Macquarie University, North Ryde, NSW 2113, Australia}
\altaffiltext{\Lowell}{Lowell Observatory, 1400 Mars Hill Rd, Flagstaff, AZ 86001, USA}
\altaffiltext{\TTU}{Department of Physics \& Astronomy, Box 41051, Texas Tech University, Lubbock TX 79409-1051, USA}
\altaffiltext{\STAR}{STAR Institute, Quartier Agora - All\'ee du six Ao\^ut, 19c B-4000 Li\'ege, Belgium}
\altaffiltext{\TexasAM}{George P. and Cynthia Woods Mitchell Institute for Fundamental Physics and Astronomy, and Department of Physics and Astronomy, Texas A\&M University, College Station, TX 77843,  USA}
\altaffiltext{\Aix}{Aix Marseille Univ, CNRS/IN2P3, CPPM, Marseille, France}
\altaffiltext{\Grenoble}{LPSC Grenoble - 53, Avenue des Martyrs 38026 Grenoble, France}
\altaffiltext{\ILNCSA}{Center for AstroPhysical Surveys, National Center for Supercomputing Applications, Urbana, IL, 61801, USA}
\altaffiltext{\Catalana}{Instituci\'o Catalana de Recerca i Estudis Avan\c{c}ats, E-08010 Barcelona, Spain}
\altaffiltext{\AstroPrinceton}{Department of Astrophysical Sciences, Princeton University, Peyton Hall, Princeton, NJ 08544, USA}
\altaffiltext{\Surrey}{Department of Physics, University of Surrey, Guildford GU2 7XH, United Kingdom}
\altaffiltext{\ANLMSD}{Materials Sciences Division, Argonne National Laboratory, 9700 South Cass Avenue, Lemont, IL, 60439, USA}
\altaffiltext{\Tohoku}{Institute of Multidisciplinary Research for Advanced Materials, Tohoku University, Sendai, 980-8577, Japan}
\altaffiltext{\Rio}{Observat\'orio Nacional, Rua Gal. Jos\'e Cristino 77, Rio de Janeiro, RJ - 20921-400, Brazil}
\altaffiltext{\SKAI}{NSF-Simons AI Institute for the Sky (SkAI), 172 E. Chestnut St., Chicago, IL 60611, USA}
\altaffiltext{\Ruhr}{Ruhr University Bochum, Faculty of Physics and Astronomy, Astronomical Institute, German Centre for Cosmological Lensing, 44780 Bochum, Germany}
\altaffiltext{\Copenhagen}{University of Copenhagen, Dark Cosmology Centre, Juliane Maries Vej 30, 2100 Copenhagen O, Denmark}
\altaffiltext{\Sussex}{Department of Physics and Astronomy, Pevensey Building, University of Sussex, Brighton, BN1 9QH, UK}
\altaffiltext{\CaseWestern}{Department of Physics, Case Western Reserve University, Cleveland, OH, 44106, USA}
\altaffiltext{\Chulalongkorn}{Department of Physics, Faculty of Science, Chulalongkorn University, 254 Phayathai Road, Pathumwan, Bangkok 10330, Thailand}
\altaffiltext{\OakRidge}{Computer Science and Mathematics Division, Oak Ridge National Laboratory, Oak Ridge, TN 37831}
\altaffiltext{\Cerca}{CERCA/ISO, Department of Physics, Case Western Reserve University, Cleveland, OH 44106, USA}
\altaffiltext{\Kerala}{Central University of Kerala, Kasaragod, Kerala, India 671325}
\altaffiltext{\BCCP}{Berkeley Center for Cosmological Physics, Department of Physics, University of California, Berkeley, CA 94720, US}
\altaffiltext{\LBNL}{Physics Division, Lawrence Berkeley National Laboratory, Berkeley, CA, 94720, USA}
\altaffiltext{\MPE}{Max-Planck-Institut f\"{u}r extraterrestrische Physik,Giessenbachstr, 85748 Garching, Germany}

\shorttitle{Galaxy Clusters Discovered in the SPT-3G Main Survey}
\shortauthors{Bleem et al., SPT-3G \& DES collaborations}

\thanks{$^{\star}$E-mail:lbleem@anl.gov}

\begin{abstract}
We report a new galaxy cluster catalog, selected using the thermal Sunyaev-Zel'dovich (SZ) effect, from 5 years of observations of the SPT-3G Main field. 
Drawn from arcminute-resolution data with white noise levels of 3.2, 2.5, and 8.9 $\muk$-arcmin at 95, 150, and 220 GHz, respectively,
the sample consists of \ncand \ cluster candidates detected above significance $\xi=4$, with an expected purity of $>82\%$ (4,480 at  $\xi\ge5$ with expected purity $>99\%$). 
Using optical and infrared (IR) data we have confirmed \nconfirm \ candidates as clusters. 
The sample spans a mass range \mlow \ $< \mass < $ \mhigh \ with a median mass of  \mmass \ = \medianmass, and a redshift range of $0.037<z\lesssim 2$ with a median redshift of $z_{\textrm{med}}$ = \medianredshift; 
\nzgtone \ clusters are at $z>1$ and \ngtonepfive \ at $z>1.5$.
Compared to previous SZ cluster samples from South Pole Telescope and Atacama Cosmology Telescope data, the SPT-3G sample presented here is highly consistent in mass and redshift but is significantly deeper, with per-cluster detection signal-to-noise 2-4 times higher and a cluster density approaching that traditionally associated with optical samples (4.5 confirmed clusters per square degree).
We cross match the SPT-3G catalog with eRASS1 cluster and point source catalogs, finding 1,279 and 1,319 matches, respectively.  The SPT and eROSITA cluster mass estimates are in relatively good agreement, and the large number of eROSITA point source matches will provide promising avenues to study both X-ray cluster active galactic nuclei and the eROSITA cluster selection function at high $z$. 
A representative set of wide-field optical/IR cluster samples recover 50-78\% of the SPT-3G confirmed clusters with generally good redshift consistency but varying levels of optical observable-SZ mass scatter. 
A number of clusters are flagged as candidate strong gravitational lenses of distant optical and/or dusty star-forming galaxies. 
We also perform a series of validation checks using both internal data splits and comparisons to external samples. 
These tests show increasing correlated (dusty) emission with redshift, with a $\sim17\times$ larger 220~GHz temperature increment for clusters at $z\sim1.5$ than $z\sim0.25$,
 but only weak evidence for correlated synchrotron emission. Previous analyses have shown the SPT-3G selection to be robust to this level of dust contamination.
Further joint sample characterizations as well as constraints on the astrophysical and cosmological properties of this new SPT-3G sample will be the focus of upcoming works. 
\end{abstract}

\keywords{Large-Scale Structure of the Universe, Galaxy Clusters}

\section{Introduction}\label{sec:intro} 
\setcounter{footnote}{0}

The past decade has seen a transformative expansion of wide-area astronomical surveys. 
Data from these surveys have enabled stringent tests of models governing the formation and evolution of the Universe both through probes of the early Universe---via  millimeter-wavelength observations of the cosmic microwave background \citep[CMB;][]{planck18-6,bicep2keck21b, louis25, camphuis26}---and of the late-time Universe via  photometric and spectroscopic galaxy and supernova surveys at optical, infrared, and X-ray wavelengths \citep{abbott22a,desi25,dalal23,ghirardini24}. 
These studies have shown strong preference for a model in which the current energy density of the Universe is dominated by a ``dark energy" ($\Lambda$) component with significant contributions from cold dark matter (CDM).

Comparisons of late- and early-time probes provide particularly powerful tests of the $\Lambda$CDM model, as they trace eras when the energy content and degree of structure formation in the Universe are markedly different. 
Thus far, such comparisons have shown remarkable agreement and support for a concordance cosmological model \citep{abbott22a, wright25}. 
However, thanks to the high statistical constraining power of these survey data, subtle tensions have been uncovered at moderate  \citep[e.g. evolving dark energy;][]{desi25,popovic26}  to high significance  (notably the discrepancy in  $H_0$  between primary CMB and some low redshift expansion measures; e.g., \citealp{riess24}, though see e.g., \citealp{hoyt26} for an alternative perspective) between the cosmological models preferred by probes drawn from the distinct eras. 

As cosmology has now firmly entered an era in which inferences from many probes are dominated by systematic errors (as opposed to statistical uncertainties), it is critical to explore these tensions with a diverse set of observables, ideally subject to very different sources of error. 
Clusters of galaxies---the largest gravitationally collapsed systems in the Universe---are a natural probe to target, as their abundance is extraordinarily sensitive to physics that governs the growth of structure.  
Indeed, clusters of galaxies could be the most constraining probe of Dark Energy, if systematic errors can be controlled \citep{dodelson16}. 
This promise has led clusters to be identified as a key pillar in dark energy science by the eROSITA, \textit{Euclid}, and LSST Dark Energy Science collaborations \citep{predehl21, mellier25, lsst18}. 

An important step towards achieving this potential is the construction of large cluster samples with well-understood selection functions \citep{allen11}. 
Cluster samples selected via intracluster-medium-derived observables---in particular those selected through the cluster thermal bremsstrahlung signature in X-ray surveys \citep{sarazin86} or the thermal Sunyaev-Zel'dovich \citep[SZ;][]{sunyaev72} effect in mm-wave observations---offer particularly significant advances in this regard. 
Progress in SZ-selected clusters in recent years has been rapid with the release of samples from the \textit{Planck} Collaboration \citep{planck15-27}, the South Pole Telescope (SPT) collaboration---including large catalogs from the SPT-SZ and SPTpol experiments \citep{bleem15b,huang20,bleem20, bleem24, klein24a} and the first deep field catalogs from the SPT-3G experiment \citep{kornoelje26, archipley26}---as well as large cluster samples from the  Atacama Cosmology Telescope (ACT) collaboration \citep{ hilton18,hilton21}, including the notable recent release in \citet{hilton26} of 10,040 clusters drawn from 16,293 square degrees of ACT DR6 data. 
The X-ray cluster field has been transformed as well by the first release by eROSITA collaboration of over 20,000 clusters found in data from the SRG/eROSITA All-Sky Survey \citep[eRASS;][]{liu22, bulbul24,balzer25}. 
New hydrodynamical simulations \citep[e.g.,][]{dolag25,frontiere25, helly26} and large volume gravity-only simulations post-processed  with sophisticated ``baryon-painting" techniques \citep[e.g.,][]{mead20,keruzore24,lau25}
have also become available to enable robust interpretation of these complex datasets. 
The recent cosmological constraints produced using  the eROSITA and SPT cluster samples have demonstrated some of the tightest constraints from growth-of-structure probes  (\citealt{bocquet24b, ghirardini24, bocquet25}---though also see the combined optical cluster, weak lensing, and galaxy clustering analysis of \citealt{Abbott25clusters}) and provide valuable points of comparison to cosmological results obtained using geometric probes such as supernovae and baryon acoustic oscillations \citep{popovic26,desi25}. 

In this work we introduce an SZ cluster catalog produced through the analysis of 5 years of survey data from the SPT-3G Main field.
This sample of over 7,190  clusters detected at a significance $\xi >4$ in 1600 \sqdeg \ of sky has a median mass of \medianmass \  and a median redshift of $z=\medianredshift$, with detections extending to $z \sim 2$. 
With its high spatial density and broad mass range and redshift reach, the SPT-3G cluster sample will enable powerful cosmological tests and astrophysical studies in its own right, and will also be highly complementary to the wider-field eROSITA and ACT samples. 

Our report on this new SPT-3G cluster sample is organized as follows: 
\begin{itemize}
\item  Section \ref{sec:obs_and_mapmaking} provides an overview of the observations and mapmaking procedure used in this work. 
\item Section \ref{sec:cluster_identification} discusses the process by which cluster candidates are identified in SPT-3G data, the measurement of our reported Compton-$y$ based observable, and our estimation of the intrinsic purity of the SZ sample. 
\item Section \ref{sec:cluster_confirmation} details the confirmation process and redshift estimation for cluster candidates using optical and infrared data. 
\item Section \ref{sec:obs_mass} describes our modeling of the SZ and optical-richness mass observable relations. Masses in this work are estimated via abundance matching with a fixed cosmology.
\item Section \ref{sec:results} combines the results of the previous sections to provide a high level summary of the sample properties as well as an estimate of completeness of the sample as a function of mass and redshift.  
\item Section \ref{sec:strong_lensing} highlights candidate strong gravitational lenses in the SPT-3G sample, both through a literature search and our own dedicated screening. 
\item Section \ref{sec:icm_comparison} summarizes high level comparisons to intracluster medium (ICM)-selected catalogs from recent SZ and X-ray surveys. Additional comparisons to X-ray source catalogs from eROSITA are provided in Appendix \ref{sec:xrayappend}.
\item Section \ref{sec:optical_ir_comparison} examines the relative completeness, redshift consistency, and scatter with SZ mass for several optical- and infrared-selected cluster samples, with a more detailed comparison to the DES \textsc{redMaPPer} Y3 sample \citep{Abbott25clusters} in  Appendix \ref{sec:opticalappend}.   
\item Section \ref{sec:internal_checks} explores consistency of the SZ properties of the cluster sample as measured using subsets of the SPT-3G bands to test for contamination from correlated dusty and synchrotron sources. 
\end{itemize}
Finally, in Section \ref{sec:conclusion} we conclude. 
Data products associated with this paper are available online\footnote{Data products: \webaddress} and we provide in the Appendix \ref{sec:tableoverview} a description of the catalog table entries. 

\textit{Conventions}: Where applicable, we assume a fiducial $\Lambda$CDM cosmology with $\sigma_8=0.80$, $\Omega_b = 0.046$, $\Omega_m~=~0.30$, $h = 0.70$, $n_s(k_s=0.002) = 0.972$, and $\Sigma m_\nu=0.06$~eV.  
This adopted fiducial cosmology enables direct comparisons across generations of catalogs from the SPT since \citet{bleem15b}. 
Cluster masses are reported in terms of  $M_{500c}$, which is defined as the mass enclosed within a radius, $r_{500c}$, at which the average enclosed density is 500$\times$ the critical density at the cluster redshift.

\section{Observations and Mapmaking} 
\label{sec:obs_and_mapmaking}

The cluster sample described in this work results from an analysis of data taken with the SPT-3G receiver between 2019 and 2023.
As detailed in \citet{sobrin22}, SPT-3G is the third-generation receiver installed on the SPT---a 10-meter, submillimeter-quality telescope located at the Amundsen-Scott South Pole Station~\citep{carlstrom11}.
SPT-3G is designed for intensity and polarization measurements at 95, 150, and 220 GHz, and the large telescope aperture results in  angular resolutions of 1\farcm{6}, 1\farcm{2}, and 1\farcm{05} \ for the three bands, respectively. 
In addition to surveys for galaxy clusters, the SPT is built for sensitive, high-resolution observations of primary  \citep{ge24, chou25, camphuis26, quan26} and secondary \citep{chaubal26,raghunathan26} CMB anisotropies and populations of mm-wave emissive point sources \citep{everett20, archipley26}, which include active galactic nuclei \citep{hood26}, dusty star-forming galaxies \citep{vieira13, reuter20}, and transient, variable, and moving sources \citep{guns21,chichura22,hood23, tandoi24,wan26}.

\begin{figure*}[th!]
\begin{center}
\includegraphics[width=0.99\linewidth]{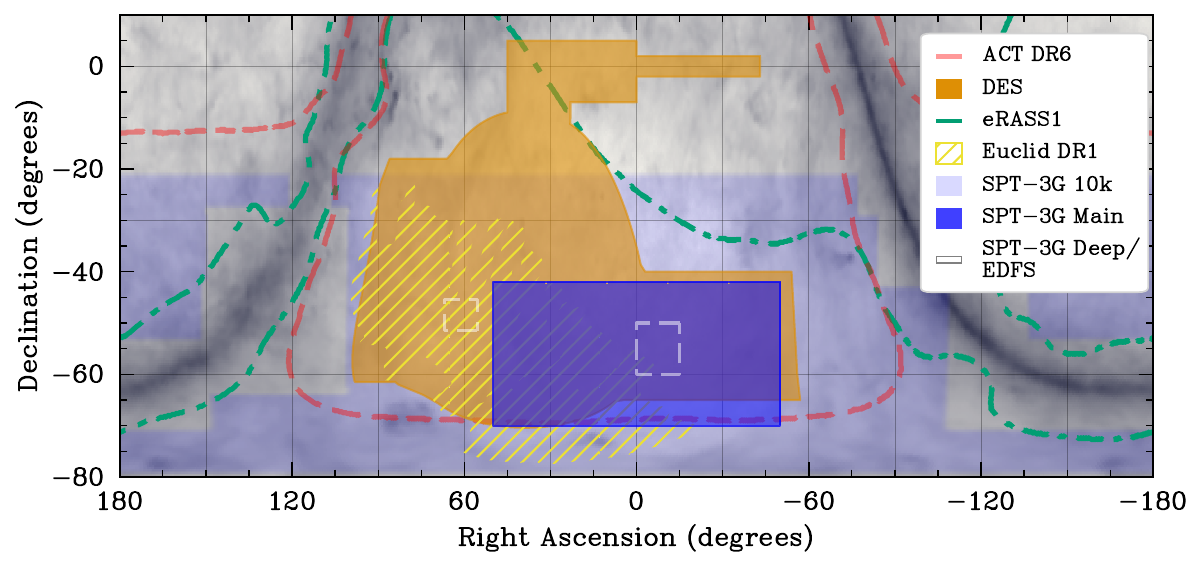}
\caption{Footprint of the SPT-3G survey plotted with those of several other multi-wavelength surveys discussed in this work. The footprints are overlaid on a dust map from \cite{planck15-10}. The full SPT-3G 10,000 deg$^2$  survey is shown in pale blue, the boundaries of the SPT-Euclid Deep field South \citep{archipley26} and SPT 100d Deep field \citep{kornoelje26} are white-dashed, and the SPT-3G Main field---the focus of this work---is plotted in bold blue.    \label{fig:survey_fields}
}
\end{center}
\end{figure*}

SPT-3G has been used to survey approximately 10,000 square degrees of the southern sky \citep{prabhu24}.  
The  cluster sample reported in this work is derived from an analysis of 1,604~deg$^{2}$ of data from the SPT-3G ``Main" field, upon which the majority of observation time of SPT-3G has been focused. 
The nominal boundaries of the SPT-3G Main field are 20$^h$40$^m$ and 3$^h$20$^m$ in right ascension (R.A.) and $-42^\circ$ and $-70^\circ$ in declination (Dec.), as depicted in Fig.~\ref{fig:survey_fields}, with borders coordinated to overlap with observations from the BICEP Array \citep{moncelsi20} for inflationary B-mode analyses \citep{bicepkeckspt21}. 
The field also has extensive sky coverage from the Dark Energy Survey \citep[DES;][]{abbott21}, eROSITA \citep{merloni24}, and, in the near future, will benefit from observations by the \textit{Euclid} mission \citep{mellier25}, 4MOST \citep{dejong19}, the Vera C. Rubin Observatory's Legacy Survey of Space and Time \citep[LSST;][]{ivezic08},  and MeerKAT \citep{jonas16}. 

\subsection{Observation Strategy}
\label{sec:obs_mapmaking}

The SPT-3G Main field is not observed as a single field, but rather is constructed from combined observations of four subfields, split along the declination axis, of $7.5$$^\circ$ in height.
These subfields are centered at Dec. $-44.75$$^\circ$, $-52.25$$^\circ$, $-59.75$$^\circ$, and $-67.25^\circ$ \citep[see Fig. 2,][]{quan26}. 
To survey a subfield, the telescope is scanned in a back-and-forth pattern across the full azimuth extent of the field at {$1^\circ$/sec} in azimuth at constant elevation, followed by a $12\farcm{5}$  \ step in elevation.
This motion repeats until the entire subfield has been observed, which typically takes $\sim$2 hours.
Owing to the telescope's location at the geographic South Pole, this observation strategy results in scan data aligned with constant lines in declination which simplifies certain aspects of our analysis as discussed below.

\subsection{Data Processing}

The procedure for translating time-ordered data (TOD) from individual SPT-3G detectors into calibrated  temperature maps closely follows that described in detail in recent SPT analyses~\citep{dutcher21, archipley26, quan26}.
Certain mapmaking choices are made to yield final maps with high sensitivity to arcminute-scale features (such as those deriving from  galaxy clusters).
We briefly summarize and highlight those choices here.

The first step in data processing involves flagging and removing data from detectors that exhibit certain well-established pathologies during survey scans. 
Next, data from locations corresponding to bright ($>30$~mJy at $95$~GHz), emissive sources is removed and interpolated over in the TOD to avoid introducing strong artifacts in the maps from the following filtering steps.

We next apply several filters to reduce contamination from instrumental noise and atmospheric signals.  
To reduce the impact of low-frequency instrumental and atmospheric noise, we apply three types of filters: First, we apply a common-mode filter that groups detectors within the same detector module and observing band together, averages their TOD, and then subtracts that averaged signal from each detector's TOD. 
Next we project out a low-order polynomial from the TOD of each detector. 
Finally, we Fourier-transform the TOD of each detector and apply a Fourier high-pass filter with a cutoff corresponding to $\ell = 500$ in the scan direction. 
In this last step, we also apply a low-pass filter with a cutoff corresponding to $\ell = 20,000$ in the scan direction, to reduce the effect of aliasing when we bin the TOD into map pixels.
Following \cite{archipley26}, corrections are then applied to the TOD to account for detector time constants (i.e., the finite temporal responsivity of the detectors) and timing offsets between detectors and telescope pointing data.

After the TOD have been filtered, data from each detector are weighted based on the inverse noise variance in the 1--4 Hz range.
Using the telescope pointing model, the weighted TOD are then binned in 0\farcm{25} pixels in a Lambert azimuthal equal-area map \citep[ZEA;][]{calabretta02} projection and averaged.
This process results in the creation of a weighted temperature map for every observation at each observing band.  

\begin{figure*}
\begin{center}
\includegraphics[width=0.95\linewidth, trim=40 0 0 0]{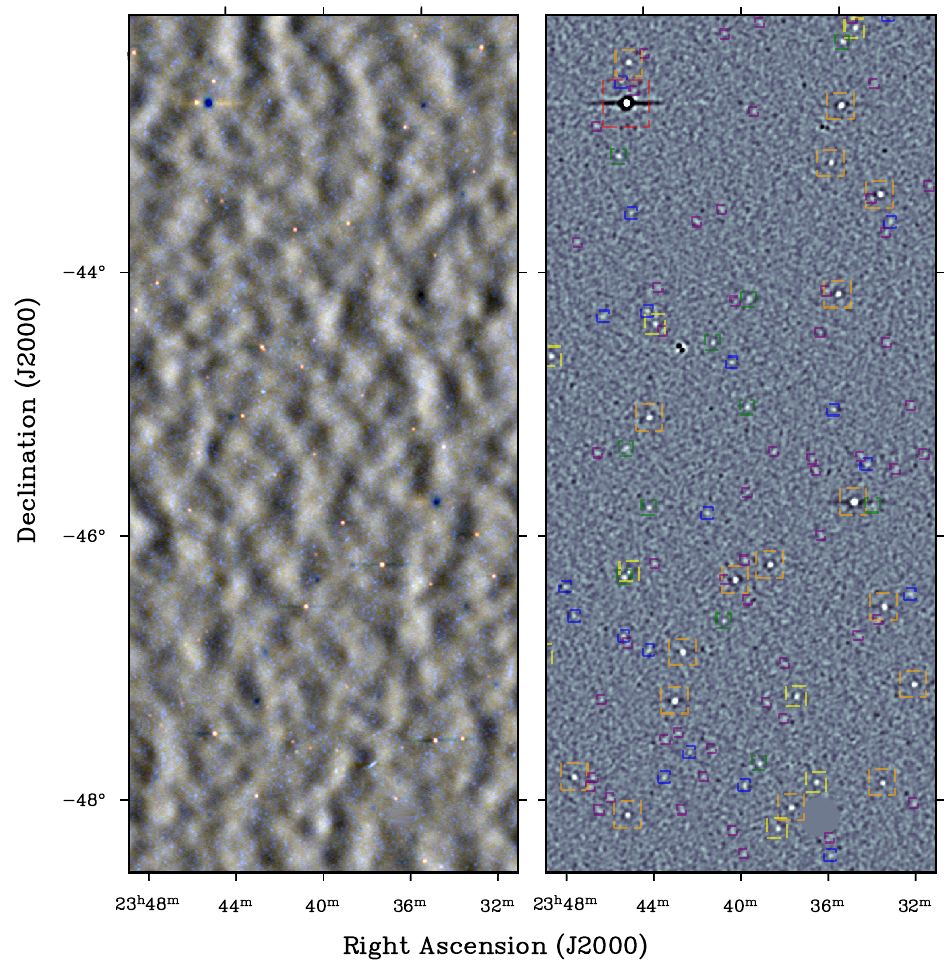}
\caption{\textit{(Left)} Cutout of $3^o \times 6.5^o$ of SPT-3G Main field survey data (approximately 1.2\% of the Main field survey area). The \textit{rgb} image is constructed using the 95 (red), 150 GHz (green), and 220 (blue) GHz data with peak-to-peak fluctuations in each channel of $\pm110 \muk$. This figure illustrates the main constituents of the SPT-3G temperature maps. As the maps are calibrated in $\mu K_\textrm{CMB}$ units, CMB fluctuations have equal intensity in all channels and thus appear gray in this image. Falling-spectrum sources (e.g., AGN) appear red, and dusty sources appear light blue. Galaxy clusters, with a strong deficit of flux in the $rg$ channels appear deep blue in this color scaling. \textit{(Right)} Cluster detection map for a matched filter with $\theta_{c} = 0\farcm{5}$ for the same region. In colored boxes of increasing size we denote the locations of cluster candidates  detected over all filter scales. We mark candidates with $4 \leq\xi<5$ (purple), $5\leq\xi<6$ (blue), $6\leq\xi<8$ (green), $8\leq\xi<10$ (yellow), $10\leq\xi<100$ (orange), $\xi>100$ (red, in this image SPT-CL~J2344$-$4243, the ``Phoenix'' Cluster). The color scaling from black to white on this image varies between $\pm7$ in detection significance. 
The SPT-3G Main field cluster sample has a high cluster density, with over 5.5 (4.5) candidates (confirmed clusters, see Sec. \ref{sec:results}) per square degree.}
\label{fig:map_cutout}
\end{center}
\end{figure*}

Following quality checks on individual observation maps, weighted maps from each subfield are then summed and deweighted to produce final coadded subfield maps. 
Absolute calibration of each subfield map to CMB temperature units is set through comparisons to \textit{Planck} \citep{quan26} and, once calibrated, the coadded subfield maps are combined into a full field  map.
The final coadded temperature maps analyzed in this work have white-noise levels of 3.2, 2.5, and 8.9 $\muk$-arcmin at 95, 150, and 220 GHz, respectively.
An example of the signal quality achieved in the final maps is shown in Fig.~\ref{fig:map_cutout}.

\subsection{Choice of Map Projection}

Analysis of small regions of sky using a flat-sky approximation can offer significant advantages in terms of analysis ease and computational complexity (e.g., allowing for two-dimensional Fourier transforms in place of spherical harmonic transforms, simplified filtering characterization, etc). 
However, analysis of curved-sky data assuming flat-sky geometry necessitates sacrifices in shape or spatial fidelity. 
The baseline ZEA projection adopted for many SPT-3G analyses has excellent distance- and shape-preserving properties. 
However, the SPT filtering and transfer function 
varies in position across the survey field in this projection, owing to changes in the angle of the telescope's scan direction relative to pixel rows in the maps~\citep{schaffer11}.
This byproduct complicates the application of an optimal matched filter for cluster detection (Sec. \ref{sec:cluster_identification}) in this map projection.
For cluster analyses, where shape preservation is less important than transfer-function fidelity (since the extended shape of clusters is not measured well by the limited resolution of the SPT), we follow previous SPT analyses and choose to work in the Sanson-Flamsteed projection~\citep{calabretta02}.
In this projection, pixel rows directly correspond to right ascension and the scan direction of the telescope, thereby guaranteeing a homogeneous transfer function across any given row of the map. 
However, the Sanson-Flamsteed flat-sky projection results in growing spatial distortions with increasing distance from the map's center.
To constrain these distortion effects near the map edges to an acceptable level, rather than examining the full field in its entirety, we re-project the full-field maps into 7 overlapping sub-patches in the Sanson-Flamsteed projection.

\subsection{Emissive Point-Source Removal}
\label{sec:point_source_removal}

Bright, emissive point sources are prominent and densely distributed across the coadded maps (as shown in Fig.~\ref{fig:map_cutout}).
Alongside these sources are filtering-wing artifacts introduced by the TOD-filtering described above.
The bright signals and their associated filtering wings can both hide detectable galaxy clusters and introduce spurious detections when the cluster-extraction algorithm is applied to the maps.
Therefore, before the cluster detection filtering is applied and the maps are examined for clusters, these point-source signals and filtering wings are removed using a procedure very similar to that used in recent SPT cluster analyses~\citep{bleem24, kornoelje26}.

For each observing band, a flux-normalized template of an emissive point source is constructed using thumbnail cutouts from the same map, guaranteeing the same transfer-function properties.
The templates are created using 2,000 of the brightest sources still present in the maps in each observing band after mapmaking.
Using these per-band, normalized templates, ``source maps" are generated with the appropriate flat-sky map distortions and flux values at every map location originally containing an emissive source with a signal-to-noise ratio (SNR) $>$5 at $95$~GHz from a dedicated point source analysis of the field.
The source maps are then subtracted from the original full-depth coadded maps at each frequency, substantially reducing the point source and filtering wing features for all but the brightest sources (discussed more in the next section).

\section{Cluster Identification} 
\label{sec:cluster_identification}

The SZ effect describes an inverse Compton-scattering phenomenon in which interacting CMB photons are up-scattered to higher energies by hot electrons in the ICM of a galaxy cluster~\citep{sunyaev72}.
The SZ signal in units of effective CMB fluctuation temperature can be expressed as:

\begin{equation}
\begin{split}
\Delta T(\mathbf{x},\nu) &= \Tcmb \ \fsz(\nu)\int n_\mathrm{e}(\mathbf{x}) \frac{k_\mathrm{B}T_\mathrm{e}(\mathbf{x}) }{m_\mathrm{e}c^{2}} \sigma_\mathrm{T} dl  \\
  &\equiv  \Tcmb \ \fsz(\nu) \ \ysz(\mathbf{x}) ,
\end{split}
\label{eqn:tsz}
\end{equation}
where $\Tcmb=2.726\pm 0.001$~K is the mean CMB temperature \citep{fixsen09b},
$\fsz(\nu)$ is a function encapsulating the frequency ($\nu$) dependence of the interaction,
$n_\mathrm{e}$ is the electron density in the gas,
$k_\mathrm{B}$ the Boltzmann constant,
$T_\mathrm{e}$ is temperature of the electron gas, 
$m_\mathrm{e} c^2$ is the electron rest mass energy,
$\sigma_\mathrm{T}$ is the Thomson cross-section of the electron,
$\textbf{x}$ is the cluster location,
$\ysz$ is the Compton $y$-parameter,
and the integral is taken along the line of sight.
SPT-3G's effective band centers for a non-relativistic SZ signal are 95.69, 148.85, and 220.15~GHz for the three bands. 
The frequency dependence results in a temperature decrement in SPT's lower-frequency channels while the 220~GHz channel is effectively at the SZ null. 

To identify galaxy clusters in SPT-3G maps, a spatial-spectral filter (commonly referred to as a ``matched filter") is applied to the maps~\citep{melin06}.
The matched filter optimally weights the data in spatial and spectral space to more clearly reveal potential clusters (i.e. signals with SZ-like frequency behavior on the angular scale of clusters) while down-weighting other signals and noise in the maps.
Here again we adopt the procedure followed in previous SPT works \citep[see most recently,][for more details]{archipley26}. 
Construction of the matched filter requires both a spectral profile (given in Eq.~\ref{eqn:tsz}) and a spatial brightness profile for the target galaxy clusters.
For the latter, we assume a projected spherical $\beta$-model as described in \cite{cavaliere76}:

\begin{equation}
\Delta T = \Delta T_{0}(1+ \theta^{2}/\theta_\mathrm{c}^{2})^{-(3\beta-1)/2} ,
\label{eqn:beta}
\end{equation}
where $\Delta T_{0}$ is a normalization free parameter,
$\beta$ is fixed to $1$,
and $\theta_\mathrm{c}$ is varied in quarter arcminute steps between $0\farcm{25}$ and $3\arcmin$ with additional filters (an expansion over previous SPT works) at $0\farcm{125}$, $4\arcmin$, $5\arcmin$, and $6\arcmin$.

\subsection{Candidate List Construction and Cleaning}
We construct our preliminary list of cluster candidates in 2 steps: a first step in which we identify the highest-significance candidates, and a second step in which candidates down to our selection threshold are identified in maps with high-significance clusters removed. This list is then cleaned of spurious detections associated with artifacts from our treatment of bright sources and massive clusters to produce the final cluster candidate sample. 

Cluster detection in each step broadly proceeds as follows. 
We create SZ-matched-filtered maps from the combination of the 95, 150, and 220 GHz maps for each of the 16 different filter scales, and we identify candidates as peaks in each of the filtered maps.
We record all candidates above a cutoff in our primary survey observable, $\xi$, which is defined for each map location as the maximum detection significance across all filter scales. 
We use a 3\arcmin \ radius to associate detections from peaks in the individual filtered maps to a common candidate.

In the first step, we identify high-significance cluster candidates with $\xi>$ 20. 
Next, before running the cluster finder to lower detection significance, we model and remove signals from the temperature maps for these high-$\xi$ detections as, 
similar to the bright emissive sources discussed in Sec. \ref{sec:point_source_removal}, compact features in the maps associated with these candidates can induce spurious detections owing to the map filtering. 
To model the cluster signals, we construct a template Compton-$y$ map by placing at the candidate locations $\beta$-profiles with amplitude and core radii matching those of the detections. 
We use SPT-3G's effective frequencies for the non-relativistic SZ spectrum to convert this template into temperature units (Eq.~\ref{eqn:tsz}), and then convolve these template maps with the SPT beams and transfer function (the latter of which encapsulates the effects of the TOD filtering during mapmaking).  
These templates are then subtracted from  the temperature maps  before identifying candidates to lower significance.  
We note that there is a well known amplitude-size degeneracy in the recovery of cluster profile parameters for experiments with arcmin-scale resolution \citep[see e.g., ][]{benson04,planck13-29} but---for our purposes of simply removing the small scale cluster signals to mitigate spurious candidates---we find this procedure works well with the exception of our highest-significance detections at $\xi\gtrsim100$ for which extra care is required (discussed more below). 

In the second step, we then apply the matched filters to the source- and cluster-subtracted maps and detect candidates down to $\xi=4$, the selection threshold for this catalog\footnote{We also tried a similar cluster template subtraction procedure at various $\xi$ thresholds to test if there was a significant change in the candidate population---e.g., if this would improve cluster recovery in a similar way that the \texttt{CLEAN} \citep{hogbom74} algorithm improves point source detection---but found that our simple source association radius of 3\arcmin \ produced comparable results at much lower computational cost.}.
We merge this list with that of the higher-significance candidates to form the full preliminary candidate list.  

\begin{figure}[htbp]
    \centering
    \includegraphics[width=0.5\textwidth]{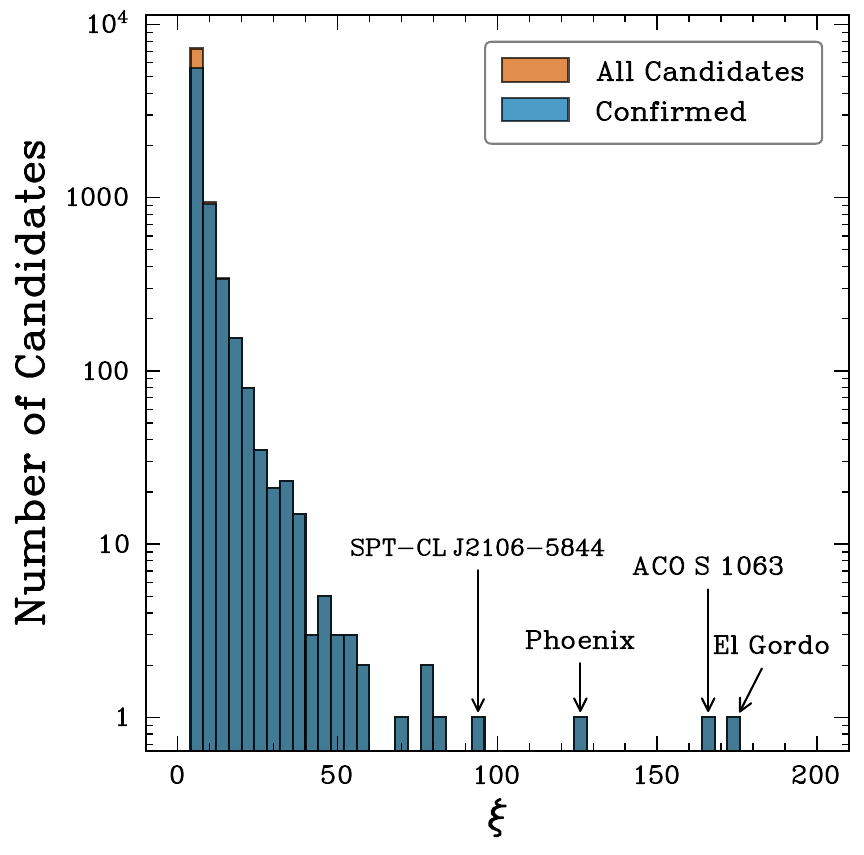}
    \includegraphics[width=0.5\textwidth]{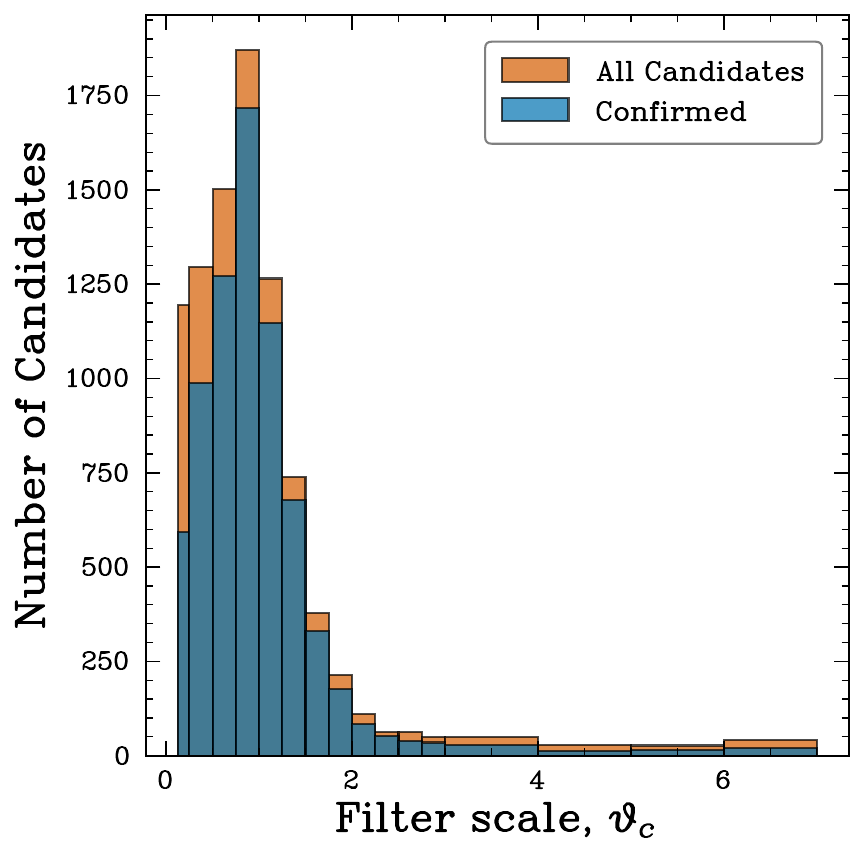}
    \caption{Distribution of cluster sample properties in terms of detection significance, $\xi$, (\textit{top}) and corresponding  filter scales (\textit{bottom}). In orange we show the full sample and, over-plotted in blue, the clusters confirmed by the optical analysis as discussed in Sec. \ref{sec:cluster_confirmation}. In the top panel we also highlight several massive well-known clusters in the SPT footprint. The distribution in $\theta_\textrm{c}$ of the predominately low-$\xi$ unconfirmed candidates---with small exceptions at high $\xi$ owing to optical incompleteness---matches expectations based on the noise properties of the maps, see Sec. \ref{sec:purity_completeness}.} 
    
    \label{fig:xi_distribution}
\end{figure}

This preliminary list is then visually inspected in both raw temperature and cluster-filtered maps.
We identify a small number of spurious detections associated with extended low-redshift galaxies \citep[typically NGC galaxies;][]{dreyer88}  or near the  centers of 
bright emissive sources detected at $>6$ mJy ($\sim 20 \sigma$) at 150~GHz. 
To remove these artifacts we add additional masks post filtering of 4\arcmin \ radius around these locations\footnote{We note this 6 mJy threshold was the same threshold adopted in previous SPT analysis, however, as our subtraction procedure has successfully accounted for the emissive source decrement wings, we are able to apply masks in postprocessing rather than prior to filtering, reducing the overall mask sizes per source from 8' to 4'.}. 
Finally, as noted above, the clusters detected at $\xi\gtrsim100$---particularly the ``El Gordo" \citep{menanteau10} and  ``Phoenix'' \citep{williamson11} clusters---still produce spurious detections following our template subtraction procedure.
We flag and remove a small number of clearly spurious detections by hand within 8\arcmin \ of these clusters. 
In total, including both the large masks over sources interpolated in mapmaking and the masks discussed here, we mask \maskfrac \ of the SPT-3G main footprint. 
The final SPT-3G cluster sample contains \ncand \ candidates detected at $\xi>4$ drawn from \skyarea~\sqdeg.  In Fig.~\ref{fig:xi_distribution} we plot the recovered $\xi$ and $\theta_\textrm{c}$ distributions of the sample. 

\subsection{Integrated Comptonization, $\bigysz^{0\farcm75}$}
While the  SPT selection observable, $\xi$, has been shown to be an excellent low-scatter mass proxy \citep{dehaan16, bocquet19}, it is not straightforward to directly combine with other datasets for constraining ICM properties \citep[e.g.,][]{ruppin21} or in comparisons to other survey and simulation measurements. 
As such, we also provide a measure of the integrated Comptonization of each cluster candidate in the SPT-3G sample. 
The reported observable, $\bigysz^{0\farcm75}$, is defined as 
\begin{equation}
\bigysz^{0\farcm75} = 2\pi \int_0^{0\farcm75} y_{0} (1+ \theta^{2}/\theta_\mathrm{c}^{2})^{-1} \theta d\theta
\label{eqn:bigysz}
\end{equation}
for our adopted pressure profile models (Eq. \ref{eqn:beta}), where $y_{0}$ is the peak Comptonization. 
We set the integration radius of the profiles to $0\farcm75$ following \citet{saliwanchik15} which analyzed mock SPT observations of maps containing thermal SZ \citep{shaw10}, astrophysical foregrounds, and realistic noise and observational effects.
The authors found this aperture to provide robust measurements of the Comptonization given SPT's beam size and that the scatter of the SZ observable $\bigysz^{0\farcm75}$ with respect to the simulated halos' virial masses is $27\pm2\%$; this is comparable to the scatter of $\xi$ with mass for the same halos.

Following  \citet{reichardt13, bleem15b} we estimate these observables for each cluster candidate using a gridded parameter search in $y_{0}$, $\theta_\mathrm{c}$, and position. 
We restrict the recovered locations to be within $\pm 1\farcm5$  from the each candidate's matched-filtered detection position and explore $\theta_\mathrm{c}$ values corresponding to physical core radii, $r_\mathrm{core}$,  ranging from  $50 \ \mathrm{kpc} \le r_\mathrm{core} \le 1 \ \mathrm{Mpc}$. 
For the conversion from physical to angular scales we use the cluster redshifts discussed below in Sec. \ref{sec:cluster_confirmation} and adopt a redshift of $z=1.5$ for unconfirmed cluster candidates.

\subsection{Purity of the SZ Candidate List} 
\label{sec:purity_completeness}
We run our cluster detection procedure on simulations to estimate expected numbers of spurious candidates and, by combining these estimates with our real detection lists, the overall purity of our sample. 
Following methods established in \cite{reichardt13, huang20, bleem24}, these simulations capture cluster signals, astrophysical and instrumental noise, as well as the effects of filtering applied to SPT-3G data. 
We refer readers to \citet{kornoelje26} for the most recent updates to this process and provide a brief overview below. 

\subsubsection{Simulation Construction}
The simulated SPT-3G maps are constructed using Gaussian realizations of the CMB, kinematic SZ \citep[kSZ;][]{sunyaev80b}, and cosmic infrared background (CIB); simulations of non-Gaussian signals from thermal SZ and radio galaxies; and instrumental noise realizations derived from the data itself.
The CMB is a realization of the $TT$ power spectrum from the best-fit lensed \(\Lambda\)CDM model from the \textit{Planck} 2018 results of \citet{aghanim20}, and maps of the kSZ and dusty sources are constructed following results from \citet{reichardt21}. 
The SZ maps are constructed using the techniques of \citet{flender16} on a halo lightcone from the Outer Rim simulation \citep{heitmann19} with maps normalized such that the SZ power spectrum amplitude is consistent with \citet{reichardt21} at $\ell=3000$. 
Radio galaxies are modeled using the $\frac{dN}{dS}$ flux distribution of \cite{lagache20} with  
source maps generated at 150~GHz, and scaled to 95~GHz and 220~GHz assuming spectral indices consistent with \citet{everett20}.  
Sources are populated up to a maximum flux of 30~mJy at 95~GHz to match the interpolation threshold of the SPT-3G maps (see Sec. \ref{sec:obs_and_mapmaking}). 
Finally, instrumental noise maps are constructed directly from SPT-3G data by creating signal-free coadds of individual map observations (by multiplying random halves of the observations by -1). Due to the intrinsic variability of AGN and telescope pointing jitter, the instrumental noise maps are masked with the same point source mask used in cluster-finding, as imperfect cancellation at the locations of bright sources can cause residuals in the noise maps leading to excess spurious detections. 
In total we constructed 3 independent simulations of the SPT-3G Main field. 

\begin{figure}
\begin{center}
\includegraphics[width=0.99\linewidth]{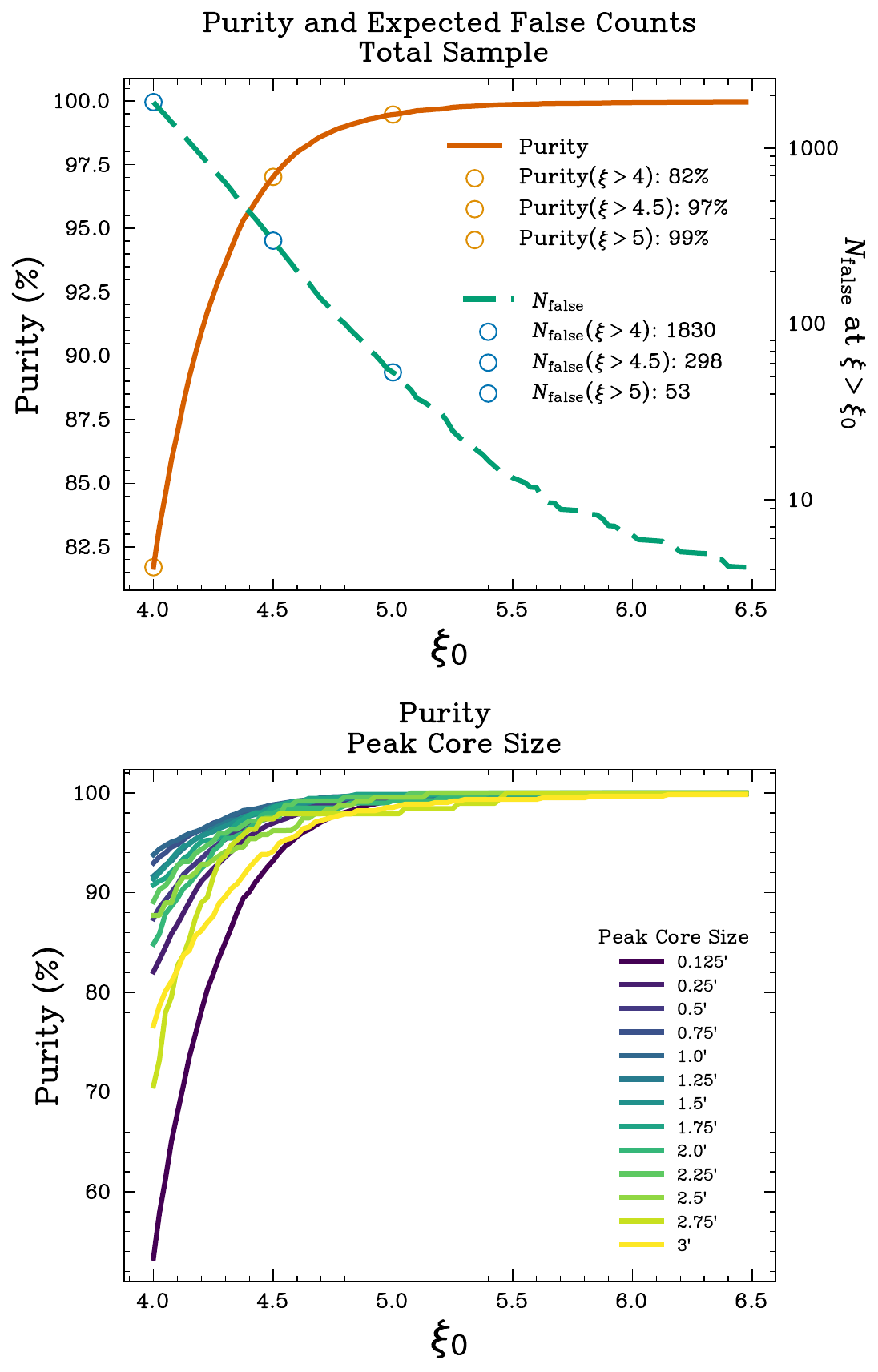}
\caption{\textit{Top}: The expected purity (orange) and number of false detections (green) for the SPT-3G Main field cluster catalog as a function of minimum $\xi$ without additional optical follow-up. We find an expected purity of $82 \%$ above $\xi_\textrm{min} = 4$, and $99 \%$ above $\xi_\textrm{min} = 5$.   
\textit{Bottom}: The expected purity of the cluster sample as a function of filter core size, $\theta_c$. Here the label ``Peak Core Size" reflects that the simulated clusters were detected at a $\xi$ corresponding to the listed $\theta_c$ (i.e., this is not plotting the purity of the  cluster finder run at a single filter scale). }\label{fig:Purity}
\end{center}
\end{figure}

\subsubsection{Purity Estimation}
The  simulated maps are then treated with the same cluster detection procedure as the data. This includes source subtraction, two-pass cluster detection with high-$\xi$ cluster subtraction, and additional post-detection point source masking. 
Once a candidate list is produced for each simulation realization, beginning with the highest-significance candidate, candidates are then cross-matched to known locations of massive halos from the Outer Rim simulation using a 2\arcmin \ association radius. 
To determine the probability of random association with a cluster of a given mass, $f_R(M)$, this process is repeated for 100,000 random sight lines. The number of false associations, $N_{\mathrm{FA}}$, above a minimum significance $\xi_\textrm{min}$, is then calculated as:

\begin{equation}
    N_{\text{FA}}(M, \xi_{\text{min}}) = N_{\text{obs}}(M) - p(M, \xi_{\text{min}})N_{\text{cand}},
\end{equation}
where $N_\textrm{obs}(M)$ is the number of candidate associations, $N_\textrm{cand}$ is the number of cluster detections for each simulated map, and $p(M,\xi_{\text{min}})$ is the fraction of true
associations to total candidates, defined as:

\begin{equation}
p(M, \xi_{\text{min}}) = \frac{N_{\text{obs}}(M) - N_{\text{cand}}f_R(M)}{N_{\text{cand}}[1 - f_R(M)]}.
\end{equation}

The estimated purity of the final SPT-3G catalog is then calculated by taking the difference between the total number of candidates in the catalog and $N_{\text{FA}}$, divided by the total number of candidates. 
Results of the purity estimates are shown in Fig.~\ref{fig:Purity}. 
We find an expected purity of $82 \%$ above $\xi_\textrm{min} = 4$, and $99 \%$ above $\xi_\textrm{min} = 5$. 
While constructed from maps of similar depth to the SPT Deep field presented in \cite{kornoelje26}, here the intrinsic purity is lower at $\xi_\textrm{min}=4$ (though comparable at $\xi_\textrm{min}=5$) owing to the inclusion of additional filter scales in this work. 
In the bottom panel of Fig.~\ref{fig:Purity} we break out the expected purity as a function of filter scale $\theta_c$. 
While the smallest filter scale adds sensitivity to smaller-scale (lower-mass/higher-$z$) clusters, there are more independent filtered pixels in the field, resulting in more noise fluctuations (and thus on balance here lower purity, though we note this ratio is dependent on the number of true clusters in the map and will vary based on map depth). The lower purity of this filter scale is also reflected in the fraction of optically confirmed clusters for this filter scale plotted in the bottom panel of Fig \ref{fig:xi_distribution}. 
Conversely, the decline in purity at the largest filter scales is driven by the relative decrease in true clusters with large angular extent. 
As we discuss in the next section, optical cleaning will increase the overall sample purity, especially at low $\xi$.

\section{Redshifts} 
\label{sec:cluster_confirmation}
Robust redshift measurements are essential for enabling scientific studies with galaxy cluster samples. 
For ICM-selected samples, optical and infrared (IR) follow-up observations are critical for providing these redshifts and additionally, for lower-significance cluster candidates, confirming the detections as bona fide galaxy clusters via the presence of significant  galaxy overdensities in color/redshift space at the candidate locations \citep{high10, song12b, klein18, kluge24}. This confirmation process can also be used to exert further control over the purity of the cluster sample as a function of detection threshold. 

To achieve this, we make use of data from the DESI Legacy Surveys DR10 \citep[LS-DR10;][]{dey19} in the optical, FourStar \citep{persson13} imaging in the near-IR, and \textit{Spitzer}/IRAC \citep{fazio04} and WISE \citep{wright10, meisner22} in the IR. We use red sequence techniques \citep{gladders00,rykoff14} to obtain redshifts and richnesses in optical and near-IR data. To estimate redshifts and confirm cluster candidates at high redshifts with WISE and \textit{Spitzer}, we use the “1.6 \textmu m Stellar Bump" method \citep{papovich08, muzzin13b, gonzalez19} that utilizes the strong color-redshift dependency of cluster galaxies in the [3.6\textmu m] -  [4.5\textmu m] color between $z=0.8$ and $z=1.6$. 

The footprint of the SPT-3G Main Survey lies completely within the LS-DR10 and the WISE survey and is, besides local masking, fully covered by both surveys.
LS-DR10 provides a uniform processing of images and photometric catalogs from survey data acquired with the DECam imager \citep{flaugher15} on the 4-meter NSF Víctor M. Blanco Telescope at Cerro Tololo Inter-American Observatory.  
The  optical data in the SPT Main field footprint was mainly acquired as part of the Dark Energy Survey \citep{abbott21} but also includes contributions from the DELVE \citep{drlica-wagner22} and DeROSITAS  \citep{zenteno25} programs in its southernmost extent.
We use the multi-component matched filter (MCMF) cluster confirmation technique \citep{klein18,klein19,klein24a} for systematic redshift estimation and confirmation using those survey data, while we use dedicated tools for the \textit{Spitzer} \citep[see most recently][]{kornoelje26} and pointed FourStar datasets. 
In the following subsections, we describe the details of the optical follow-up for each of the datasets in greater detail.

\subsection{MCMF}
\label{sec:MCMF}
The MCMF algorithm was created to confirm cluster candidates identified in large X-ray or SZ surveys using wide-field optical survey data. We closely follow the approach used in previous work  \citep{klein24a,bleem24,archipley26}, which consists of two separate runs of MCMF, one using optically selected galaxies and one using IR-selected galaxies. The optical and IR-based measurements are then combined based on their probabilities of being the true cluster match based on redshift and richness as well as on the separation between SZ-based center and galaxy-based center. Here, MCMF measurements from optically selected galaxies provide the majority of the cluster redshifts and confirmation with a redshift reach typically extending out to redshift $z\sim1.4$, while the WISE-based IR-selected MCMF run is optimized for redshifts $z>1$. The WISE-based  selection typically becomes advantageous over the optically selected LS-DR10 data sets at redshift $z\sim1.1$.

\subsubsection{MCMF Confirmation on LS-DR10 Dataset}
The MCMF pipeline using the LS-DR10 dataset is identical to that used in \citet{klein24b} to identify galaxy clusters in the ACT-DR5 \citep{naess20} dataset. It utilizes redshift- and magnitude-dependent color filters, in $g-r$, $r-z$, and $z-W1$, to search for overdensities of red-sequence galaxies to estimate redshifts and richnesses of candidate optical clusters. In regions with additional $i$-band data available, $r-i$ and $i-z$-band colors are used as well.
Besides galaxy colors, MCMF also employs radial weighting using a projected Navarro–Frenk–White \citep[NFW;][]{navarro96} profile centered at the SZ-selected candidate position, as well as a specific magnitude range depending on redshift.

At a given redshift, MCMF calculates, for each cluster candidate, the color and radially weighted, background-subtracted richness, $\lambda(z)$, within a radius of $R_{500c}$, derived from the SZ-based estimate of the cluster mass. The richness is measured at 300 redshifts separated by an offset of $\Delta z=0.005$ providing a richness distribution as function of redshift, $\lambda(z)$, along the line of sight of the cluster candidate. Clusters are identified as characteristic peaks in the $\lambda(z)$ data, and are subsequently fit by ``peak profiles'' obtained from MCMF measurements towards known galaxy clusters with recorded spectroscopic redshifts. The usage of peak profiles allows the code to deblend individual clusters with overlapping red sequence colors, provide direct calibration to spectroscopic redshifts, and to study higher redshifts where red sequence galaxies are only weakly detected in LS-DR10.
For examples of peak profiles we refer the interested reader to previous MCMF analyses \citep[e.g., Figs. 4 and A2,][]{klein19}.
The so-found peaks with associated redshift and richness are then recorded by MCMF as potential optical counterparts.

To confirm or reject a cluster candidate as a genuine cluster, we calculate the probability that a given optical counterpart is a coincidental overlap with the SZ-based candidate. This can be done if the richness and redshift distributions of the contaminants in the SZ-selected cluster candidate list can be measured.
The MCMF approach uses random positions within the SPT-3G footprint and mock estimates of the SPT detection significance to mimic false detections within the SPT-3G candidate list. This mock catalog is then used as input to MCMF and treated the same way as the true candidate list. In a final step, close matches in position and redshift with the real catalog are excluded from the catalog. This last step ensures that the richness distribution of contaminants and not a mix of true and contaminating optical overdensities is measured. The catalog of redshifts and richnesses along random lines of sight is called the \textit{random catalog}.

Using the richnesses and redshifts of the true candidate list and the random catalog, we define the quantity \fcont, where ``cont'' is short for contamination. To obtain a measurement of \fcont\ for candidate $i$ with redshift $z_i$ and richness $\lambda_i$, we calculate the fraction of optical overdensities from the random catalog around $z_i$ with richness $\lambda\ge\lambda_i$ and divide by the fraction of SPT-3G candidates above the same richness $\lambda_i$,
\begin{equation}\label{eq:fcont}
   f_{\mathrm{cont}}(\lambda_i,z_i)=\frac{\int_{\lambda_i}^{\infty} f_\mathrm{rand}(\lambda,z_i) d\lambda}{\int_{\lambda_i}^{\infty}
f_\mathrm{obs}(\lambda,z_i) d\lambda}.
\end{equation}
The \fcont\ parameter is calculated for each richness peak associated with a candidate. In the case of multiple peaks for the same SPT-3G candidate, the peaks are sorted by \fcont.

\begin{figure}
\begin{center}
\includegraphics[width=\linewidth, ]{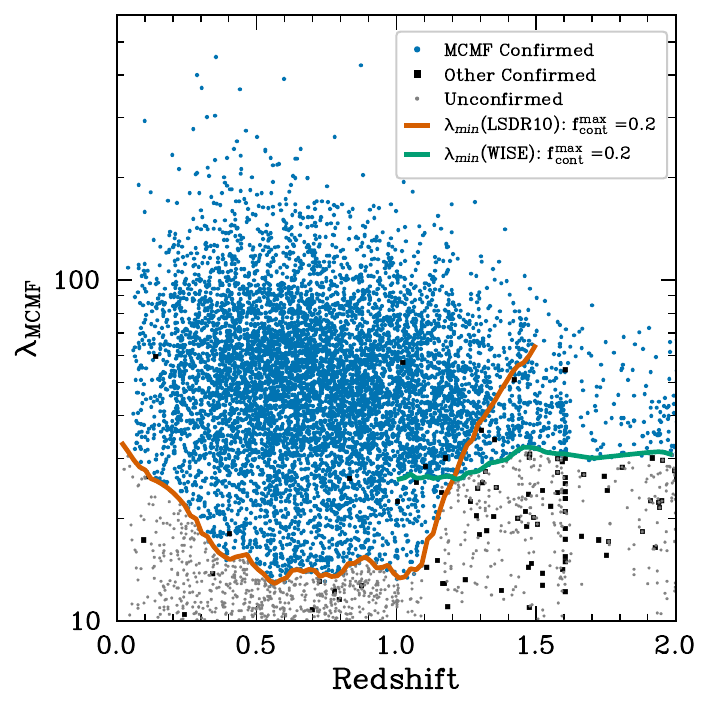}
\caption{MCMF richness $\lambda$ vs. redshift for the SPT-3G candidate list at $\xi>4$. In solid lines we overplot the richness confirmation thresholds of $f_\mathrm{cont}^\mathrm{max}$=0.2 for the LS-DR10 and WISE datasets. In blue we plot clusters confirmed by MCMF in the sample, in black we mark clusters confirmed using other datasets (see Sec. \ref{subsec:rules}.), and in gray SPT candidates with optical associations below our confirmation threshold. The pile-up of ``Other Confirmed" candidates at $z=1.6$ is caused by the assignment of this redshift as a lower limit owing to color-redshift degeneracies in the IR data, see discussion at the end of Sec. \ref{subsec:rules}. }\label{fig:fcont_redshift}.
\end{center}
\end{figure}

A sample of confirmed galaxy clusters can then be obtained by selecting cluster candidates below a threshold value $f_\mathrm{cont}^\mathrm{max}$.
Here the adopted value of $f_\mathrm{cont}^\mathrm{max}$ corresponds to the fraction of contamination remaining in the cluster sample with respect to the initial amount of contamination of the candidate list. If the initial contamination is known, the remaining contamination of the sample after imposing a cut at $f_\mathrm{cont}^\mathrm{max}$ is then simply the product of the two values.
Our baseline catalog generally adopts a contamination fraction $f_\mathrm{cont}^\mathrm{max}$=0.2, except in the specific cases noted below. 
In Fig.~\ref{fig:fcont_redshift} we plot the redshift-$\lambda$ distribution of the SPT cluster sample and overplot the \fcont =0.2 threshold in this plane. Beyond the intrinsic cluster sample properties, the richness-redshift dependence of \fcont \ depends on a number of observational factors (such as the depth of the photometric catalogs for red-sequence/IR galaxies) and astrophysical factors (such as the color-luminosity distribution of  unassociated field galaxies, and evolution in the luminosity functions of clusters). 

\subsubsection{MCMF on WISE}\label{sec:HIGHZ}

The MCMF implementation using WISE-selected galaxies and additional LS-DR10 optical photometry
is identical to that used in the previous work on ACT-selected and SPT-3G-selected clusters \citep{klein24b,archipley26}.

As the input galaxy catalog we use a version of the unWISE catalog \citep{meisner22}, which incorporates eight years of $W1$ and $W2$ band WISE imaging data from the NEOWISE-R phase in addition to data from the main mission \citep{wright10}. The WISE data cover the entire sky, enabling the matching of unWISE and LS-DR10 catalogs for optical/IR photometry of WISE sources over $\sim25,000$~deg$^2$ of extragalactic sky, fully including the SPT-3G survey area. Besides the WISE-only color ($W1-W2$), optical-to-WISE galaxy colors (e.g., $z$-$W1$) also exhibit a strong redshift dependence, allowing the rejection of foreground galaxies and providing cluster photometric redshift estimates at high redshifts. 

Estimation of cluster richness is more complicated in the WISE+LS-DR10 dataset than in the optical dataset alone. 
At high redshifts, not all IR galaxies are detected in LS-DR10. Additionally, for cluster galaxies, the $W2$ data is typically shallower than $W1$ leaving a significant fraction of $W1$ detections without a measurement in the $W2$ band. LS-DR10 and both WISE bands also do not have identical survey masks, necessitating additional corrections in combined richness estimations. To calculate the richness we therefore have to trace the coverage of the data for different combinations of datasets, in decreasing order of available photometry per galaxy: LS-DR10+$W1$+$W2$ (hereafter ALL), LS-DR10 + $W1$, $W1$+$W2$, and $W1$ only. Every galaxy is assigned to one of these combinations. The final richness, $\lambda_\mathrm{HZ}$, is calculated as the sum of the richnesses of the different subsets of galaxies with different data coverage. It is given as
\begin{equation}\label{eq:richnto}
  \begin{split}
    \lambda_\mathrm{HZ}(z) =& \lambda_\mathrm{ALL}(z) + \lambda_\mathrm{LS-DR10+W1}(z) \\
    &+ \lambda_\mathrm{W1+W2}(z) + \lambda_\mathrm{W1}(z),
  \end{split}
\end{equation}
where the individual richnesses are defined in the same manner as the standard MCMF richness \citep[see][]{klein18,klein19}, with the color-weights depending on the availability of the bands ($i,z,W1,W2$ for $\lambda_{\mathrm{ALL}}$, $i,z,W1$ for $\lambda_{\mathrm{LSDR10+W1}}$, $W1$, $W2$ for $\lambda_{\mathrm{W1+W2}}$, and no color weight for $\lambda_{\mathrm{W1}}$).

Similar to the optically based MCMF code using LS-DR10, the richness is calculated at equally spaced redshifts, but here in the redshift range $0.63<z<2.0$. Peaks are then subsequently identified in the $\lambda(z)$ data in a similar way as for the standard approach using LS-DR10. The WISE-based MCMF implementation is run on all candidates as well as the same random catalog used for the MCMF run on LS-DR10 data, and measurements of \fcont\ are obtained. Although there is significant overlap in redshift between the LS-DR10 and WISE implementations of MCMF, we only use the WISE implementation at redshifts $z>1$.

\textit{Limitations of WISE}:
The large point-spread function of the WISE imaging ($\sim6^{\prime\prime}$) is one of the key limitations in the use of WISE data for cluster confirmation. The unWISE sample features improved modeling of crowded regions compared to previous work \citep[such as  ALLWISE,][]{wright19} but source blending, especially in the dense regions of clusters at high redshift still causes significant limitations.
Blending reduces richness counts and (at some level, though the background is also impacted) cluster contrast at high redshift,  increasing contamination rates and impairing cluster confirmation. We leverage the limited targeted higher-resolution observations from \textit{Spitzer} and FourStar data discussed below for additional cluster confirmation as well as to assess the WISE confirmation completeness at the highest redshifts. 

\subsubsection{Combining MCMF on LS-DR10 and WISE}

The MCMF measurements using LS-DR10 and WISE are combined based on their measurements of \fcont \ for individual optical/IR counterparts. We use here the run on LS-DR10 as a baseline and update the best counterpart with the WISE-based measurement if the counterpart exceeds a redshift of $z=1$ and shows a measured \fcont \  below the best counterpart from LS-DR10. Further complications arise as massive clusters at low redshift ($z\lesssim0.2$) can create spurious signals in the WISE-based MCMF measurements as faint low-$z$ cluster members leak into the WISE galaxy sample. To avoid overriding the LS-DR10 MCMF measurement by a spurious high-$z$ measurement, the high-$z$ measurement is not considered as the primary counterpart if a massive low-$z$ cluster ($f_\mathrm{cont}<0.14$) is found or if a similar good counterpart ($\Delta f_\mathrm{cont}<0.05$) is found to be 0.5 arcmin closer to the SZ position than the high-$z$ counterpart.

\subsection{Spitzer/IRAC} 
For a subset of the SPT-3G candidates there is additional deeper and higher-resolution IR data from \spitzer /IRAC \citep{fazio04}. 
These data are sourced from two separate efforts: targeted follow-up campaigns of high-redshift cluster candidates from SPT-SZ and SPTpol \citep{bleem15b, bleem24} and overlap with the 94 \sqdeg \  \textit{Spitzer}-SPT deep field \citep[SSDF;][DOI:10.26131/IRSA432.]{ashby13}. 
Redshifts from \textit{Spitzer} data also are derived using the 1.6~$\mu$m bump method.  
We do not measure new redshifts in this work, but rather use redshifts previously measured in prior SPT catalog releases. For more details on the estimation of these redshifts, see \citet{bleem24} and \citet{kornoelje26}.

\subsection{FourStar}\label{subsec:fourstar}
\begin{figure}[th!]
\begin{center}
\includegraphics[width=\linewidth, ]{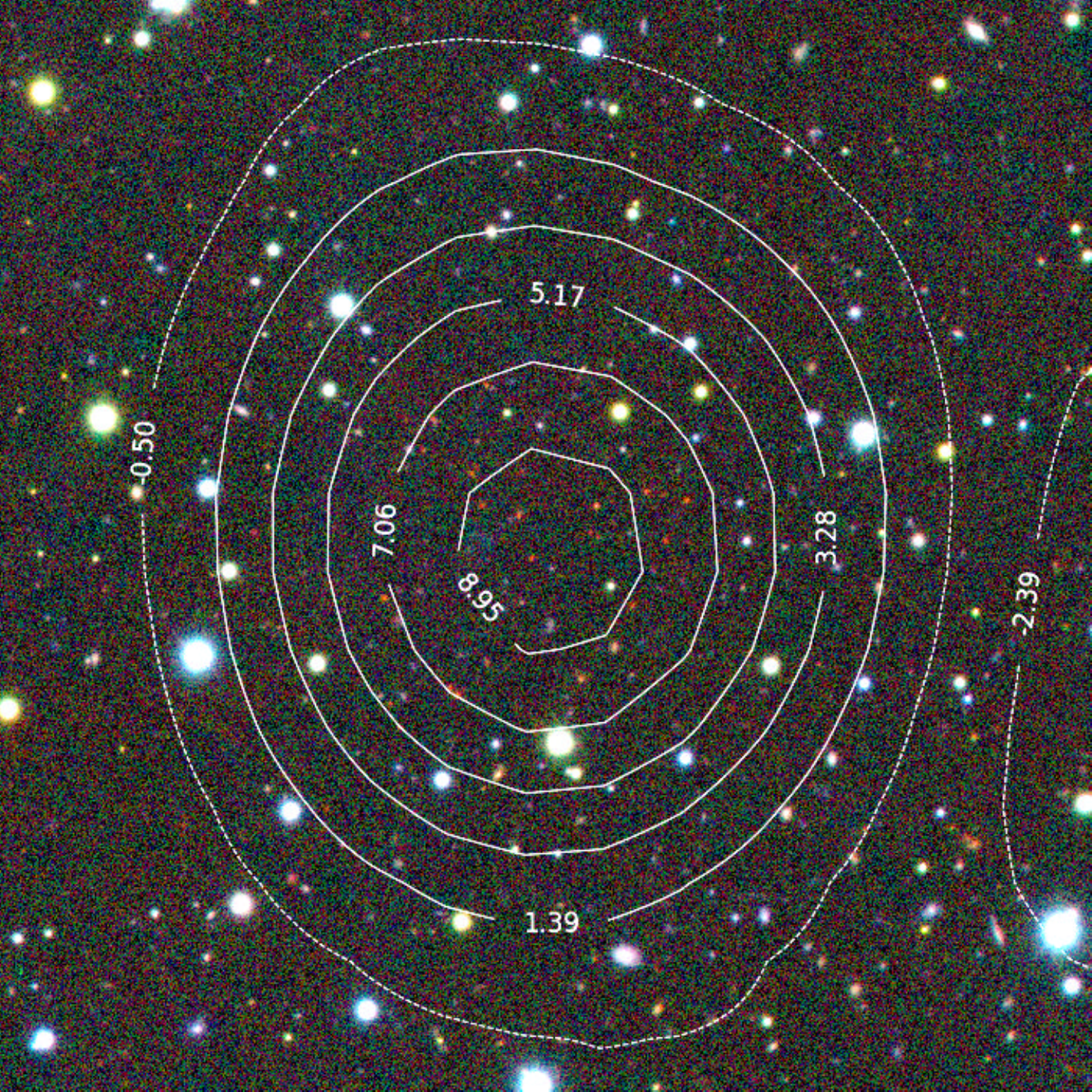}
\caption{SPT-CL~J2128.4$-$5658.0, a $\xi=10$ cluster at $z\gtrsim1.6$ confirmed using the combination of WISE and Magellan/FourStar data. The 3\arcmin \ diameter \textit{rgb} image is constructed from FourStar/J-band and LS-DR10 \textit{z-} and \textit{r-}band data; SZ detection countours are overlaid on the image. }\label{fig:FourStar}
\end{center}
\end{figure}
The depth of the SPT-3G survey results in  significant numbers of high-redshift clusters detected at high confidence. 
As we discuss in the next subsection, a non-negligible number of these clusters are not (or only weakly) detected in LS-DR10 and WISE data, necessitating additional follow-up observations for cluster confirmation. 
We used the FourStar \citep{persson13} near-IR imager on the 6.5 m Magellan/Baade telescope to obtain observations of a subset of such SPT-3G cluster candidates.

The FourStar observations targeted exposure times of 900 s in $J$-band, sufficient to detect luminous red-sequence cluster members at $z \sim 1.5$ in bright time at better than 0\farcs{7} seeing. 
While weather conditions over a series of observing nights in 2021-2025 were variable, we obtained usable data for $\sim150$ candidates (including our primary targets at typically $\xi>5$ based on preliminary-to-final versions of the SZ candidates list and  lower-significance candidates also within the 10\farcm{9} $\times$ 10\farcm{9} field of view).

For data acquired prior to May 2023, we reduced the images using the \texttt{PHOTPIPE} pipeline \citep{rest05a, garg07, miknaitis07} following the process detailed in \citet{bleem15b}.
For observations acquired after May 2023 we use the FourStar Reduction (\texttt{FSRED}) pipeline provided by Carnegie Observatories\footnote{\url{https://instrumentation.obs.carnegiescience.edu/FourStar/SOFTWARE/reduction.html}} to reduce the imaging data. The pipeline performs linearity correction, flat fielding, dark subtraction, background subtraction, astrometric calibration, image resampling, and final co-addition. Individual exposures are grouped into $\sim$10-minute intervals, with background subtraction applied to each interval. A global astrometric solution is solved for by \texttt{SCAMP} \citep{bertin06} based on source positions extracted using \texttt{SExtractor} \citep{bertin96}.  Images are then resampled and reprojected onto a common grid using \texttt{SWarp} \citep{bertin10}. Images are co-added with inverse-variance weighting by \texttt{IRAF} \citep{tody93} using \texttt{imcombine}.  Photometric zero points are determined by matching stars detected in the final co-added mosaic to the 2MASS point source catalog \citep{skrutskie06}.

 Following reduction of the images, we construct \textit{rgb} images combining the \textit{J}-band data with \textit{r-} and \textit{z-}band data from LS-DR10. Images with SZ detection contours  overlaid are inspected by multiple team members for clear signatures of high-$z$ clusters. We flag clusters with pronounced over-densities of red galaxies. We then cross-match these systems against the MCMF redshift results, finding most of the FourStar flagged systems to have high-redshift counterparts in MCMF, but at  contamination fractions higher than our adopted threshold of $f_\mathrm{cont}^\mathrm{max}$=0.2.
 Following this process, we include   \fourstarconfirmed \ additional clusters as confirmed, and adopt the MCMF WISE-based redshifts for these systems; the clusters are tagged as being confirmed by a combination of the datasets. Beyond this additional confirmation, the FourStar data provides higher confidence in the MCMF-redshift assignment for a number of high-$\xi$ systems near the $f_\mathrm{cont}^\mathrm{max}$ threshold and is being used for targeting in ongoing spectroscopic follow-up efforts. 
 We show in Fig.~\ref{fig:FourStar} an example of one of these newly-discovered high-$z$ clusters confirmed with the additional FourStar data. 

\subsection{Characterization of WISE High-$z$ Confirmation}\label{subsec:wise_complete}
While the relative depth of WISE data is variable across the SPT-3G footprint \citep[see e.g., ][]{meisner22}, we provide a rough assessment of high-$z$ confirmation with WISE by comparing MCMF results using LS-DR10 and WISE to an analysis based on \textit{Spitzer} confirmations at high $z$. 
For this comparison, we use a sample of 442 confirmed galaxy clusters detected in the 100 \sqdeg \ SPT Deep field with redshifts estimated using DES and the \textsc{redMaPPer} \citep{rykoff14} algorithm at low $z$ and \textit{Spitzer} data in the SSDF \citep{ashby13} at high $z$ \citep{kornoelje26}. 
The SPT Deep field is of comparable depth to the data used in this work and so samples the population of high-$z$, low-mass clusters explored here. 
Of the 442 confirmed SPT Deep clusters, there are matches in our SZ candidate list for 392  within 1$\farcm{5}$. Of these, 366 (93\%) of the clusters meet the MCMF confirmation threshold. Two of these clusters (0.5\%) were flagged as having incorrect lower-$z$ counterparts assigned by MCMF. Of the 7\% unconfirmed, 25/26 were at $z>1$, with 17 of these at $z>1.4$. 
The remaining low-$z$ system was not confirmed by MCMF as the algorithm excluded a large fraction of the optical/IR data in the cluster's proximity owing to potentially corrupted imaging data.\footnote{Such corruption typically occurs near saturated sources or other artifacts in optical images resulting in local regions of problematic data, with different analysis tools having different criteria for data exclusion. 
Both the MCMF and \textsc{redMaPPer} algorithms generate maps of such regions and we use these maps to quantify the fraction of sky with corrupt data around candidates within fixed angular apertures. We use the term \textit{optical mask fraction} to describe these values, and can refer to cluster candidates whose centers fall in optically corrupted regions as \textit{optically masked}.}

The SPT Deep field is roughly 1/16 the area of the Main field.
Using the 26 clusters at $z>1$ unconfirmed by MCMF in the Deep field (5 at $\xi>6$ where the SZ sample is expected to be $\sim100\%$ pure), we would expect 416$\pm{81}$ high-$z$ unconfirmed clusters at $\xi>4$ (80$\pm35$ at $\xi>6$) without additional follow-up. More closely examining the $\xi>6$ subset, our additional dedicated (N)IR analyses confirm 72 systems leaving 51 $\xi>6$ candidates to be confirmed. Of these, 20 have optical mask fractions within 1 arcmin $>0.5$ (impeding confirmation at any redshift), leaving 31 additional high-confidence high-$z$ candidates to be confirmed. This total (72+31=103) is in agreement with our extrapolation of WISE incompleteness from the SPT-3G Deep field.   

\subsection{Final Redshift Assignment}\label{subsec:rules}
\begin{figure*}[th!]
\begin{center}
\includegraphics[width=\linewidth, ]{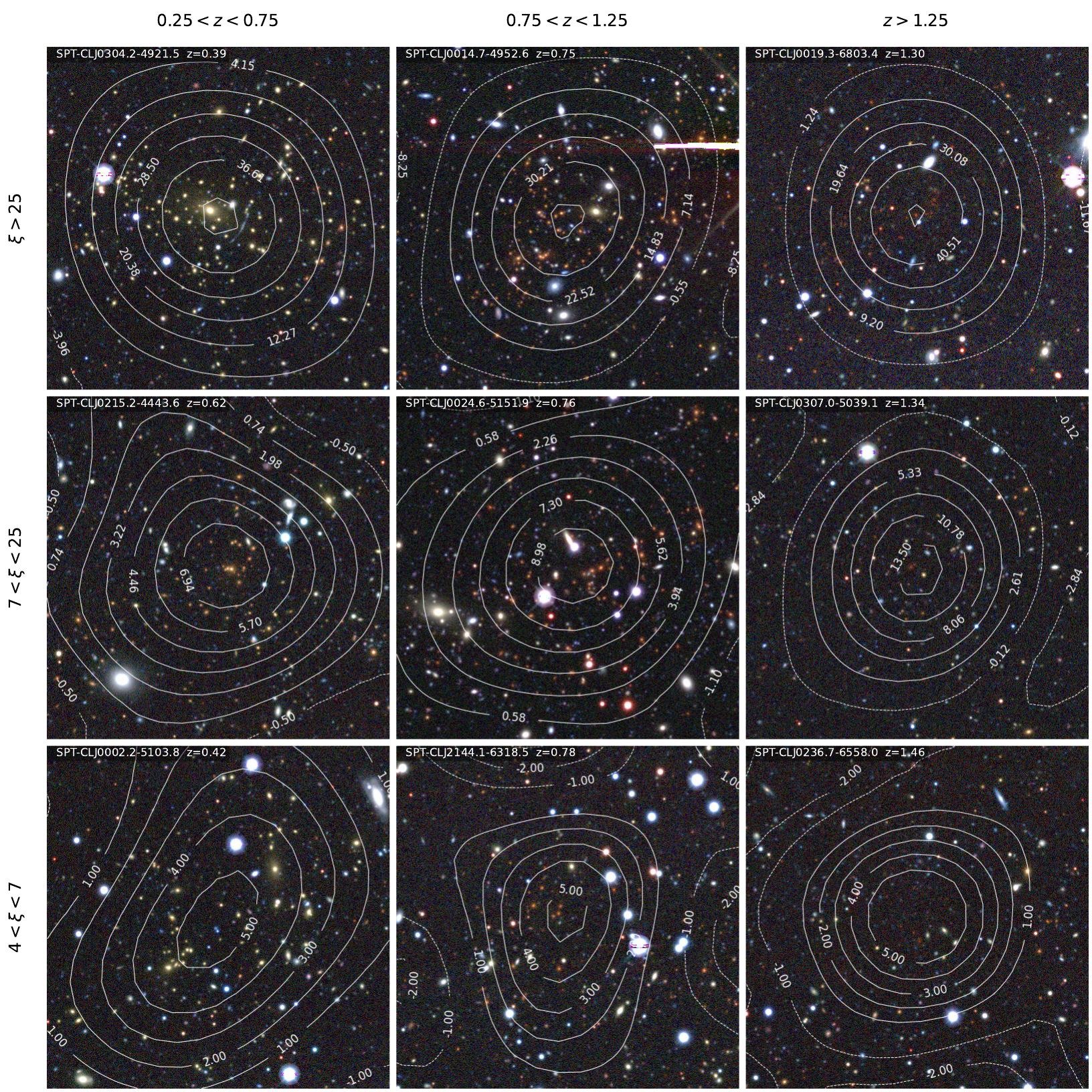}
\caption{DECam imaging of SPT-3G clusters from the LS-DR10 release \citep{dey19} with SZ detection contours overlaid. The \textit{rgb} images were created from $z-,r-,g-$band data and are 3\farcm{8} in diameter. Clusters were selected to illustrate the SZ significance and redshift reach of the sample. The highest-redshift clusters were confirmed with additional observations from \textit{Spitzer}, FourStar (Sec. \ref{subsec:fourstar}), and WISE. }\label{fig:optical_panel}
\end{center}
\end{figure*}

We display a panel of LS-DR10 images with SZ detection contours overlaid for clusters over the $\xi$ and redshift range of the sample in Figure \ref{fig:optical_panel} and plot the redshift distribution of the confirmed clusters in Fig.~\ref{fig:redshift_histogram}. An  entry in the cluster catalog file (see catalog overview in the Appendix \ref{sec:tableoverview}) provides the redshift source for each cluster.  
Final redshifts and cluster confirmation status are determined as follows.  

\begin{figure}
\begin{center}
\includegraphics[width=\linewidth, ]{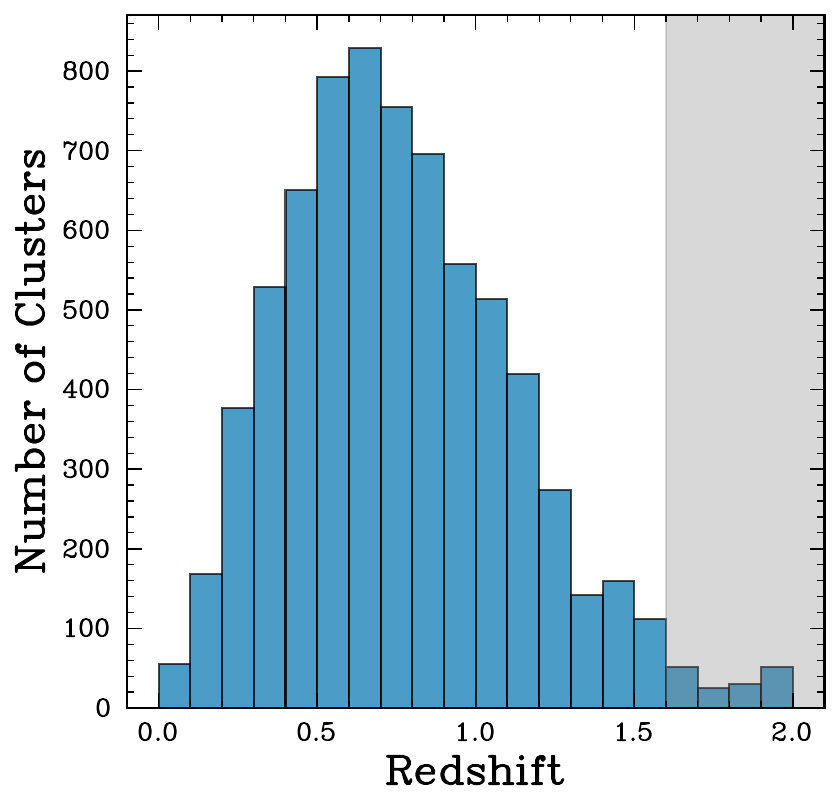}
\caption{Redshift distribution of the \nconfirm \ confirmed clusters in the SPT-3G Main field sample confirmed using optical-IR data as discussed in Sec.~\ref{sec:cluster_confirmation}. As noted in Sec.~\ref{subsec:rules}, owing to color degeneracies in the IR, we assign clusters with redshifts estimated at $z>1.6$ using the 1.6~$\mu$m bump method to a value of $z=1.6$ in the catalog table, but we use their raw best-fit redshifts in the histogram here to illustrate the reach of the SPT-3G sample. This high-$z$ region is denoted by the gray-shaded band.}\label{fig:redshift_histogram}
\end{center}
\end{figure}

\begin{itemize}
\item \textit{MCMF:} Our primary confirmation and  redshift assignment is based on MCMF. We begin by denoting as \textit{confirmed} candidates with $f_\mathrm{cont}$ less than our maximum contamination threshold of 0.2 in either the LS-DR10 or LS-DR10+WISE analysis. Given the expected purity of the SZ catalog (see Sec.~\ref{sec:purity_completeness}), this results in an expected overall sample purity of $>96\%$ at $\xi=4$ and  $>99.8\%$ at $\xi>5$.  

\item \textit{Multiple Counterparts:} For candidates with multiple optical/IR counterparts below the $f_\mathrm{cont}^\mathrm{max}$ threshold, we select the primary optical counterpart based on a combination of matching criteria including the value of $f_\mathrm{cont}$, the distance to the SZ center, and visual inspection of complex systems. We found that only using the lowest $f_\mathrm{cont}$ for this assignment leads to a bias against properly selecting some of the high-$z$ WISE counterparts. We flag systems in the catalog with multiple significant counterparts along the line-of-sight. 

\item \textit{(Near) Infrared:} Our next set of redshift assignments leverage the high-resolution data from FourStar and \textit{Spitzer}. We cross-match SPT-3G candidates to clusters from  previous SPT cluster samples \citep[particularly, ][]{bleem24, kornoelje26} within 1$\farcm{5}$. For systems unconfirmed in MCMF but confirmed by \textit{Spitzer} data in these prior works,\footnote{Note, in these prior works we have adopted a lower $f_\mathrm{cont, Spitzer}^\mathrm{max}$=0.1 given the limited (i.e., SSDF) wide field area for contamination estimation, see \citet{bleem24} Sec 5.2.} we assign the clusters the \textit{Spitzer} redshift, richness, and contamination values previously computed. We also check for systems with significantly discrepant redshifts (e.g., typically \mbox{low-$z$} \ $z<0.8$  assignment from MCMF and \mbox{high-$z$} $z>1$ in clusters with \textit{Spitzer} data). We reinspect each of these systems and assign associated properties from prior works when appropriate \cite[see][Fig. 4, for an image of one such cluster misidentified by the limitations of LS-DR10+WISE data, SPT-CL~J2331.6-5737.2 at $z=1.5$]{bleem24}. 
We note that we choose to keep MCMF+WISE-based richnesses and redshifts when redshifts are consistent to decrease the number of clusters reported with properties differing from our baseline method.  In total 34 clusters have redshifts assigned in this step.   Finally, as discussed above in Sec \ref{subsec:fourstar}, \fourstarconfirmed  \ clusters were confirmed using a combination of FourStar data and MCMF.

\item \textit{Regions with Significant Masking in Optical Data:}
 We loosen our optical data masking criterion and also check an archival optical cluster sample with different optical masking choices around bright stars to enable confirmation of a small number of additional candidates. For this we make use of the \cite{WenAndHan24} catalog (hereafter WH24 and discussed further below in Sec. \ref{sec:optical_ir_comparison}).  We match the optical clusters to SPT candidates using an association radius of 1\farcm{5} and compute contamination fractions for the sample using a million random locations unmasked in DES data. We visually inspect optical images of all candidates for which (1) the MCMF masking fraction exceeds 50\% within 1\arcmin \ or (2) the contamination estimated from MCMF exceeds 0.2 but is below this threshold in the WH24 match.  We mark as confirmed and adopt either the WH24 (or MCMF where available) redshift when there is clear evidence of a rich cluster whose formal confirmation is impeded by the optical mask.  The inspection process adds \nwenhan \ clusters  in areas masked by MCMF and 8 additional MCMF confirmations. We also include for all candidates the total fraction of area masked in LS-DR10 data in MCMF within 1$\arcmin$ and 2$\arcmin$ of the SZ candidate position. 

\item \textit{Spectroscopic Redshifts and Additional Archival Redshifts:} Next, we search the literature and tag those systems with spectroscopic redshifts, in total spectroscopic redshifts are sourced for \nspecz \ clusters. 
Many of these spectroscopic redshifts come from previous dedicated follow-up observations of SZ-selected clusters \citep[e.g.,][]{sifon13, ruel14, bayliss16, khullar19}. 
As an additional check on the high-$z$ confirmation, we repeat our matching and contamination estimation procedure, this time using the clusters from the Massive and Distant Clusters of WISE Survey 2 \citep[MaDCoWS2;][also discussed further in Sec.~\ref{sec:optical_ir_comparison}]{thongkham24}. We inspect candidates with $f_\textrm{cont}<0.2$ in MaDCoWS2 (but unconfirmed in MCMF), adopting redshifts and confirmations for \nmadcows \ additional clusters. 

\item \textit{Scaling Relation Cross Check:} As a final check on our counterpart assignment, we examine systems that are significant outliers in our mass-richness scaling fit, discussed in the next section. Most of these clusters were determined to have low richnesses because of optical masking, but for 5 we found misassignment of a lower-$z$ counterpart to the high-$z$ detection. For 4 of these systems we could adopt a detection from MCMF, but for one (SPT-CL~J0249.6-5138.7 at $z>1.6$) a high-$z$ counterpart was not detected in MCMF but was found in the MadCOWS2  sample. 

\item \textit{IR-Color Redshift Degeneracy: }
Finally, given the flattening of color-redshift relation in IR-based redshifts in the range $1.6 \lesssim z \lesssim2$  \cite[see e.g., Fig. 2 in][]{muzzin13b}, we choose to set redshifts as lower limits at $z=1.6$  for systems with redshifts estimated from this color combination at $z>1.6$.
\end{itemize}

\section{Observable-Mass Relations}
\label{sec:obs_mass}

In this section, we discuss how we relate the SPT cluster detection significance, $\xi$, to the underlying halo mass.
We model the cluster sample accounting for its selection and present constraints on the $\xi$–mass relation and the optical richness-to-mass relation.
For simplicity, we restrict this analysis to cluster redshifts $0.25<z<1$ to avoid modeling the transition from LS-DR10 to WISE at redshift $\sim1$ when performing the optical confirmation (see Sec.~\ref{sec:MCMF}).
Based on these relations, we present mass estimates for all clusters in our catalog.

\subsection{SZ-Mass Relation}

We establish the relationship between the detection significance $\xi$ and halo mass. We then construct a likelihood function for the cluster sample that, together with an assumed fiducial cosmology, allows us to constrain the parameters of the $\xi$-mass relation.

Following previous SPT work \citep[e.g.,][]{vanderlinde10, bleem24}, we model the relationship between the detection significance $\xi$ and the unobserved, unbiased significance $\zeta$ as a normal distribution
\begin{equation}
    P(\xi|\zeta)=\mathcal{N}\left(\sqrt{\zeta^2 + 3}, 1\right).
\end{equation}
The mean of the relation (between $\zeta^2$ and $\xi^2$) is offset by three to account for the three degrees of freedom of the maximization process during cluster detection (two positional degrees of freedom and the choice of filter scale).

Still following previous SPT work, we model the mean unbiased significance $\zeta$ as a power law in mass $M$ and in the Hubble parameter $H$ at the cluster redshift
\begin{equation}
  \label{eq:zeta-mass}
  \begin{split}
    \langle\ln\zeta\rangle =& \ln A_\mathrm{SZ} + B_\mathrm{SZ}\ln\left(\frac{M_{500c}}{10^{14}~h_{70}^{-1}M_\odot}\right) \\
    &+ C_\mathrm{SZ}\ln\left(\frac{H(z)}{H(z=0.6)}\right).
  \end{split}
\end{equation}
Note that because SPT-3G probes the cluster population down to lower masses than previous SPT generations, we lower the pivot mass from $3\times10^{14}\,h^{-1}M_\odot$ to $10^{14}\,h_{70}^{-1}M_\odot = 7\times10^{13}\,h^{-1}M_\odot$. We assume normal scatter in $\ln(\zeta)$ of width $\sigma_{\ln\zeta}$.

We model the subset of the cluster sample with $\xi>5$ and $0.25<z<1$. The Poisson likelihood of this sample of clusters is
\begin{equation}
  \label{eq:abundance_likelihood}
  \begin{split}
    \ln \mathcal L(\vec p) =& \sum_i\ln\frac{\mathrm{d}^2 N(\vec p)}{\mathrm d\xi\,\mathrm d z}\bigg|_{\xi_i,z_i} - \iint d\xi\,\mathrm d z\, \Theta_\mathrm{s}(\xi,z)\,\frac{\mathrm{d}^2 N(\vec p)}{\mathrm d\xi\,\mathrm d z} \\
    &+ \mathrm{const.}
  \end{split}
\end{equation}
with the survey selection function $\Theta_\mathrm{s}(\xi-5,z-0.25)$ and with the astrophysical and cosmological model parameters $\vec p$.
The differential cluster abundance is
\begin{equation}
  \label{eq:dN_dxi_dz}
  \begin{split}
    \frac{\mathrm{d}^2 N(\vec p)}{\mathrm d\xi\,\mathrm d z} = \int \mathrm d\Omega_\mathrm{s}\iint\mathrm{d}M\,\mathrm{d}\zeta \Big[&P(\xi|\zeta)P(\zeta|M,z,\vec p)\\
    & \frac{\mathrm{d}^2 N(\vec p)}{\mathrm d M\,\mathrm dV} \frac{\mathrm d^2 V(\vec p)}{\mathrm dz\,\mathrm d\Omega_\mathrm{s}}\Big]
  \end{split}
\end{equation}
with halo mass function $\frac{\mathrm{d}^2 N}{\mathrm d M\,\mathrm dV}$ and the differential volume element $\frac{\mathrm d^2 V}{\mathrm dz\,\mathrm d\Omega_\mathrm{s}}$ within the survey footprint $\Omega_\mathrm{s}$. For more details, we refer the reader to, e.g., Sec.~VII in \cite{bocquet24a}.

\begin{figure*}[t!]
  \includegraphics[width=\textwidth]{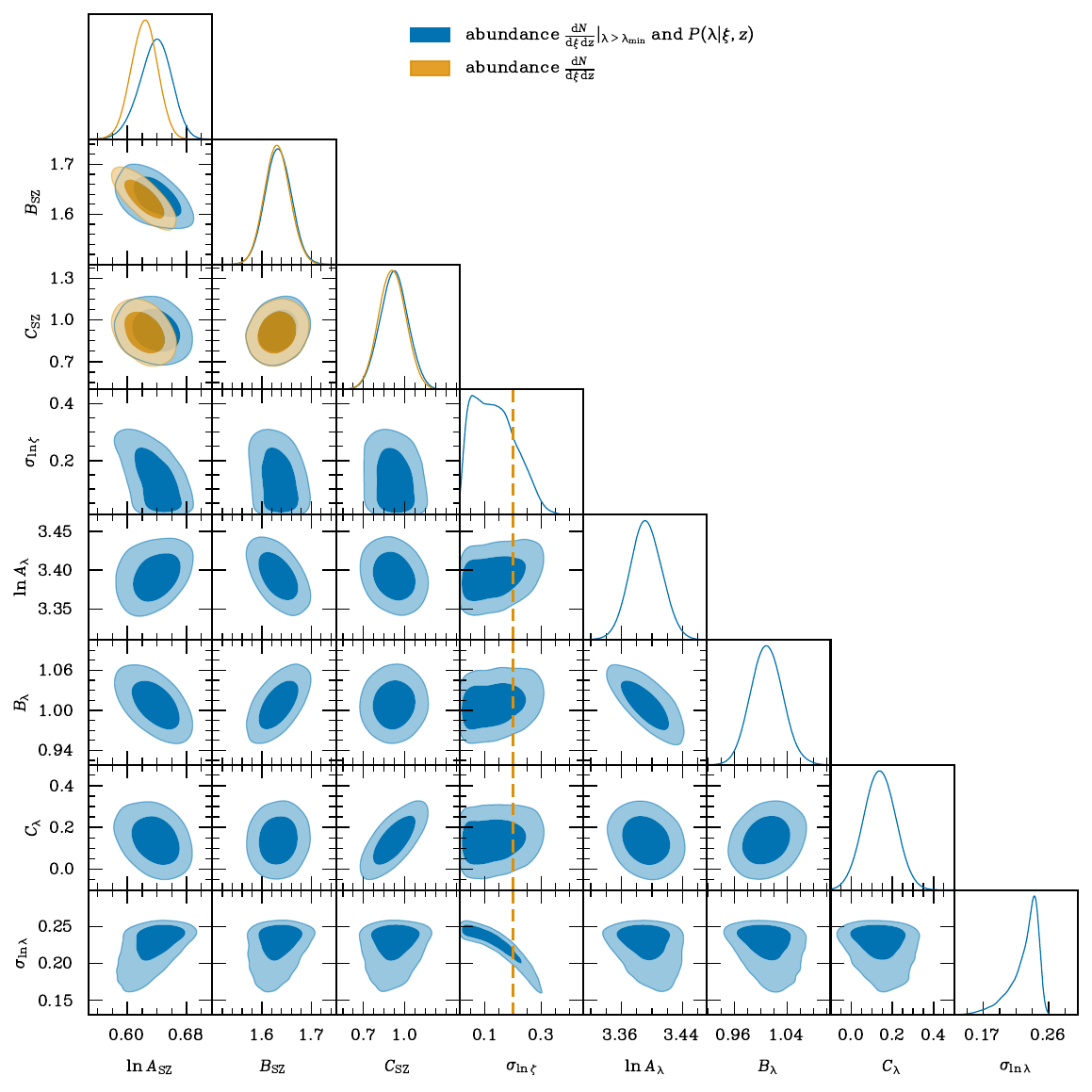}
  \caption{Parameters of the SZ and richness observable-mass relations assuming a fixed reference cosmology (68\% and 95\% credibility). Gold contours show an analysis of the cluster counts where we model the sample selection in detection significance $\xi$ and redshift $z$ (we fix $\sigma_{\ln\zeta}=0.2$ in this case as shown by the dashed line). Blue contours show an analysis of the cluster counts where we model the sample selection in $\xi$ and $z$ but where we also model the optical selection $\lambda>\lambda_\mathrm{min}$. To calibrate the parameters of the $\lambda$-mass relation, this analysis also includes the likelihood of the measured optical richness $\lambda$ given the SZ significance $\xi$.}
  \label{fig:GTC}
\end{figure*}

To remain consistent and comparable with previous SPT cluster catalog publications, we assume the fiducial cosmology described in Sec. \ref{sec:intro}, we fix $\sigma_{\ln\zeta}=0.2$, and we infer the SZ-mass scaling relation parameters $\ln A_\mathrm{SZ}$, $B_\mathrm{SZ}$, and $C_\mathrm{SZ}$ using the likelihood function (Eq.~\ref{eq:abundance_likelihood}) and assuming wide uniform priors.
The recovered parameter constraints are summarized in Table~\ref{tab:obs-mass} and shown in gold color in Fig.~\ref{fig:GTC}.

In previous SPT cluster analyses, for each SPT field, the amplitude $A_\mathrm{SZ}$ was multiplied by the effective field depth $\gamma_\mathrm{field}$ such that the observable-mass relations in all fields were normalized with respect to a common reference (namely the depth of the \texttt{ra5h30dec$-$55} field from the 150~GHz-only SPT-SZ analysis of \citealt{vanderlinde10}).
In this work, we do not apply such a correction (that is, we set $\gamma_\mathrm{SPT-3G Main}=1$), and we recover $A_\mathrm{SZ}=1.86$ (see Table~\ref{tab:obs-mass}). When we adopt a pivot mass $3\times10^{14}~h^{-1}M_\odot$ as in previous works, we obtain $A_\mathrm{SZ}=20.1$. Comparing this number to $A_\mathrm{SZ}=4.08$ for pre-3G surveys, we empirically determine $\gamma_\mathrm{3G~main}=4.99$.
This number closely matches the effective depth of the SPT-Deep field $\gamma_\mathrm{Deep}=4.97$ \citep{kornoelje26}.
For a given cluster, the typical detection significance in the SPT-3G Main data is thus over twice as high as it was in SPTpol 500d data ($\gamma_\mathrm{500d}=2.23$) and over four times as high as in full SPT-SZ data ($\langle\gamma_\mathrm{SPT-SZ}\rangle=1.2$).

\begin{table}[]
  \centering
  \begin{tabular}{l c c}
    Parameter & SZ only & SZ and richness \\\hline
    $\ln A_\mathrm{SZ}$ & $0.62\pm0.02$ & $0.64\pm0.02$ \\
    $B_\mathrm{SZ}$ & $1.63\pm0.03$ & $1.63\pm0.03$ \\
    $C_\mathrm{SZ}$ & $0.91\pm0.10$ & $0.92\pm0.10$\\
    $\sigma_{\ln\zeta}$ & 0.2 & $0.13\pm0.07$ \\\hline
    $\ln A_\lambda$ & \dots & $3.39\pm0.02$\\
    $B_\lambda$ & \dots & $1.01\pm0.02$\\
    $C_\lambda$ & \dots & $0.14\pm0.08$\\
    $\sigma_{\ln\lambda}$ & \dots & $0.23\pm0.02$
  \end{tabular}
  \caption{Constraints on the parameters of the observable-mass relations defined in Eqs.~\eqref{eq:zeta-mass} and \eqref{eq:richness-mass} (mean and standard deviation).}
  \label{tab:obs-mass}
\end{table}

\subsection{Modeling the Optical Confirmation and the Richness-Mass Relation}
\label{sec:richnessmassrelation}

In the previous section, we modeled the cluster sample as being selected based on cuts in significance $\xi$ and redshift $z$.
In a more complete approach, we now also model the optical confirmation via a cut in \fcont\ (see Sec.~\ref{sec:MCMF}).
To do so, we compute a redshift-dependent threshold in optical richness $\lambda_\mathrm{min}(z)$ that reproduces a sample with the given \fcont \ (see also Fig.~\ref{fig:fcont_redshift}).
The survey selection function now reads $\Theta_\mathrm{s}\left[\xi-5, \lambda-\lambda_\mathrm{min}(z), z-0.25\right]$.
We define a richness-mass relation
\begin{equation}
  \label{eq:richness-mass}
  \begin{split}
    \langle\ln\tilde\lambda\rangle =& \ln A_\lambda + B_\lambda\ln\left(\frac{\mass}{10^{14}~h_{70}^{-1}M_\odot}\right) \\
    &+ C_\lambda\ln\left(\frac{H(z)}{H(z=0.6)}\right)
  \end{split}
\end{equation}
with the intrinsic (true and unobservable) richness $\tilde\lambda$.
We assume lognormal intrinsic scatter in $\tilde\lambda$ of width $\sigma_{\ln\tilde\lambda}$.
The true richness is related to the observed richness $\lambda$ via a normal distribution of width $\sigma_{\tilde\lambda}$
\begin{equation}
  P(\lambda|\tilde\lambda) = \mathcal N\left(\tilde\lambda, \sigma_{\tilde\lambda}^2\right).
\end{equation}
If richness was measured independently of background galaxies, then one would expect Poisson errors $\sigma_{\tilde\lambda}=\sqrt{\tilde\lambda}$. We go beyond this approximation and empirically determine $\sigma_{\tilde\lambda}$ by accounting for the impact of the redshift-dependent local galaxy background on the richness measurement. We establish an approximate relation based on measurements from the LS-DR10 data of:
\begin{equation}
  \sigma_{\tilde\lambda}(z) = \sqrt{\tilde\lambda+10}\times\big[1.08+0.45(z-0.6)\big].
\end{equation}
We now expand the likelihood from Eq.~\eqref{eq:abundance_likelihood} and the survey selection function $\Theta_\mathrm{s}$ to incorporate richness
\begin{equation}
  \label{eq:abundance_3_likelihood}
  \begin{split}
    \ln \mathcal L(\vec p) =& \sum_i\ln\frac{\mathrm{d}^3 N(\vec p)}{\mathrm d\xi\,\mathrm d\lambda\,\mathrm d z}\bigg|_{\xi_i,\lambda_i,z_i} \\
    &- \iiint d\xi\,\mathrm d\lambda\,\mathrm d z\, \Theta_\mathrm{s}(\xi,\lambda,z)\,\frac{\mathrm{d}^3 N(\vec p)}{\mathrm d\xi\,\mathrm d\lambda\,\mathrm d z} \\
    &+ \mathrm{const.}
  \end{split}
\end{equation}
The differential cluster abundance in two observables
\begin{equation}
  \label{eq:dN_dxi_dlambda_dz}
  \begin{split}
    \frac{\mathrm{d}^3 N(\vec p)}{\mathrm d\xi\,\mathrm d\lambda\,\mathrm d z} = \int \mathrm d\Omega_\mathrm{s}\iiint&\mathrm{d}M\,\mathrm d\tilde\lambda\,\mathrm{d}\zeta \bigg[P(\lambda|\tilde\lambda)\,P(\xi|\zeta)\\
    & P(\tilde\lambda,\zeta|M,z,\vec p)\,\frac{\mathrm{d}^2 N(\vec p)}{\mathrm d M\,\mathrm dV} \frac{\mathrm d^2 V(\vec p)}{\mathrm dz\,\mathrm d\Omega_\mathrm{s}}\bigg]
  \end{split}
\end{equation}
is analogous to the single-observable case in Eq.~\eqref{eq:dN_dxi_dz}.\footnote{Equation~\eqref{eq:dN_dxi_dz} can be obtained from Eq.~\eqref{eq:dN_dxi_dlambda_dz} via marginalization over richness $\lambda$
\begin{equation*}
    \frac{\mathrm{d}^2 N(\vec p)}{\mathrm d\xi\,\mathrm d z} = \int\mathrm d\lambda\, \frac{\mathrm{d}^3 N(\vec p)}{\mathrm d\lambda\,\mathrm d\xi\,\mathrm d z}.
\end{equation*}}
The SZ and richness observable-mass relation $P(\ln\tilde\lambda,\ln\zeta|\ln M,z,\vec p)$ incorporates the mean relations (Eqs.~\ref{eq:zeta-mass} and~\ref{eq:richness-mass}), and the intrinsic scatter $\sigma_{\ln\zeta}$ and $\sigma_{\ln\tilde\lambda}$.

We assume our fiducial cosmology and wide uniform parameter priors; for the scatter parameters, we additionally set a hard lower bound $\sigma>0.01$. We then use the likelihood (Eq.~\ref{eq:abundance_3_likelihood}) to infer the SZ and richness scaling relation parameters.
Because we now consider two mass proxies, we can empirically calibrate the intrinsic scatter parameters, and we do not fix the SZ scatter $\sigma_{\ln\zeta}$ as we did in the previous section.
The richness of clusters that are affected by the optical mask are likely to be mis-estimated. Therefore, we marginalize over the richness of clusters with more than 20\% of the area within 1\arcmin \ of the SZ centroid masked.
The results are presented in Table~\ref{tab:obs-mass} and in blue color in Fig.~\ref{fig:GTC}.
The mean recovered constraints on $B_\mathrm{SZ}$ and $C_\mathrm{SZ}$ are almost identical to the ones obtained from the more simplistic SZ-only analysis. The data-driven measurement of the SZ scatter exhibits a shift toward smaller values than the fiducial $\sigma_{\ln\zeta}=0.2$ (probability to exceed $p=0.23$, corresponding to $1.2\sigma$). Due to the degeneracy between the SZ scatter and amplitude, treating $\sigma_{\ln\zeta}$ as a free parameter leads to broader constraints on $\ln A_\mathrm{SZ}$ than obtained in the previous section (gold contours in the figure) and also implies a shift of the amplitude $\ln A_\mathrm{SZ}$ ($+0.7\sigma$).
In summary, the overall agreement confirms that, while the \fcont\ cut is crucial to remove spurious cluster detections, it does not lead to a significant removal of true clusters (otherwise the two sets of results would be inconsistent). 
Note that compared to the richness-mass relation of SPT-selected clusters presented in \cite{bocquet24b}, we now use an updated richness measurement scheme that produces higher estimates.
With these new richness measurements, we recover a mass trend that is consistent with unity within $1\sigma$.
We report a hint for an evolution with redshift ($C_\lambda>0$ at $<2\sigma$).

\begin{figure}
  \includegraphics[width=\columnwidth]{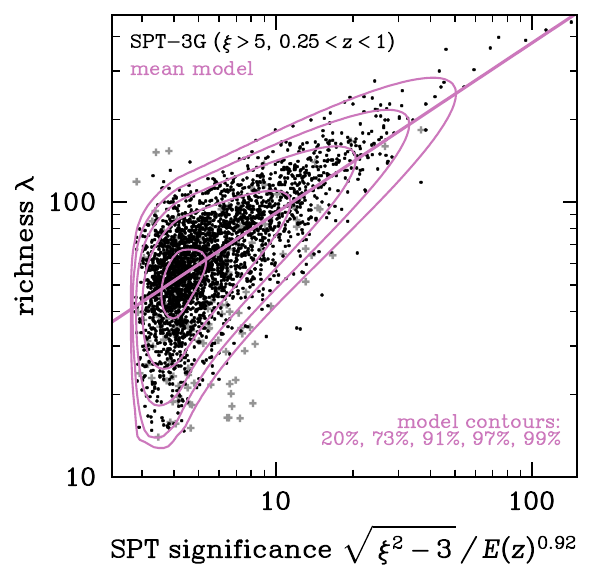}
  \caption{Distribution of the ($\xi>5,\,0.25<z<1$) subsample of SPT-3G clusters in the space of SPT significance and optical richness. The straight line shows the mean model and contours show the expected distribution. We find 27 out of 3,005 clusters outside of the 99\% contour, in line with the expectation. Clusters whose richness estimates are likely to be affected by masking are shown with gray crosses and are not used in the fit.}
  \label{fig:richnessSZ}
  \vspace{0.1cm}
\end{figure}

In Fig.~\ref{fig:richnessSZ}, we show the cluster sample in significance-richness space, along with the prediction of the mean relation and of the full distribution.
In particular, we check if the number of potential outliers is consistent with the model expectation.
In a first iteration, we noticed a population of clusters with SPT significance around 7 and richness around 20 that did not seem to be well described by our model (gray points in Fig.~\ref{fig:richnessSZ}). Upon closer inspection, we found issues in these clusters' redshift and/or richness assignments owing to masking or incorrect assignment of a higher-$z$ cluster to a lower-$z$ system (see Sec.~\ref{subsec:rules}). As discussed above, we do not use these partially optically masked clusters' richness measurements in the fit, even if including them only leads to minor shifts in the recovered parameters (mainly increasing $\sigma_{\ln\tilde\lambda}$).

\subsection{Mass Estimation}

Assuming the fixed cosmology and the mean SZ scaling relation parameters we just obtained, the parameters $\vec p$ are set and we compute the probability distribution for the mass of each cluster $i$ as
\begin{equation}
    \frac{\mathrm dP}{\mathrm d\mass}\bigg|_{\xi_i,z_i} = \frac{\int\mathrm d\zeta\,P(\xi_i|\zeta)\,P(\zeta|\mass,z_i,\vec p)}{\iint\mathrm d\mass\,\mathrm d\zeta\,P(\xi_i|\zeta)\,P(\zeta|\mass,z_i,\vec p)}.
\end{equation}
The width of the recovered distributions captures the noise due to observational and intrinsic scatter but not the systematic uncertainties in the model parameters themselves. 
The latter will be assessed in the context of a mass-calibrated cosmological analysis \citep[such as][]{bocquet24b}. In Appendix~\ref{sec:tableoverview}, we present the mean recovered mass for each cluster, along with the 68\% credible interval.

\begin{figure*}[th]
    \centering
    \includegraphics[width=\textwidth]{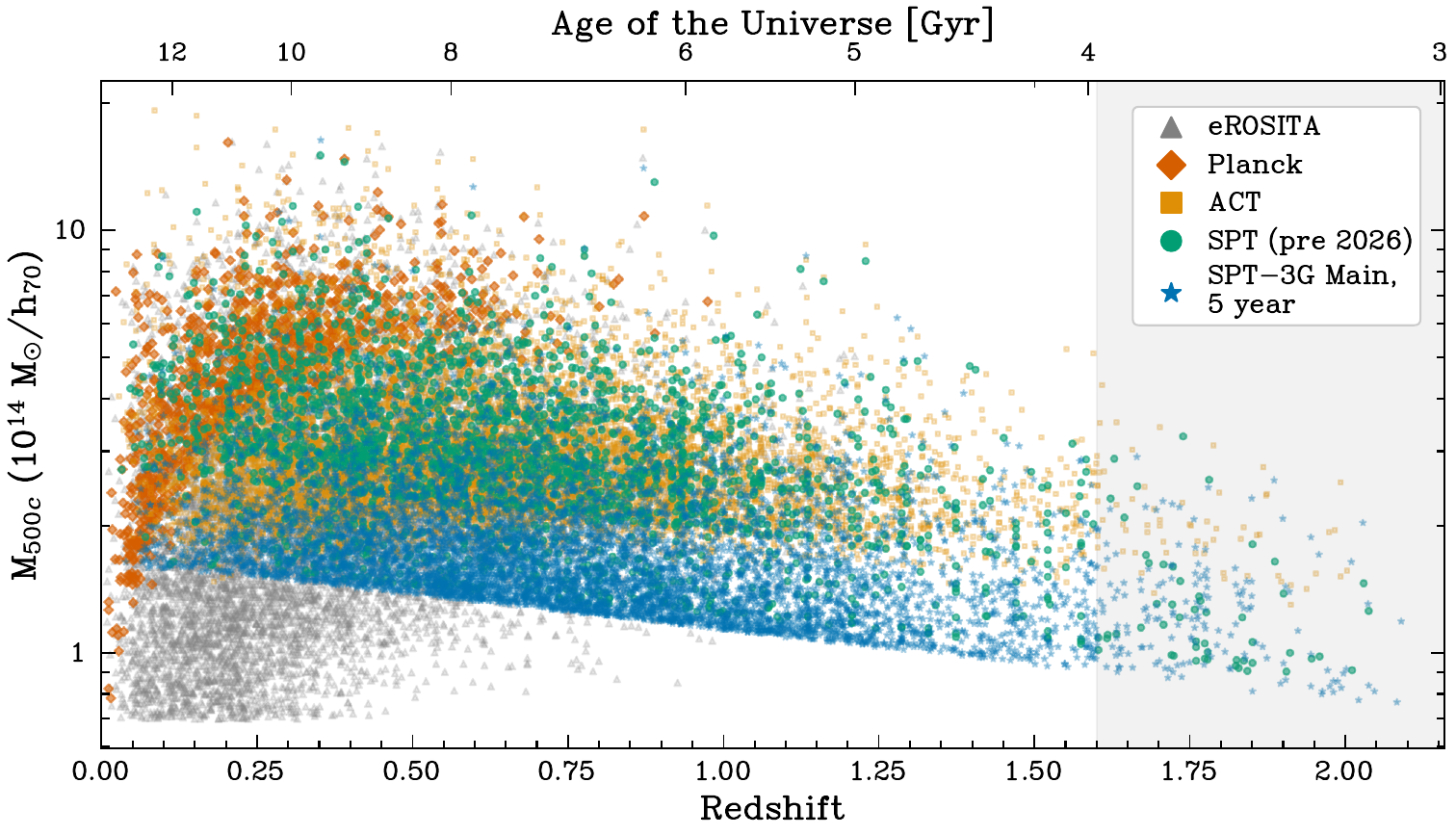}
    \caption{The mass-redshift distribution of the five-year SPT-3G Main field sample  compared to other recent large ICM-selected cluster samples, including those from ACT \citep{hilton26}, eROSITA \citep{bulbul24}, \textit{Planck} \citep{planck15-27}, and previously published samples from the SPT-SZ, SPTpol, and SPT-3G collaborations \citep{bleem15b, bleem20,bleem24, klein24a, kornoelje26, archipley26}. As noted in Sec. \ref{sec:cluster_confirmation}, above $z=1.6$ there is a large degeneracy in the color-redshift relation used for IR-confirmed clusters. Here, for display purposes, we plot the confirmed clusters at $z>1.6$ at redshifts obtained from sampling a realization of the cluster mass function for our fiducial cosmology to better illustrate the expected distribution of the SPT-3G sample. } 
    \label{fig:mz_plot}
\end{figure*}

\section{The SPT-3G Main Field Cluster Sample} 
\label{sec:results}

Combining the results of the previous sections, the SPT-3G Main field cluster sample consists of \ncand \ cluster candidates above a detection significance $\xi=4$, of which \nconfirm \ are confirmed as clusters using optical-IR data discussed in Sec. \ref{sec:cluster_confirmation}. The sample extends in redshift from $z= 0.037$ to $z\sim 2$, and in mass from $\mass = $ \mlow \ to $ \mass = $ \mhigh. 
The  median mass of the sample is \mbox{$M_{500c}^\textrm{median}$=~\medianmass}, and  the median redshift of the sample is \mbox{$z_\textrm{median}=\medianredshift$}, with  \nzgtone \ clusters ($\sim25\%$) at $z>1$ and \ngtonepfive \ ($\sim4\%$) at $z\geq1.5$. 
In  Fig.~\ref{fig:mz_plot} we plot the  mass-redshift distribution for the SPT-3G Main field sample as well as samples from other ICM-selected catalogs, including samples from multiple generations of experiments on the SPT \citep{bleem20, bleem24,klein24a, archipley26, kornoelje26}, the X-ray selected eRASS1 cluster sample from \cite{bulbul24},  and the most recent SZ cluster releases by the ACT \citep{hilton26} and \textit{Planck} \citep{planck15-27} collaborations. 
The SPT-3G cluster sample has a significantly lower mass threshold and higher cluster density (4.5 confirmed clusters/deg$^2$) than the other wide-field surveys. 
The increased sensitivity and the steepness of the cluster mass function also results in comparable cluster counts between this sample (\nconfirm \ confirmed clusters) and the most recent ACT sample (10,040 confirmed clusters)---including more high-redshift ($z>1$) clusters (\nzgtone \ in this sample vs.~1,171 in ACT) despite a factor of 10 difference in sky area.
The sample is particularly complementary to the eRASS1 selected sample, which probes similarly low-mass clusters at low redshifts, where SPT is less sensitive.    

\subsection{Completeness}

\begin{figure}
  \includegraphics[width=\columnwidth]{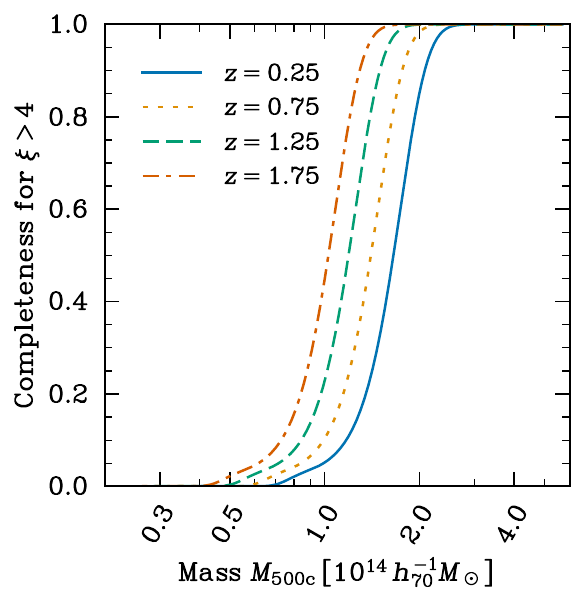}
  \caption{Estimated completeness of the SPT-3G Main cluster sample above $\xi=4$ at several different redshifts. As discussed in Sec.~\ref{sec:results}, the sample has a declining mass threshold as a function of redshift. The sample is expected to be $>90\%$ complete for masses $M_{500c}\gtrsim 2 \times 10^{14}~M_\odot/h_{70}$ at $z>0.25$.}
  \label{fig:completeness}
\end{figure}

As noted in previous works \citep[e.g., ][]{huang20}, signal-to-noise-selected SPT cluster catalogs have an effective mass threshold that declines as a function of redshift owing to (1) the characteristics of noise in the maps---i.e., ``redder''  residual CMB and atmospheric signals on larger scales transitioning to lower noise at small scales results in more effective detection of (more compact) high-$z$ systems---and (2) self-similar evolution in clusters that results in hotter clusters at fixed mass at higher redshift \citep{kaiser86b}. 
This trend is clearly seen in the completeness of the SPT-3G sample (Fig.~\ref{fig:completeness}). 
We model the completeness as a Heaviside function in significance, $\Theta(\xi - 4.0)$, which reflects the hard cut in SZ significance used to select cluster candidates. Using the $\xi - M$ relation, the completeness in significance is converted to completeness as a function of mass and redshift to represent the probability of a cluster of a given mass at a given redshift to be found in the SPT-3G catalog. At redshift $z=0.25$, the sample is expected to be $>90\%$ complete for masses $M_{500\mathrm c}>2.1\times10^{14}~M_\odot/h_{70}$ and $>50\%$ complete for $M_{500\mathrm c}>1.6\times10^{14}~M_\odot/h_{70}$. As discussed, the mass threshold drops with increasing redshift, and at the median redshift of the sample $z=0.726$, $>90\%$ completeness is expected for $M_{500\mathrm c}>1.8\times10^{14}~M_\odot/h_{70}$. Finally, for $z>1$, the sample is $>90\%$ complete for $M_{500\mathrm c}>1.6\times10^{14}~M_\odot/h_{70}$. We note that the completeness of the SPT-3G catalog below a threshold of $z < 0.25$ becomes difficult to model owing to the filters applied during the SPT mapmaking process which remove large-scale Fourier modes, and hence large-angular-scale signals from clusters. 
Additionally, this modeling does not account for biases from correlated contaminants such as radio-loud cluster AGN (discussed more in Sec.~\ref{sec:internal_checks}). 

In the sections below we further explore properties of the sample, particularly focusing on strong lensing clusters in the SPT-3G sample (Sec. \ref{sec:strong_lensing}), comparisons to other ICM-selected (Sec. \ref{sec:icm_comparison}) and optically selected (Sec. \ref{sec:optical_ir_comparison}) cluster samples, and internal systematic checks with the SPT-3G data (Sec. \ref{sec:internal_checks}).

\section{Strong Gravitational Lenses}\label{sec:strong_lensing}

Owing to their immense masses and high mass densities, galaxy clusters can produce some of the most spectacular strong gravitational lensing phenomena in the Universe (see e.g., \citealt{postman12, coe19, pontoppidan22} and the recent review of \citealt{natarajan24}).
By cleanly identifying the most massive clusters, SZ surveys have proven highly useful to search for such systems \citep[e.g., ][]{mahler25, cerny26}. 
Over 110 clusters from previous generations of SPT samples have been identified as strong lenses, both as new discoveries \citep{staniszewski09, high10, song12b, bleem15b, bleem20} and through associations with lenses reported in the literature. 
 
In this section we examine the strong lensing properties of the SPT-3G sample in two ways: first through cross-matching cluster locations with previously identified strong lenses  and lens candidates from the literature, and second, through visual inspection of the confirmed SPT-3G cluster sample using DECALS and archival SPT follow-up data.  
We additionally flag potential lenses in Sec.  \ref{subsec:dsfg} based on associations with bright dusty sources detected in SPT data.

\subsection{Archival Lensing Search}

Over the last decade, there has been a tremendous expansion of strong lenses and strong lens candidates in the literature owing to the availability of high-quality wide-area optical survey data coupled to new crowd-sourcing- and machine-learning-based lens identification tools \citep[e.g.,][and many others]{more24}. 
Data from DECALS and the Dark Energy Survey have proven particularly valuable given the deep, relatively uniform coverage in common filter bands. 
These data have enabled a wide variety of searches for lensed galaxies and quasars at galaxy \citep[e.g.,][]{diehl17, rojas22, he23, gonzalez25} and cluster scales \citep[e.g.,][]{odonnell22,mork25}. 
Here we leverage the lens and candidate lists provided in the Strong Lensing Database \citep[SLED;][]{vernardos25} in the SPT-3G footprint augmented by our own literature compilation of cluster and galaxy lenses;  in total we cross-match against lenses from 39 different publications for this search. 

We conduct our cross-match within apertures of 1\arcmin \ (2\arcmin) to account for statistical uncertainty in the SPT position as well as astrophysical offsets between ICM centroid tracers and cluster centers of mass. 
We find that 242 (285) or 3.3\% (4\%) of confirmed clusters have matches in the archival strong lensing compilation within 1\arcmin \ (2\arcmin).   
There is naturally significant redundancy in lenses in this compilation: SPT clusters match 370 entries within 1\arcmin, with e.g.,  3 clusters (ACO~S1063, SPT-CL~J2138-6007, and ACT-CL~J0034.4-4551) matching reported lenses or lens candidates in 4 different publications, 20 clusters matching candidates in 3 publications, 79 clusters in 2, and 140 clusters with a single match to a reported lens or lens candidate. 

\subsection{Visual Inspection in Optical Follow-up Data }

We next visually inspect images of confirmed clusters constructed using \textit{z-, r-}, and \textit{g-}band imaging data from LS-DR10 \citep{dey19}.  
Using a new tool, \texttt{Image Marker}, which was developed for the efficient classification of astronomical images \citep{walker25}, we scored cluster images from 0-3 where 0 denotes no evidence of strong gravitational lensing and 3 records strong evidence  (e.g., multiple images, arcs, etc.) of lensing. 
We also examined and scored observations from earlier SPT follow-up campaigns, particularly from Magellan/PISCO \citep{stalder14, bleem20} and Hubble Space Telescope observations (PIs: Stubbs, High, Schrabback, Gladders) of SPT-SZ \citep{bleem15b} and SPTpol \citep{huang20,bleem24} clusters which are also in the SPT-3G sample.
In total we tag \nstronglensing\ clusters as having strong evidence (score=3) for gravitational lensing, \nnewstronglens \ newly reported here.

\section{Comparison to Other ICM-selected Cluster samples}
\label{sec:icm_comparison}

\begin{table*}
    \centering
    {
    \renewcommand{\arraystretch}{1.15}
\begin{tabular}{lcccccc}
Catalog & N$_\textrm{total}$ & N$_\textrm{Footprint}$ & \shortstack{Fraction \\Matched}  & \shortstack{Median Sep \\ (arcmin)} & Median Mass Ratio & Reference \\
\hline
SPT-SZ & 811 & 495 & 0.89 & 0.32 $\pm 0.02$ & 1.03 $\pm 0.01$ &\citet{bleem15b,klein24a} \\
SPTpol 500d & 544 & 542 & 0.92 & 0.25 $\pm 0.01$ & 0.97 $\pm 0.01$ & \citet{bleem24} \\
SPT-3G Deep & 469 & 461 & 0.85 & 0.10 $\pm 0.01$ & 1.00 $\pm 0.01$ & \citet{kornoelje26} \\
ACT DR6 & 10040 & 1094 & 0.93 & 0.28 $\pm 0.01$ & 1.04 $\pm 0.01$ & \citet{hilton26} \\
eRASS1 & 10442 & 1126 & 0.62 & 0.36 $\pm 0.02$ & 1.09 $\pm 0.02$ & \citet{bulbul24} \\
eRASS1 (Cosmology) & 5019 & 564 & 0.78 & 0.32 $\pm 0.02$ & 1.05 $\pm 0.02$ & \citet{bulbul24} \\
\end{tabular}

}
\caption{Summary statistics from the angular association of SPT-3G Main field cluster candidates to other ICM-selected cluster samples. \textnormal{For this comparison we restrict to clusters in the external samples with provided redshifts and, when masses are supplied,  estimated masses greater than 0. Here N$_\textrm{total}$ is the total number of such clusters in the ICM catalogs, N$_\textrm{footprint}$ is the number of clusters within the SPT-3G Main field area used in the cluster search  (see Sec. \ref{sec:cluster_identification}), and the fraction matched is determined using a 3\arcmin \ matching radius to candidates in this footprint. Mass ratios are reported as the median of the SPT-3G catalog mass over the comparison catalog mass.  More details on these matches are provided in Sec. \ref{sec:icm_comparison}. We perform additional matches to other eROSITA source catalogs in Appendix \ref{sec:xrayappend}.} } 
\label{tab:table_icm_comparison}
\end{table*}

We next cross-match the SPT-3G catalog with other ICM-selected catalogs from SPT, ACT, and eROSITA. 
As optical confirmation significantly improves the purity of the samples---especially near their respective selection thresholds---for each of these samples we conduct our primary comparison to clusters with provided masses and redshifts (that is, not including ``unconfirmed" candidates). 
Balancing the angular density (highest for the SPT-3G sample at 4.5 clusters/deg$^2$) and the various positional uncertainties of the surveys (median 0.3\arcmin \ for SPT-3G),  we adopt a 3\arcmin \ matching radius. A summary of the matches is provided in Table \ref{tab:table_icm_comparison}. 

\subsection{Comparisons to Previous SZ-Selected Samples from SPT}

From SPT we compare to samples from  SPT-SZ \citep{bleem15b, klein24a}, SPTpol 500d,  \citep{bleem24}, and the 100 deg$^2$ SPT Deep field \citep[][hereafter K26]{kornoelje26}; the latter sample was constructed from combined SPT-3G and SPTpol data. We find excellent agreement in both positional recovery and mass estimation for the SPT-selected samples, with the new SPT-3G sample containing a strong majority of all of the previously published clusters. 
Exploring further the unmatched clusters, from SPT-SZ we have 57 clusters in the SPT-3G footprint not matched in the new sample.  The median $\xi_\textrm{SPT-SZ}$ of these unmatched clusters is 4.2, with only 1 cluster detected at $\xi>5$ (the 95\% purity threshold for SPT-SZ without additional optical follow-up).  For the unmatched clusters,  the median optical richness is low ($\lambda=17$ vs. the median of the matched sample of $\lambda=58$) with a corresponding increase in optical contamination $f_\textrm{cont}$ (median of 0.10 for the unmatched candidates vs.  $5\times10^{-4}$ for the matched candidates). 
At the unmatched cluster locations, we recover a median $\xi$=1.6 in matched-filtered SPT-3G maps using a forced photometry approach we will use for the rest of this section when comparing to other unmatched cluster samples.
Given the SPT-3G maps are  $>4\times$ deeper for cluster detection than those from the SPT-SZ survey (Sec. \ref{sec:obs_mass}), the combined evidence supports that noise boosted the detection significance of the unmatched clusters high in this earlier survey. 

A similar conclusion holds for the SPTpol 500d sample. 
Here there are 42 unmatched clusters in SPT-3G data with a median $\xi_\textrm{SPTpol}=$4.3, richness $\lambda=$17, and contamination $f_\textrm{cont}=$0.05 compared to 5.1/39/0.008 for the respective statistics in the matched sample. Only 2 unmatched clusters had $\xi_\textrm{SPTpol}>5$ with modest richnesses (both  $\xi_\textrm{SPTpol}\sim5.3$ and $\lambda=$9  and 21). The median force-photometered SPT-3G value for all the unmatched clusters was $\xi=$2.2.

The last SPT sample we consider is from the SPT Deep field analyzed in K26. 
Results from this previous sample are especially correlated with this work as Main field data from SPT-3G (without time constant corrections in mapmaking, an improvement in this work, see Sec. \ref{sec:obs_mapmaking}) provide the most constraining CMB survey data used in that analysis, though additional data from the SPTpol 100d \citep{huang20}, 500d \citep{bleem24}, as well as from \textit{Herschel}/SPIRE \citep{griffin03} was also used. 
As seen in Table \ref{tab:table_icm_comparison},  these shared data greatly reduce the typical angular offsets between  the SPT-3G Main and Deep field matches. 
Cross-matching the two samples we find 68 unmatched clusters (16/68 of these from an alternative analysis in which an estimated model for correlated dust emission was removed from the filtered SPT maps, see K26 Sec. 8.2), with 2 again at $\xi>5$. Inspection of the maps shows these 2 centroids in the Deep field on the borders of point source masked regions which differ slightly between the runs. For the remainder of the unmatched clusters, the median Deep field significance is slightly higher, $\xi_\textrm{Deep}=4.2$,  compared to $\xi=$3.6 in force-photometered SPT-3G  maps here. 

\subsection{Comparison to the ACT DR6 SZ-Selected Catalog}

We next cross-match to the SZ-selected  cluster catalog produced from 16,293 deg$^2$ of ACT DR6 survey data \citep{hilton26}. 
The optically confirmed sample contains over 10,000 systems with a median mass of {2.8$\times 10^{14} \ M_\odot$} and a median redshift of $z=0.58$; 1,171 clusters are at $z>1$. 
From this full sample, 1,127 clusters are in the SPT-3G Main footprint. 
Cross-matching to the SPT-3G cluster sample, here again we find a high fraction (93\%) of the ACT confirmed clusters in the SPT-3G sample, with good agreement in the mass and position estimates. 
Matched clusters are typically detected at 2$\times$ higher signal-to-noise in the SPT-3G data. 
There are 77 ACT clusters not matched to the SPT-3G catalog, with 69 (of 77) detected at ACT SNR less than 5.  
The median unmatched cluster has a ACT SNR$=4.3$, and forced photometry of the SPT-3G maps at those locations gives a median $\xi=1.9$.  
Of the 8 unmatched systems detected in ACT at SNR$>$5, 3 were in the border regions of point source masks, 
and 2 of the clusters had force-photometered $\xi$ values above the $\xi=4$ threshold (ACT-CL J0003.3$-$5254 was detected at $\xi=17.5$ and ACT-CL J0057.3$-$5222 at $\xi=6.2$) but had SPT recovered centroids in the point source exclusion region and so are excluded from this catalog. 
One additional higher SNR ACT cluster (ACT-CL J2249.1$-$4426, SNR=6) is 6 arcmin away from the massive cluster ACO 1063 which was detected at $\xi=166$ in the SPT-3G maps. 
As noted in Sec. \ref{sec:cluster_identification}, match-filtering can cause secondary artifacts in ground-based CMB survey maps around massive clusters; visually inspecting the SPT maps at the ACT cluster location we do not see clear signs of a missed high-significance cluster. 
The remaining 5 higher-significance unmatched ACT clusters had median force-photometered $\xi$=3.4 in SPT-3G data, with values ranging from 2.2 to 3.8.  

\subsection{X-ray Selected Clusters}\label{subsec:erosita_comparison}
Our final ICM sample comparisons focus on X-ray-selected clusters from the first data release of the  eROSITA survey \citep[eRASS1;][]{merloni24}. 
While cross-comparisons with prior SZ surveys allow validation of our selection in mm-wave survey data and mass estimation methods, comparisons with eROSITA additionally allow us to probe other systematics in cluster sample construction---particularly sources of selection bias that may strongly impact one survey's selection observable but not the other. 
While detailed comparisons and validation of the eROSITA and SPT selection functions are beyond the scope of this work \citep[see e.g.,][for a detailed study using earlier SPT survey data]{clerc24}, this section and Appendix \ref{sec:xrayappend} briefly highlight the potential of such studies with the new SPT-3G dataset. 

We first compare to the first galaxy cluster and group catalog produced from eROSITA western-hemisphere data \citep{bulbul24}. 
This cluster sample, selected as sources of extended X-ray emission in 13,116 \sqdeg \ of survey data,  contains 12,247 optically confirmed systems in the redshift range $0.003<z<1.32$. We restrict our comparisons to clusters with redshifts and positive mass estimates.  
Of this sample, 1,126 systems fall within the SPT cluster search region. 
The SPT-3G sample matches 62\% of the eRASS1 sample in this region, leaving 430 unmatched. 
Of these, 161 are at $z>0.25$ where the SPT sample is expected to be more complete (see Sec. \ref{sec:purity_completeness}), for an overall matching percentage at $z>0.25$ of 75\%. 
The story is complicated, however, by the necessity of adopting a mixture model in the X-ray confirmation process to account for spurious associations with X-ray AGN and other sources \citep[see e.g., discussion ][ Sec. 2.4]{bulbul24}. 
Examining the $P_\textrm{cont}^\textrm{AGN}$ parameter which records the probability of random X-ray associations compared to cluster candidates (and was set using a combination of richness, redshift, and X-ray count data), we see a marked difference in this statistic between our matched and unmatched population, with 92\% of the matched clusters having $P_\textrm{cont}^\textrm{AGN}<0.1$ at $z>0.25$, but only 30\% of the unmatched sample.

\begin{figure}
    \centering
    \includegraphics[width=0.5\textwidth]{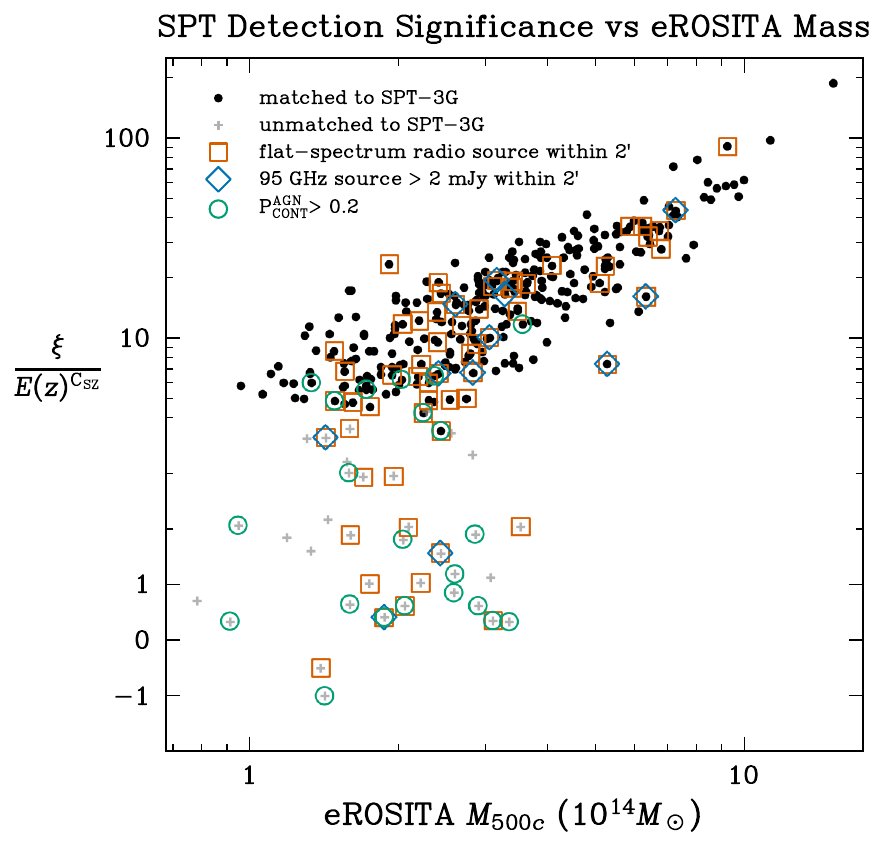}
    \caption{Redshift-scaled (Sec. \ref{sec:obs_mass}) SPT detection significance, $\xi$, vs. X-ray mass for clusters at $z>0.25$ in the eRASS1 cosmology sample \citep{bulbul24}. In bold we plot eRASS1 clusters that have counterparts within the SPT-3G catalog (89\%), while in pale gray we plot force-photometered $\xi$ values for clusters without matches; note that the y-axis scaling transitions from log to linear below the SPT detection threshold of $\xi=4$.  Green circles mark eRASS1 clusters with higher probability of spurious X-ray AGN contamination  and blue and orange symbols mark clusters with nearby radio sources detected in SPT and/or low-frequency ASKAP \citep{duchesne24,hale21} data. While mm-bright radio sources can bias the recovered SZ significance low, the majority of unassociated eROSITA clusters in unmasked areas in the SPT footprint do not have radio counterparts. }
    \label{fig:erosita_comparison}
\end{figure}

We move to the eRASS1 cosmology sample to increase the intrinsic purity of the X-ray catalog.  
This sample has a stricter X-ray extension likelihood threshold ($L_\textrm{ext}>6$ compared to 3 for the full sample) and is expected to be 95\% pure with no additional cleaning (compared to 86\% for the full sample). 
Starting from the full eRASS1 cosmological catalog, we find 586 clusters in the SPT cluster footprint, with 564 of these with provided X-ray masses (96\%).
Adopting a 3\arcmin \ matching radius, 76\% of these X-ray clusters have matches in the SPT-3G catalog and 87\% of the 331 clusters at $z>0.25$ are matched.
This match rate rises slightly to 89\% when considering only eRASS1 clusters which also have positive reported X-ray masses. 
These association rates are comparable to the completeness of the SPT-3G catalog vs. the other SZ surveys. 

We expect the SPT-3G sample to be 90\% (50\%) complete at $\mass = 2.0\times10^{14}\msun$ ($1.6\times10^{14}\msun$) at $z=0.41$, the median redshift of the full eRASS1 sample at $z>0.25$. 
The 35 unmatched eRASS1 clusters with positive X-ray masses above this redshift have a median mass of $1.87\times10^{14}\msun$; 15 of these systems fall below SPT's 50\% completeness mass. 
Overall SPT matches 93\% of the eRASS1 clusters with $\mass > 2\times10^{14}\msun$ at $z>0.25$, roughly consistent with expectations.  
In Fig.~\ref{fig:erosita_comparison} we plot the X-ray masses compared to the clusters' SPT-3G $\xi$ values; for clusters unmatched in the SPT catalog we force-photometer $\xi$ at the X-ray location.
We indeed see most of the unmatched clusters are at the low-mass end of the X-ray sample; we will further examine unmatched clusters in the context of potential radio source contamination of the SZ signal below in Sec. \ref{subsec:radiopart2}.
Finally, while this section focused on cross matches to the main eRASS1 cluster sample of \citet{bulbul24}, in Appendix \ref{sec:xrayappend} we extend these comparisons to both X-ray clusters selected from a reanalysis of the eRASS1 point source catalog \citep{balzer25} as well as cross matches to the point sources \citep{merloni24} themselves.

\section{Comparison to Optical- and Infrared-Selected Cluster Catalogs}
\label{sec:optical_ir_comparison}

Following the approach adopted in the \citet{bleem24} analysis, we next associate the SPT-3G Main field cluster catalog with a set of optical- and IR-selected cluster catalogs. Specifically, we consider the DES~Y3 \textsc{redMaPPer} catalog \citep{Abbott25clusters} and the WISE/LS DR10-based cluster catalogs of \citet{WenAndHan24}, and MaDCoWS2 \citep{thongkham24b}. These catalogs are representative of galaxy-based cluster detection techniques across optical and IR wavelengths, providing a complementary comparison to the SZ-selected SPT-3G sample.
We particularly focus on the fraction of SPT clusters detected in each external catalog, consistency of redshift estimates, and the measured scatter between SZ mass and optical/IR richness for the associated counterparts. 
We detail the measurement of these quantities in the following subsections and summarize these properties in Table \ref{tab:optical_table}.

\begin{table}
    \centering
    {
    \renewcommand{\arraystretch}{1.15}
\begin{tabular}{lcccccc}
Catalog &  \shortstack{Ext. vs. MCMF \\ Confirmed}  & $\sigma_{\Delta z}/(1+z)$ & $\sigma_{\mathrm{ln}(\lambda/M_{500,\textrm{SZ}})}$ \\
\hline
RM Y3 & 0.66 & 0.009 & $0.3 \pm 0.01$ \\
MaDCoWS2 & 0.52 & 0.013 & $0.53 \pm 0.03$ \\
WH24 & 0.72 & 0.012 & $0.44 \pm 0.04$ \\ 
\end{tabular}

}
\caption{Summary statistics from the association of SPT-3G Main field cluster candidates to optical/IR cluster samples. \textnormal{Using a  $f_\mathrm{cont}$-based association (Sec. \ref{sec:matching_procedure}),  we report the fraction of SPT-3G clusters confirmed by each catalog compared to our baseline MCMF analysis, the scatter in redshift estimates between MCMF and the external catalogs at $z<1$ (0.95 for RM), and a simplified estimate of the scatter between the optical/IR mass proxy and our SZ-based mass estimates for clusters with masses greater than $M_{500c}=3\times 10^{14}$\msun \ (Sec. \ref{sec:comparingrichness}).  We have corrected for a small bias between the MaDCoWS2 and MCMF redshifts when reporting the scatter. SPT-3G cluster recovery is primarily limited by different effects for each catalog, namely (1) the redshift reach of RM, (2) the relatively high richness selection threshold of MaDCoWS2, and (3) the large scatter between richness and SZ observable for WH24. }} 
\label{tab:optical_table}
\end{table}

\subsection{Matching Procedure}
\label{sec:matching_procedure}

We adopt a more sophisticated matching procedure for associating clusters from the higher-density optical/IR catalogs with SPT-3G counterparts. 
The matching strategy is identical for all the comparisons in this section and consists of two approaches. The first approach is 
MCMF-like in which we estimate $f_\mathrm{cont}$ using a maximum cross-match radius of 2\arcmin \ between external and SPT positions to select potential candidates. To generate random locations, we shift the positions in the external catalog randomly by more than 0.5$^\circ$ and repeat the matching. We then follow the steps in Sec.~\ref{sec:MCMF} to find the best counterpart showing the lowest measurement of $f_\mathrm{cont}$.

The second approach is a more simplistic positional matching widely used in literature. We first rank the clusters in each optical/IR catalog according to their respective mass proxies and then perform a ranked positional cross-match with the SPT-3G catalog within a maximum separation of $1\farcm{5}$. Each SPT-3G cluster is associated with the optical/IR counterpart of highest mass proxy found within the matching radius. To avoid duplicate associations, each optical or IR cluster can be matched only once and is removed from subsequent matching iterations.

To quantify the purity and completeness of the resulting positionally matched sample, we repeat the matching procedure for several control configurations. These include: (1) increasing the matching radius beyond $1\farcm{5}$, and (2) performing matches to catalogs whose positions have been artificially shifted by $\pm0.5$ and $\pm1.0$~deg in right ascension and declination (8 diagonal permutations). The nominal $1\farcm{5}$ matches are used for direct catalog-to-catalog comparisons, while the enlarged and shifted configurations serve to estimate the fraction of chance associations.

Following the methodology described in \citet{klein24b}, we model the offset distributions to separate genuine and random matches. The distribution of separations for the unshifted catalogs represents a mixture of real and chance matches, while the shifted catalogs yield purely random associations. We normalize the shifted-offset distributions to match the amplitude of the unshifted distribution at large separations ($>3^{\prime}$), where the contribution from real matches is negligible. The scaled distribution of shifted offsets then provides an empirical estimate of the expected number of chance coincidences as a function of separation. Integrating this function up to $1\farcm{5}$ yields the expected number of false matches within the default radius. Subtracting the chance component from the total offset distribution isolates the contribution from true physical associations. The integral of this true-match distribution out to $1\farcm{5}$ gives the number of genuine matches, while extending to $5^{\prime}$ provides a measure of all possible real associations between the two samples. 

The positional matching method provides estimates of completeness and purity for a more commonly used approach, while the MCMF-like approach makes use of richness and redshift information to provide overall better counterparts.
The selection cuts used in the individual external catalogs complicate the sample purity estimate compared to a pure $f_\mathrm{cont}$ selection, as they miss low-richness systems that would have made the selection. We therefore use $f_\mathrm{cont}$-informed counterparts for redshift and richness comparisons, while providing purity and completeness estimates from the positional matching method.

\subsection{Comparison to the DES~Y3 \textsc{redMaPPer} Catalog}
\label{sec:redmapper_comparison}

The DES~Y3 \textsc{redMaPPer} (RM) catalog was recently employed in the cosmological analysis of \citet[]{Abbott25clusters}. 
The sample consists of $>$869,000 groups and clusters detected above richness $\lambda_{RM}=$5 in the $\sim$5,000 \sqdeg \ DES observing region. The sample covers a redshift range of $0.1 < z < 0.95$ and overlaps approximately 85\% of the SPT-3G survey footprint.

Following \citet{Abbott25clusters}, we first limit the catalog to $\lambda_\mathrm{RM}>$20, the 99\% purity threshold of the sample \citep{costanzi19}.  
Applying the angular matching procedure described above,  we identify 3,129 SPT-RM matches ($\sim52\%$ of the SPT sample matched in the common sky area), of which $99.6\%$ are confirmed by MCMF ($f_{\mathrm{cont}} < 0.2$).
We find that 99\% of all true associations lie within $1\farcm{5}$, and estimate a $5\%$ chance of random matches within the same radius.
Lowering the richness threshold to $\lambda_\mathrm{RM}>$10 increases the match fraction to $\sim65\%$, but doubles the fraction of chance matches. 
If we instead adopt the redshift-dependent  $f_\mathrm{cont}$-selected RM catalog down to $\lambda_\mathrm{RM}=$5 and impose a $f_\mathrm{cont}^\mathrm{RM}<0.2$, we find 4,003 matched clusters. The confirmation fraction of the $f_\mathrm{cont}$-selected sample between $0.1<z<0.8$ reaches 92\% of that of MCMF, showing very similar performance over the redshift range well probed with DECam data.

Restricting the analysis to $z<0.95$, we find a mean scatter between the MCMF and RM redshifts of $\sigma_{\Delta z}/(1+z) = 0.009$. The scatter shows a clear redshift dependence, increasing beyond $z \approx 0.6$ and reaching $\sigma_{\Delta z}/(1+z) = 0.017$ at $z = 0.9$. We identify 39 systems with $|\Delta z| > 5\sigma$, corresponding to  $\sim1$\% of the sample. All but five cases have a match to the second-ranked optical counterpart. After visual inspection of all systems with discrepant redshifts, we find that the main cause of redshift mismatches are multiple structures along the line of sight. 
In cases where no match with the best or second-best MCMF candidate was found, redshift mismatches were caused either by local masking or the blending of two clusters with small ($\Delta z<0.15$) redshift differences.

\subsection{Comparison to the MaDCoWS2 Catalog}
\label{sec:madcows_comparison}

We next compare the SPT-3G Main catalog to the DECaLS/WISE-based MaDCoWS2 cluster catalog \citep{thongkham24b}. The MaDCoWS2 sample provides a complementary view of the massive cluster population and extends to higher redshifts ($0.1\le z \le 2)$ than RM. 
The sample has an estimated $\sim84$\% overlap with the SPT-3G survey footprint and 
contains 133,036 galaxy clusters detected at SNR $\ge5$ from  6,498 \sqdeg \ of imaging data. 

Using the simplified angular matching procedure described above, we identify 3,032 matched systems within $1\farcm{5}$ ($\sim50\%$ of SPT clusters in the common sky area), of which only 36 ($\sim$1\%) are not confirmed by MCMF; 4\% of the matches within this radius are expected to occur by chance. 
Turning to the $f_\mathrm{cont}$-associated MaDCoWS2 catalog, we find 3,169 clusters below $f_\mathrm{cont}^\mathrm{MaDCoWS2}=0.2$. Restricting to $0.1<z<0.8$ the confirmation rate reaches 64\% of the MCMF value. This is well below that found for the RM catalog and likely caused by the significantly higher selection threshold of MaDCoWS2 catalog.

We find a bias between MCMF and MaDCoWS2 redshifts above $z\sim0.25$, increasing to $\delta z = z_\mathrm{MCMF}-z_\mathrm{MaDCoWS2}\approx0.035$ at $z=0.4$ and then being approximately constant until $z=0.8$.
To obtain a less biased measurement of the scatter between redshifts, we correct the MaDCoWS2 redshifts using the median redshift offset of the 100 closest clusters in redshift.
We find a redshift scatter of $\sigma_{\Delta z}/(1+z) = 0.013$ for clusters at $z<1$, increasing to $\sim0.05$ for systems in the range $1 < z < 2$. 
Overall, 97.4\% of all matched clusters agree within $5\sigma$ in redshift. Among the remaining 2.6\% outliers, approximately 80\% show consistency with the second-ranked optical counterpart, leaving only $\sim0.55\%$ of the systems without agreement among the two best matches.

\subsection{Comparison to the LS-DR10-Based Cluster Catalog of Wen \& Han 24}
\label{sec:wen24_comparison}

As a third comparison, we match the SPT-3G main catalog to the LS-DR10-based optical/IR cluster catalog of \citet[][WH24]{WenAndHan24}. This catalog provides one of the largest homogeneous optical cluster samples to date, covering a wide redshift range ($0.1 \le z \lesssim 1.5$) and nearly the full SPT-3G Main footprint. 
The sample consists of 1.58 million clusters with richness $\lambda_\mathrm{WH24}>5$ drawn from the $\sim$24,000 \sqdeg \ LS-DR10 release. 

We identify 5,931 matches within $1\farcm{5}$ to SPT-3G clusters (78\% of confirmed SPT-3G clusters in the common footprint), finding that 97\% of the true SPT-WH24 associations lie within $1\farcm{5}$, while approximately 11\% of the matches within this radius are expected to occur by chance. Among these, 5.5\% have $f_{\mathrm{cont}}>0.2$ and are thus not confirmed by MCMF, implying that 94.5\% of all matched systems are independently confirmed by our baseline analysis.
Using the $f_\mathrm{cont}$-selected catalog, we find 5,154 clusters below $f_\mathrm{cont}^\mathrm{WH24}=0.2$. 
Comparing the confirmation rate from $0.1<z<0.8$, we reach 87\% of the MCMF value, better than MaDCoWS2 but below the low-richness version of the RM catalog.

The comparison of redshift estimates shows good overall agreement. At $z<1$, we measure a redshift scatter of $\sigma_{\Delta z}/(1+z)=0.012$, which increases to $\sim0.04$ at $z>1$. Approximately 5\% of the matches deviate by more than $5\sigma$ in redshift; of these, 75\% agree with the second-ranked optical counterpart, suggesting ambiguous associations or projection effects. 

\begin{figure*}[t]
\begin{center}
\includegraphics[width=0.98\linewidth]{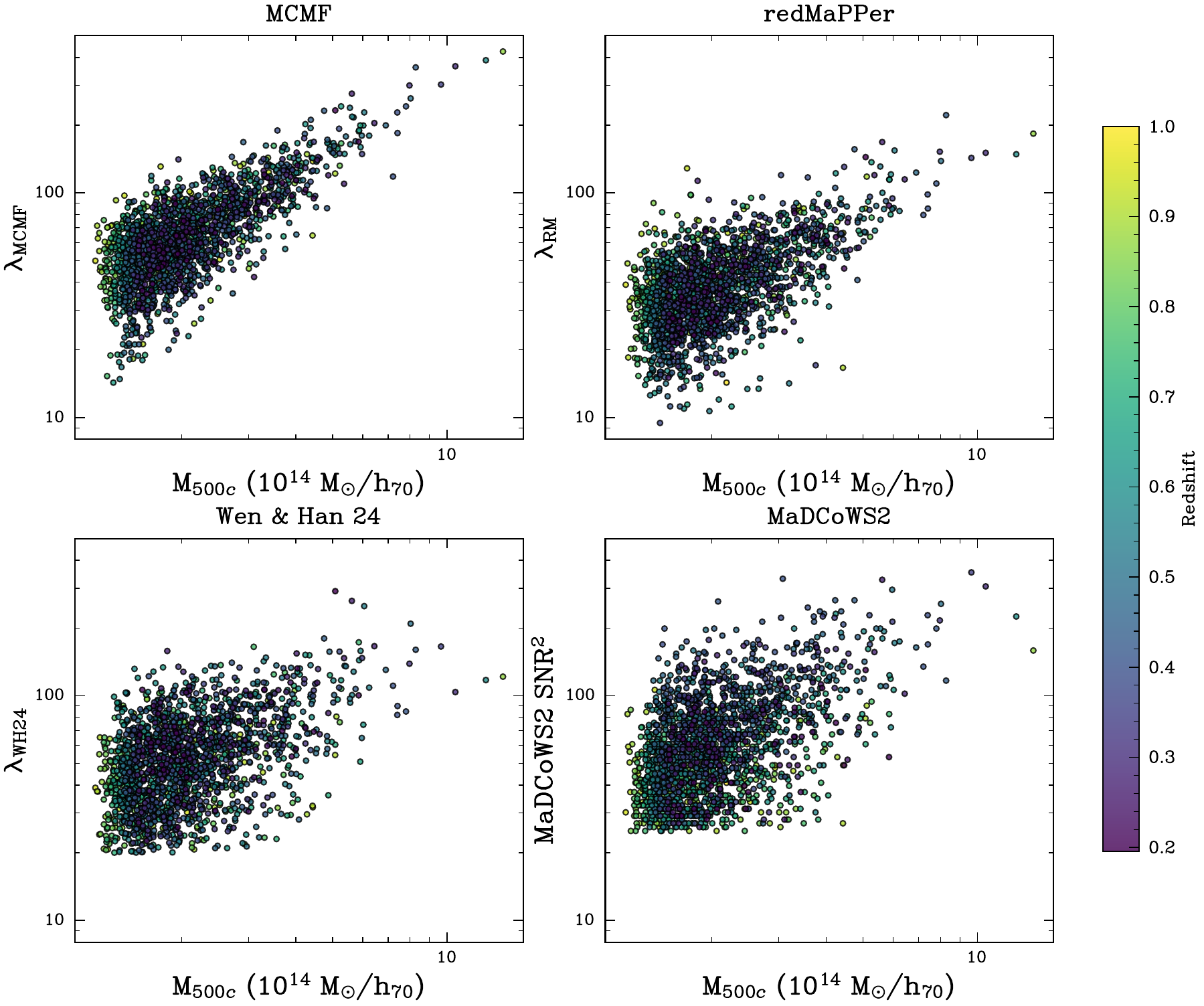}
\caption{Optical richness vs. SZ-inferred mass for the subset of 1,909 clusters at $0.2<z<0.95$ found in all four (SPT-3G,  \textsc{redMaPPer} Y3, MaDCoWS2, and WH24) cluster samples. MCMF, our baseline confirmation method (Sec. \ref{sec:cluster_confirmation}) shows the smallest scatter with SZ mass, followed by the \textsc{redMaPPer} sample. The WH24 sample, while having large scatter, is the most complete ($\sim78$\%) of the external catalogs. } \label{fig:richnessmass}
\end{center}
\end{figure*}

\subsection{Comparing Optical Mass Proxies with SZ-based Mass}
\label{sec:comparingrichness}
In the previous subsections we compared confirmation fractions, purity, and redshifts of three external optical/IR cluster catalogs. In this section we compare the performance of the individual optical mass proxies. For that we create a subsample containing all confirmed clusters in common between surveys. From the SPT-3G Main catalog we select all clusters with $f_\mathrm{cont}<0.2$ in MCMF, RM, MaDCoWS2, and WH24. We further restrict the redshift range to $0.2<z<0.95$ and redshift offsets to be $\Delta z/(1+z)<0.1$ from the MCMF value.
The final sample for this test contains 1,909 clusters.
We use richness as the optical mass proxy in all cases except for the MaDCoWS2 sample (where it is not provided); for that sample we use the square of the provided SNR assuming Poisson background noise as the mass proxy. 
Under the assumption of Poisson noise the square of the SNR should scale with the richness. 

Optical mass proxies like richness typically scale approximately linearly with cluster mass; assuming this holds for the samples considered here, one can use the scatter in the richness-to-mass ratio to evaluate the performance of different optical mass proxies.
In Fig.~\ref{fig:richnessmass}, we show richness vs. SZ-inferred mass for the optical catalogs and MCMF. The richness-to-mass ratio is well described by a lognormal distribution. From the external catalogs we find the smallest scatter of $\sigma_{\mathrm{ln}(\lambda/M_{500,\textrm{SZ}})}=0.36\pm0.01$ for the RM catalog, followed by MaDCoWS2 and WH24 each with $\sigma_{\mathrm{ln}(\lambda/M_{500,\textrm{SZ}})}=0.44\pm0.01$.

In Fig.~\ref{fig:richnessmass}, the MaDCoWS2 and WH24 samples exhibit incompleteness at low masses due to richness selection thresholds of 25 (SNR = 5) and $\sim$20 (corresponding to $f_\mathrm{cont}^\mathrm{WH24}=0.2$), respectively. As a result, the richness-to-mass distributions are partially truncated and the inferred scatter is underestimated. Above $M_{500c}=3\times 10^{14}$\msun \ , the richness distribution at fixed mass is significantly less affected by these thresholds, yielding more reliable but noisier scatter estimates.
Restricting to masses above this threshold we find $\sigma_{\mathrm{ln}(\lambda/M_{500,\textrm{SZ}})}=0.3\pm0.01$ for RM, $\sigma_{\mathrm{ln}(\lambda/M_{500,\textrm{SZ}})}=0.44\pm0.04$ for WH24, and $\sigma_{\mathrm{ln}(\lambda/M_{500,\textrm{SZ}})}=0.53\pm0.03$ for MaDCoWS2. Overall the RM richness appears to perform significantly better than the mass proxies of the other two catalogs. For MCMF richness we find $\sigma_{\mathrm{ln}(\lambda/M_{500,\textrm{SZ}})}=0.29\pm0.01$ over all masses and $\sigma_{\mathrm{ln}(\lambda/M_{500,\textrm{SZ}})}=0.20\pm0.01$ above $M_{500c}=3\times 10^{14}$\msun, outperforming RM richness, but note that the MCMF confirmations were constructed with the advantage of using SZ prior information.
In Appendix \ref{sec:opticalappend} we examine the RM Y3 mass-richness relation in more detail, using the more sophisticated modeling of Sec. \ref{sec:obs_mass}.

\section{Mm-wave Contamination}\label{sec:internal_checks}
The multi-frequency data of SPT-3G  allows us to conduct several systematic checks for contamination in our recovered cluster properties. 
 We particularly  focus on consistency checks between Compton-$y$ measured from our baseline three-band combination and that estimated using subsets of these bands. 
The main sources of contamination are expected to be from dust emission correlated with cluster galaxies and the cluster environment and synchrotron emission from radio-loud cluster members. 

\begin{figure*}[t!] 
    \centering
        \includegraphics[width=0.48\textwidth]{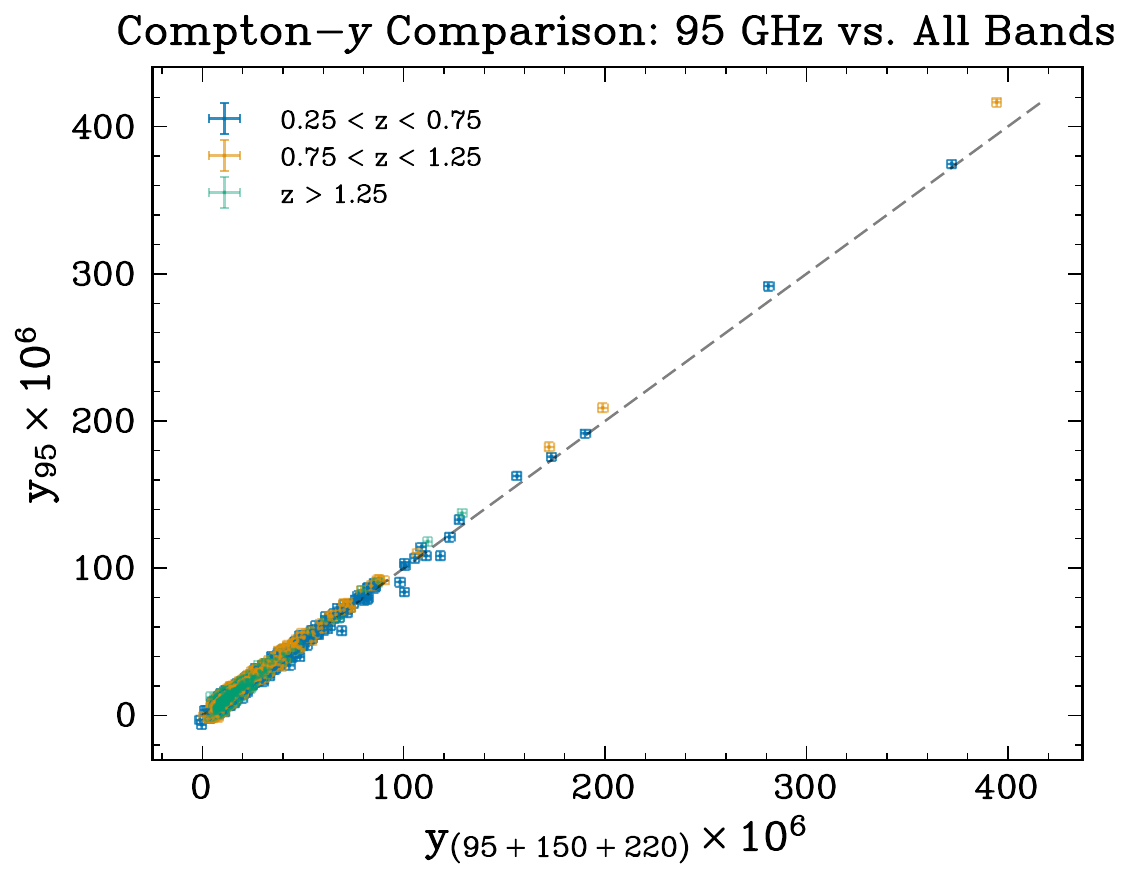}
        \includegraphics[width=0.48\textwidth]{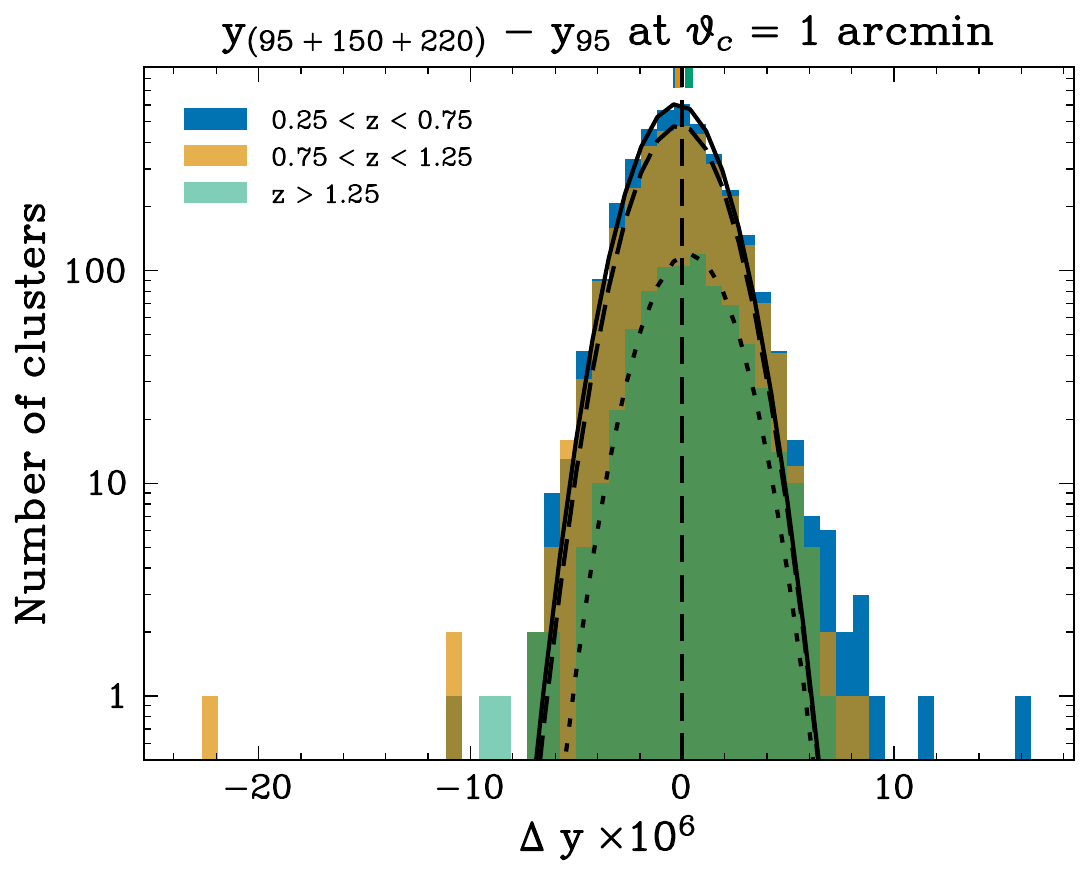}
        \includegraphics[width=0.48\textwidth]{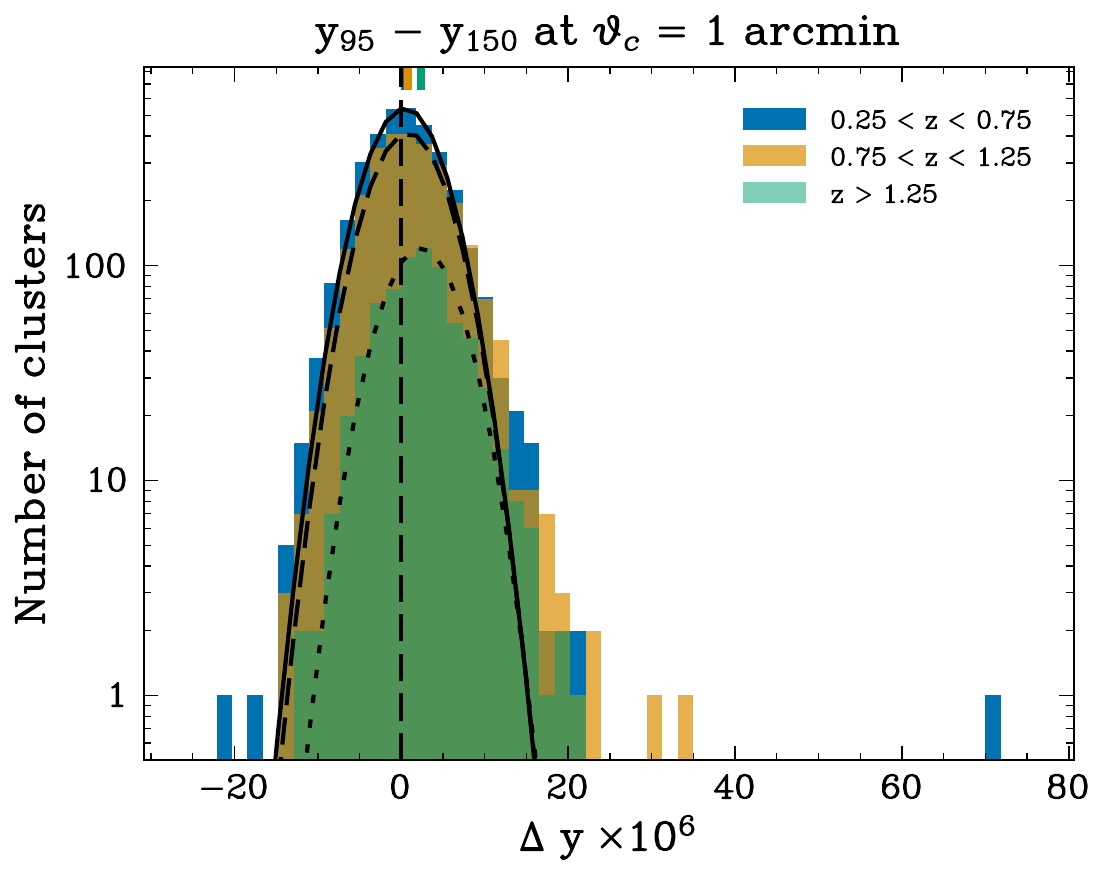}
        \includegraphics[width=0.48\textwidth]{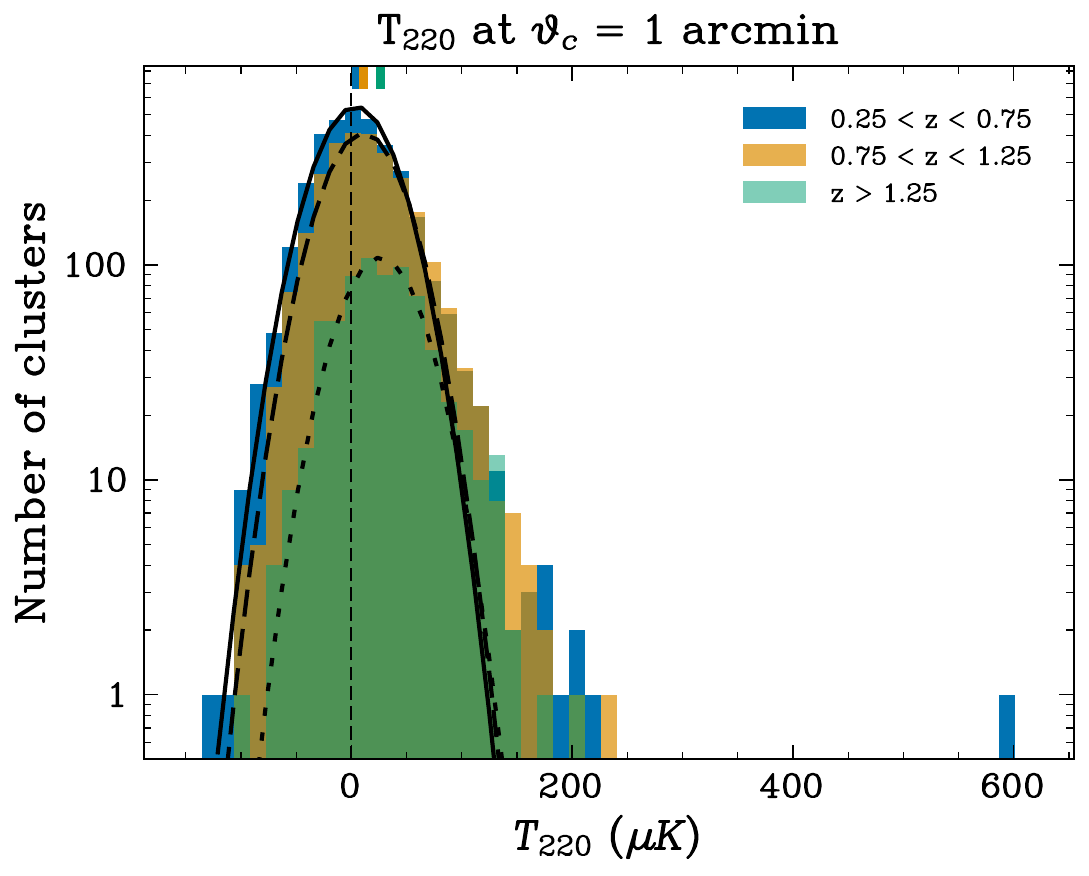}

    \caption{Internal frequency characterization of the SPT-3G Cluster sample. In all panels we plot the profile amplitudes ($y_{0}$ or $T_{0}$, see Eq.~\ref{eqn:beta}) for our matched filter with $\theta_c=1\arcmin$. In the \textit{top} panels we compare the Compton-$y$ measured at 95~GHz to that from the full 3 band combination. In the \textit{top-left} panel we show a scatter plot of these quantities, with a 1-1 line over-plotted (dashed-black). In the \textit{top-right} panel we plot the distribution of  measured differences with Gaussian curves reflecting the noise levels of the difference maps over-plotted centered at the median differences (traced by colored bars at top) for 3 different redshift bins. Overall the Compton-$y$ values are in good agreement, though we do see a small systematic shift in the recovered Compton-$y$ at 95 GHz with increasing redshift, reflecting a small level of dust contamination. The distributions are also highly Gaussian with scatter tracing the noise in the fields; outliers are discussed in Sec. \ref{subsec:dsfg}, \ref{subsec:outliers}. To examine contamination with different spectral indices, in the \textit{bottom-left} panel we plot the difference between measurements at 95 and 150~GHz. Clusters contaminated by sources with rising (e.g.,  dusty) flux will have positive differences, while sources with sufficiently steep negative indices (see Sec. \ref{subsec:radio}) can lead to negative differences. Here again we see a clear increasing level of positive contamination and skewness with redshift, pointing to increased correlated dusty emission with redshift.  This signal is also traced by the filtered temperature measurements at 220~GHz (\textit{bottom-right}); this channel is at the SZ null so signals here arise from astrophysical emission correlated with cluster locations.}
    \label{fig:internal_comparison}
\end{figure*}

\subsection{Correlated Dust Emission in the SPT-3G Main Sample} \label{subsec:cib}

While cluster environments at low redshift and/or high mass are typically dominated by quiescent galaxies, studies have found that the typical star formation rates of cluster members significantly increase at $z>1$ \cite[see the recent review of][and references therein]{alberts22}. 
There are strong correlations between star formation and emission at mm-wavelengths, as the dust enshrouding young stars is heated by their UV and optical photons and re-emits this energy at longer wavelengths, peaking in the sub-mm/far IR \citep{blain02}.  
For the purposes of SZ cluster selection, such correlated emission is a contaminant that can partially fill in the SZ decrement with a signal that,  in the SPT bands, rises with frequency and is brightest in the 220~GHz band. 
K26 conducted a detailed study of the impact of this bias on cluster identification with SPT. 
Combining data from the SPT-3G Deep field (see  description in Sec. \ref{subsec:wise_complete}) with sub-mm data from \textit{Herschel}/SPIRE \citep{griffin10}, K26 showed a factor of $\sim3\times$ increase in the fractional dust contamination of cluster SZ signals over the redshift range of $0.25 < z \lesssim2$, reaching up to $18\pm4\%$ at 150~GHz ($4\pm1\%$ at 95~GHz) in their highest redshift bin of $z>1$; the average correlated emission at 220 GHz was also found to be dominated by this signal.  
K26 also demonstrated the ability of multi-frequency data---particularly through inclusion of the 220~GHz channel---to mitigate this bias, resulting in robust cluster samples using the minimum-variance approach we also adopt in this work. 
Here we do not repeat the detailed exercise explored in that work, but rather briefly examine this contamination in the context of the greatly enhanced sample size ($\sim16\times$) of the SPT-3G Main survey. 

To characterize this bias we filter each of the individual frequency maps and various multi-frequency combinations with a filter scale of $\theta_c$=1\arcmin \  (close to 0\farcm{81}, the mean $\theta_c$ of the confirmed clusters) and extract the profile amplitudes (Eq.~\ref{eqn:beta}) at the locations of cluster candidates from the multi-frequency catalog. 
For 95~GHz, 150~GHz, and the two- and three-band combinations, we report the  amplitudes in Compton-$y$ units (converted using the SPT effective band centers for a non-relativistic SZ spectrum, Sec. \ref{sec:cluster_identification}), but leave 220~GHz in temperature units as there is expected to be negligible SZ signal in this band. 
We also record the robust standard deviation \citep{hoaglin00} in each of the filtered maps, as well as from differences between the filtered map combinations. The relative noise levels illustrate the relative strength of each frequency contribution (not accounting for correlated signals) to the overall SPT detections. 
In Compton-$y$ units, the noise levels for the profile amplitudes on the $\theta_c$=1\arcmin \ scale are listed in Table \ref{tab:filternoise}.

\begin{table}
    \centering
    \renewcommand{\arraystretch}{1.15}
    \begin{tabular}{l|c}
    Frequency Combination & \shortstack{Noise Level \\ (Compton-$y$ units)}\\
    \hline
    95~GHz                & $2.9\times 10^{-6}$ \\
    150~GHz               & $5.1\times 10^{-6}$ \\
    150+220~GHz           & $3.1\times 10^{-6}$ \\
    \textbf{95+150+220~GHz} & $\bm{2.3 \times 10^{-6}}$
    \end{tabular}
    \caption{Filtered map noise levels in Compton-$y$ units for the $\theta_c$=1\arcmin \ filter. \textnormal{In bold we highlight the noise for the three band combination used in cluster identification at this filter scale. The 95~GHz data alone have comparable weight to the combined 150+220~GHz data and only $1.26\times$ the noise of the full combination.}}
    \label{tab:filternoise}
\end{table}

In Fig.~\ref{fig:internal_comparison} we plot recovered values from this filtering in three different redshift bins: $0.25<z<0.75$ (3370~systems), $0.75 < z< 1.25$ (2711 systems), and $z>1.25$ (702 systems).  In the histograms of Compton-$y$ differences/temperature distributions we also over-plot Gaussian curves centered on the median signal in each bin with standard deviations measured as discussed above. 

We first compare  95~GHz-extracted Compton-$y$ values to those from the full multi-frequency combination. 
As in K26, we find good agreement between the two, with a small suppression of the 95~GHz-only signal in the highest redshift bin.  The distribution, with a small number of outliers, is also extremely Gaussian, providing evidence that there is not a large population of clusters with strong unmodeled correlated signals in the sample relative to the noise levels probed in the SPT-3G dataset. 
The largest outliers are from massive clusters; the 22 clusters at $z>0.25$ with absolute differences $>4\sigma$ have a median detection significance of $\xi=30$. 
Some of these clusters are in close proximity to dusty star-forming galaxies (DSFGs), including the most negative (-$12.7\sigma$) outlier: SPT/ACT-CL J0102.9$-$4915.6, $z=0.87$ \citep[with the cluster previously identified as strongly lensing a DSFG in][]{sun22}  and the most positive ($9\sigma$): SPT-CL J2332.4$-$5358.5, $z=0.4$  \citep[reported as a lens in][]{vanderlinde10, andersson11}. 
Such outliers are expected from lensing of DSFGs, which can boost or suppress the recovered SZ signal \citep{hezaveh13b}. 

Moving to the bottom panels of Fig.~\ref{fig:internal_comparison}, we plot the differences between the Compton-$y$ measured at 95 and 150~GHz as well as the 220~GHz temperature distribution at the location of clusters. We plot the 95-150~GHz difference to more clearly highlight differential contamination caused by sources with rising (e.g., dusty) spectral indices which will lead to positive residuals, and falling spectral indices which could lead to negative differences. As the SPT 220~GHz effective frequency is approximately at the SZ null, significant deviations from a null signal here trace (generally dusty) correlated emission from the clusters. 
In these panels we can clearly see evidence of the  dusty emission explored in K26; the single-temperature Compton-$y$ recovered at 95~GHz is systematically higher at high redshift than that of 150~GHz, and the distribution---while still largely Gaussian---shows a clear skew in the direction of higher contamination suppressing the 150~GHz flux  [or Compton-$y$] at high redshift. 
This skewness is even more pronounced in the plot of the temperature distribution at 220~GHz in which signals from dust  emission are significantly brighter.  

\begin{figure}[t]
    \centering
    \includegraphics[width=0.5\textwidth]{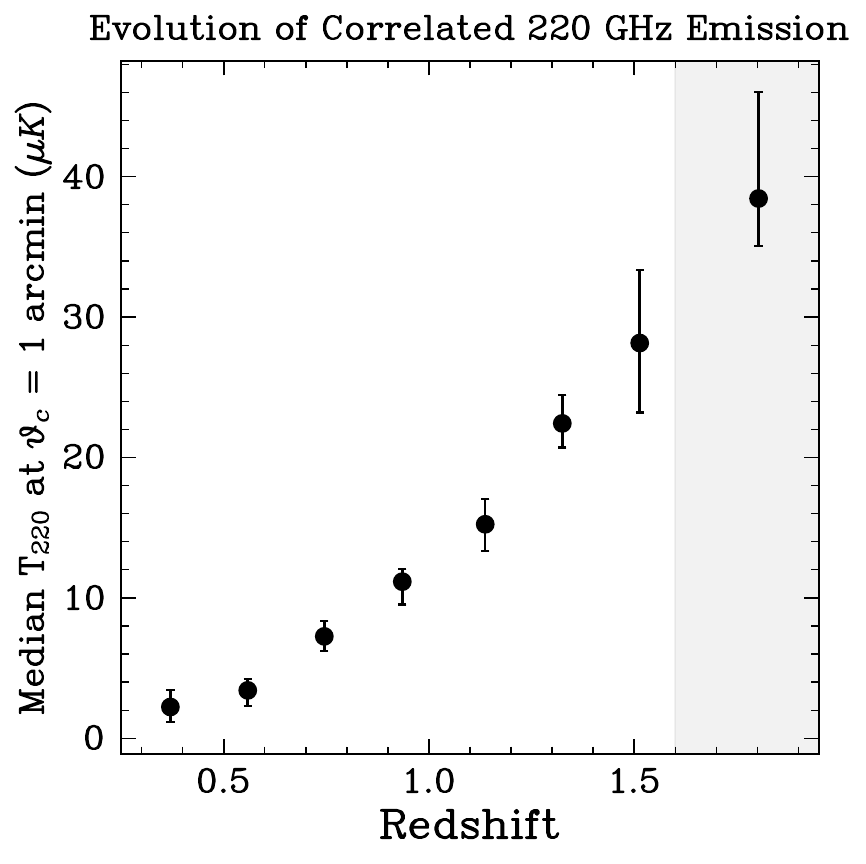}
    \caption{Redshift evolution of the median correlated 220 GHz emission at SPT-3G cluster locations; this emission increases by over an order of magnitude over the redshift range probed by the SPT sample ($0.25 < z \lesssim 2$), while the median Compton-$y$ value in each bin is comparable. As above, the region $z>1.6$ is shaded to denote higher redshift uncertainties; the plotted redshifts in each bin are the median of the best fit values in the bin. }
    \label{fig:t220_redshift}
\end{figure}

In Table \ref{tab:internal_stats} we report both the median filtered 220 GHz emission as well as the skew, as traced by the Fisher-Pearson coefficient, for the SPT-3G sample in several redshift bins; we also plot the redshift evolution in Fig.~\ref{fig:t220_redshift}.
The typical 220~GHz emission increases by greater than $17\times$ from our lowest ($0.25<z<0.45$) to highest ($z>1.6)$ bin, while the median filtered Compton-$y$ drops only by 3\% between the redshifts. 
Our highest-$\xi$ \textit{unconfirmed} clusters also show similar behavior to the higher redshift bins, as would be expected for clusters beyond the redshift reach of our current optical/IR confirmation data. 
The distribution in all bins has a small positive skew and is generally consistent across all bins (outside of the lowest-$z$ bin; excluding the large outlier SPT-CL J2332.4-5358.5, the skew in this bin drops to $0.5\pm0.1$). 
Fisher-Pearson coefficients, in the range measured here between 0.38 and 0.65, generally indicate that the distributions are approximately symmetric with mild to moderate skewness. 

We can estimate the median suppression of the 95 and 150~GHz SZ signals from correlated dusty emission by modeling it as a modified black body and scaling the 220~GHz emission to the lower frequencies. 
We assume a dust temperature of 22 K and dust emissivity index $\beta=2$ (consistent with the results of K26, see Eq.~14 and Table 2 in that work). 
We find that the Compton-$y$ recovery at 95~GHz is only mildly biased low (0-5\% from low to high-$z$).
Conversely, the single band estimates of Compton-$y$ from 150 GHz can be highly biased, with a median bias rising from 2\% 
for clusters $0.25 < z < 0.45$ to $\sim30\%$ for $z\gtrsim 1.6$ clusters. Adopting wider bins, our results are consistent with K26, 
with the larger sample size in this work enabling us to more finely trace the increase in dusty contamination with redshift.

While the redshift evolution of the correlated dusty emission is striking, it is important to reemphasize the findings of K26 that 
the construction of the SPT-3G sample is robust to biases from this signal. 
As discussed in Sec. \ref{sec:cluster_identification}, clusters are identified using a multi-band matched filtering process designed 
to maximize sensitivity to SZ signal in the presence of noise.  
While the process is not explicitly tailored to remove correlated dusty signals, at small scales (most relevant for high-$z$ cluster identification) 
the CIB is a significant noise contribution in the SPT-3G maps at higher frequencies. 
As the CIB has a similar spectral dependence to the dusty contaminant, and one that is spectrally distinct from the SZ signal in the SPT-3G bands, 
the filter optimization naturally down-weights contributions from dust-like signals; the 
220~GHz channel (at the SZ null) effectively acts as a high-fidelity dust tracer enabling clean mitigation of this bias. 
Additionally, as discussed above, the 95~GHz data---the most important contributor to the combined detection 
significance (i.e., Table \ref{tab:filternoise} and similar at other filter scales)---is 
only weakly impacted by this bias at the mass scales probed by the SPT-3G sample.

\begin{table}
    \centering
    \renewcommand{\arraystretch}{1.3}
\begin{tabular}{|c|c|c|c|}
\hline
Redshift Bin & N$_\textrm{Clusters}$ & \~{T}$_{220}$ ($\mu$K) & Skew \\
\hline
\textbf{$0.25 < z < 0.45$} & 1047 & 2.2 $^{+1.1}_{-1.2}$ & 2.59 $^{+1.26}_{-2.16}$ \\
\hline
\textbf{$0.45 < z < 0.65$} & 1546 & 3.4 $^{+1.1}_{-0.8}$ & 0.65 $^{+0.11}_{-0.12}$ \\
\hline
\textbf{$0.65 < z < 0.85$} & 1473 & 7.3 $^{+1.0}_{-1.1}$ & 0.60 $^{+0.11}_{-0.12}$ \\
\hline
\textbf{$0.85 < z < 1.05$} & 1209 & 11.2 $^{+1.6}_{-0.9}$ & 0.53 $^{+0.13}_{-0.15}$ \\
\hline
\textbf{$1.05< z < 1.25$} & 806 & 15.2 $^{+1.9}_{-1.8}$ & 0.44 $^{+0.09}_{-0.10}$ \\
\hline
\textbf{$1.25< z < 1.45$} & 353 & 22.4 $^{+1.7}_{-2.0}$ & 0.49 $^{+0.14}_{-0.15}$ \\
\hline
\textbf{$1.45< z < 1.6$} & 190 & 28.2 $^{+4.9}_{-5.2}$ & 0.08 $^{+0.14}_{-0.15}$ \\
\hline
\textbf{$z > 1.6$} & 159 & 38.4 $^{+3.4}_{-7.6}$ & 0.65 $^{+0.24}_{-0.34}$ \\
\hline
\textbf{unconfirmed} & 1702 & 13.4 $^{+1.1}_{-1.3}$ & 0.61 $^{+0.08}_{-0.09}$ \\
\hline
\textbf{unconfirmed $\xi>6$} & 51 & 29.0 $^{+2.8}_{-18.1}$ & 0.33 $^{+0.27}_{-0.32}$ \\
\hline
\end{tabular}
\caption{Median temperature and skew at 220 GHz for the SPT-3G sample. \textnormal{This filtered signal at $\theta_c$=1\arcmin \  primarily traces correlated dust emission from cluster members and the cluster environment. We note that, though the number of candidates is small, the median of our highest-$\xi$ (high purity) unconfirmed candidates is similar to values in our higher redshift bins, as would be expected for real clusters beyond the confirmation reach of current follow-up data.}} 
    \label{tab:internal_stats}
\end{table}

\subsubsection{Cluster-Correlated DSFGs}\label{subsec:dsfg}

While Sec. \ref{subsec:cib} primarily focuses on overall population trends in correlated dusty emission, 
some of this excess emission is detectable at modest-to-high SNR in individual systems. 
Re-examining the above recovered Compton-$y$ values from 95 GHz and 150 GHz alone, 
we find 159 candidates (1.8\%) with differences $>3 \sigma$ pointing to dusty contamination and 35 with differences $>4 \sigma$.  
Of these 159 candidates, 116 were confirmed (38 with redshifts $z>1$) and 43 were 
unconfirmed (6 of these with $\xi>5.5$, i.e., at high SZ purity, and thus likely true very high-$z$ clusters). 
The three largest outliers in the differences between Compton-$y$ measured at 95 and 150 GHz were 
all previously flagged as cluster-scale lenses: SPT-CL J2332.4$-$5358.5 (17$\sigma$ outlier) and 
SPT-CL J0102.9$-$4915.6 (7$\sigma$), discussed above, and  SPT-CL J2240.0$-$6117.4, an 8$\sigma$ outlier flagged in \cite{bleem24}. 

DSFGs---either those lensed by the cluster (as in the case of SPT-CL J2332.4-5358.5), 
members of the cluster themselves \citep[e.g.,][]{kumar26}, or just chance line-of-sight field projections---can also be detected individually at 
high significance in SPT-3G data and can lead to positive skew in the 220 GHz temperature distributions. 
We next cross-match cluster detections with 220~GHz sources from a dedicated point source analysis of the SPT-3G data. 
The SPT-3G source catalog was constructed using point-source-filtered versions of the same single-band maps as we use for the cluster analysis, resulting in $5 \sigma$ point source detection thresholds of roughly 1, 1, and 5 mJy at 95, 150, and 220~GHz, respectively; here we focus on sources detected at 220~GHz at a SNR $\ge 5$ with a positive spectral index as measured between 150 and 220~GHz. 
A total of 225 cluster candidates have such a source within 1\arcmin \ (compared to 63.5 expected from random associations, a 3.5$\times$ excess). 

The SPT-3G source catalog is a mixture of both dusty and AGN sources. 
As the presence of the cluster could bias internal source classification with SPT data alone, we leverage low frequency radio survey data to remove potential radio contaminants from our list. 
The 1.4~GHz Rapid ASKAP Continuum Survey survey \citep{duchesne24} has a high source density (over 135,000 sources in the SPT-3G Main field footprint), with a population dominated by steeply falling radio sources that will have little remaining signal at SPT's frequencies \citep{dezotti10}. 
Flat-spectrum AGN, however, can still have appreciable signal in SPT's bands.
Using data from 887.5~MHz \citep{hale21} and 1.4~GHz to provide a crude estimate of the spectral index, $\alpha$, we trim the initial ASKAP catalog by requiring  sources to have a detection SNR $>5$ at 1.4~GHz and an estimated 95~GHz flux (scaled from 1.4~GHz, flux $\sim$ $\nu^\alpha$) in excess of 2 mJy. 
We will refer to these ASKAP sources here and in later sections (esp. Sec. \ref{subsec:radio} where we test for contamination from these sources) as ``flat-spectrum" radio sources. 
Removing SZ candidates with such flat-spectrum sources within the matching radius from our 220 GHz matched list drops the matches to 218 SZ candidates (61 expected). 
Reducing the matching radius to $0\farcm{5}$ to further reduce random association probabilities, we find 72 observed (16 expected, a 4.5$\times$ excess) or 70 (15.2 expected) for candidates without nearby flat-spectrum sources, with 14 observed (3 expected) at $z>1$.

The two different ways of checking for dusty outliers have high overlap but are not identical. 
The Compton-$y$ difference test probes fluctuations at a slightly larger scale than the 220~GHz point source search and at lower frequencies. For the high ($>4 \sigma$) differences in single-band Compton-$y$, 26/35 have 220~GHz sources within 1\arcmin, while 4 of the 9 remaining are at $z>1$, with the optical/IR data for an additional unconfirmed candidate (1/9) providing evidence of a high-$z$ cluster, but not sufficient for  confirmation here.  
We flag both systems with dusty sources within  $0\farcm{5}$ in the cluster sample table as well as candidates with single band Compton-$y$ differences $>4\sigma$.

\subsection{Correlated Radio Emission}\label{subsec:radio}

Unlike the correlated dust emission from cluster members probed in Sec. \ref{subsec:cib}, stringent internal constraints on radio contamination are a challenge for SZ-selected samples from arcminute-scale observations. 
For these surveys, the modest resolution prevents angular isolation of correlated radio sources and the similarity of the spectral index of AGN and SZ signals in key signal bands (95 and 150 GHz) prevents clean spectral separation.  
Low-frequency (1.4~GHz) radio surveys such as ASKAP \citep{duchesne24} can provide high-fidelity catalogs of potential contaminant sources, but source variability and varying spectral indices and break frequencies (at which these indices steepen sharply) make it challenging to connect low-frequency observations to mm-wavelength observations. 

The cleanest way to constrain this bias in SZ surveys would be to characterize cluster samples with similar properties (i.e., redshift-independent samples with a low-scatter mass proxy) with arc-second scale mm observations. 
These cluster sample properties are, however, a unique result of SZ selection and so contamination studies are generally biased to lower redshift \citep[e.g., X-ray samples as in ][]{coble07, lin09, gupta17, hilton26} or use optical/IR samples \citep{gralla11, gralla14, orlowski21} the latter of which can suffer from significant impurities owing to projection effects \citep[e.g.,][]{grandis25}.  
While these studies have generally found that above $z=0.25$---the lowest redshift to which we quantify our sample completeness---the impact of very bright sources is expected to be small, much remains uncertain at high redshift where there is evidence at other wavelengths for significant increases in AGN activity in clusters \citep[][amongst many others]{mo20, alberts22,hashiguchi23}.  
The increased sensitivity of the new generation of SZ surveys to lower mass systems also increases the importance of characterizing the impact of lower-brightness radio sources. 

In this work we briefly look for radio contamination in the SPT-3G cluster sample in two different ways. 
First, we use data from low-frequency radio surveys and our own single frequency Compton-$y$ measurements to look for signs of biased Compton-$y$ measurements in  clusters detected by SPT-3G.  
Second, we re-examine the detection of eRASS1 clusters \citep{bulbul24} in SPT-3G data looking for signs of clusters missed in the SPT-3G sample owing to radio sources.

\subsubsection{Tests with 1.4~GHz Radio Surveys}\label{subsec:outliers}

For the tests in this subsection we focus on the ratio of Compton-$y$ measured at 95~GHz alone compared to 150~GHz. 
A synchrotron source with a falling spectral index will more strongly impact 95~GHz, but the 150~GHz data may also be impacted, and so these ratios provide lower limits to the typical fractional contamination.  
We first examine the measured Compton-$y$ properties of clusters with nearby ($<1\arcmin$) radio sources detected at SNR $>5$ in 1.4~GHz ASKAP data \citep{duchesne25}.  
We adopt this conservative matching radius in this test to capture the impacts of positive synchrotron emission on our Compton-$y$ measurements and avoid spurious SZ signal boosting from the sources' filtering wings (Sec. \ref{sec:point_source_removal}). 

There are 1,917 confirmed clusters (27\% of all confirmed systems) with such nearby sources, and we find the median ratio of $y$(95~GHz)/$y$(150~GHz) is close to unity at $0.98^{+0.005}_{-0.01}$, reflecting the expected steep fall off in flux at high frequencies \citep{tucci11, lagache20} of the majority of synchrotron sources in low-frequency radio surveys. 
If we instead examine clusters with candidate ASKAP flat-spectrum sources (Sec. \ref{subsec:dsfg}) within $1\arcmin$, we find 219 candidates (210 confirmed clusters, or 3\% of the confirmed sample) with 23 clusters at $z<0.25$.  
The median ratio of Compton-$y$ measured at 95 GHz compared to 150 GHz in this subsample of clusters is $0.89^{+0.02}_{-0.03}$ reflecting on average $>10\%$ bias from sources with falling spectral indices for these clusters.
We can further isolate potentially biased systems by placing harsher cuts on the radio spectral indices and expected fluxes at high frequencies.
If we restrict the cluster sample to those with sources with estimated fluxes $>2$ mJy at 95~GHz, adopting an assumed spectral index $\alpha=-0.7$ \citep[approximately the median spectral index of high-frequency radio sources found in clusters,][]{coble07}, we find 98 confirmed clusters (1.3\% of the sample with 25/98 at $z<0.25$) with a median Compton-$y$ ratio between 95 and 150 GHz of $0.83^{+0.02}_{-0.015}$. 

These results are broadly consistent with previous characterizations of SZ-selected clusters. 
Scaling low-frequency data from NVSS \citep{condon98} and SUMSS \citep{mauch03} to SPT bands using the results of \cite{coble07}, \cite{bleem20, bleem24} estimated that the SZ signals were reduced  $>10$\% ($>20$\%) by radio contamination for $\sim5$\% (2\%) of SPTpol clusters.  
\cite{dicker21,dicker24} used high-resolution (9\arcsec) imaging at 90~GHz from the MUSTANG 2 \citep{dicker14} camera on the Green Bank Telescope to characterize the high-frequency radio source population in ACT-selected clusters. 
In \citet{dicker24},  the authors observed 243 ACT DR5 clusters, finding 24 sources (in 22 clusters fields) brighter than 1~mJy within 4\arcmin \ of the cluster centers. 
A source at this threshold could reduce the recovered SZ signal for a typical ACT cluster by 12\%. 
Owing to filtering of the ACT data (much like the filtering wings discussed in Sec. \ref{sec:point_source_removal}) these sources can both suppress (within 1\arcmin) or boost ($>1\arcmin$) the recovered SZ flux. 
In total, 6/243 clusters (2.5\%) had a source within 1\arcmin; this more direct measurement at 90~GHz indicates slightly lower contamination than estimated from low-frequency data with the SPTpol sample but is consistent within uncertainties given the sample sizes.

\subsubsection{Radio Sources in eROSITA Clusters}\label{subsec:radiopart2}

\begin{table}
    \centering
    {
    \renewcommand{\arraystretch}{1.15}
\begin{tabular}{lll}
\ & eRASS1-SPT-CL & eRASS1-No SPT-CL  \\
\hline
N$_\textrm{clusters}$ & 285 & 35 \\
mm-wave source  \ & 8 (3\%) & 3 (8.5\%) \\
Flat Spec. ASKAP & 45 (16\%) & 14 (40\%) \\
Steep Spec. ASKAP & 15 (5\%) & 2 (5\%) \\
$P_\textrm{cont}^\textrm{AGN}>0.2$ & 9 (3\%) & 14 (40\%) \\
\end{tabular}

}
\caption{Summary statistics for X-ray clusters with positive masses from the eRASS1 cosmology sample at $z>0.25$. \textnormal{We divide the sample into X-ray clusters with/without an SPT-3G counterpart. We list the number and fraction of clusters matching to sources within 2\arcmin \ from (1) an SPT-selected mm-wave sample brighter than 2 mJy at 95~GHz, (2) low-frequency radio sources from ASKAP with extrapolated 95~GHz fluxes $>2$ mJy based on spectral indices measured in ASKAP data, and (3) low-frequency radio sources from ASKAP with projected 95~GHz fluxes $>2$ mJy but now adopting a spectral index of $\alpha=-0.7$ which is more typical for low-frequency cluster radio sources also detected at higher frequencies \citep{coble07}. As X-ray AGN can bias the eRASS1 selection, we also provide the number and fraction of X-ray clusters with estimated higher AGN contamination (here $P_\textrm{cont}^\textrm{AGN}>0.2$).}}
\label{tab:erosita_rad_table}
\end{table}

We next consider the correlation of radio sources with the recovery of eROSITA clusters by SPT-3G. 
We focus on the higher purity eRASS1 cosmology catalog to reduce ambiguities between SPT-3G incompleteness and spurious X-ray detections. 
As discussed in Sec. \ref{subsec:erosita_comparison}, SPT matches 93\% of the eRASS1 clusters with $\mass > 2\times10^{14}\msun$ at $z>0.25$---roughly consistent with our expectations for the SPT-3G sample completeness---so a strong signal is not expected in this test.
 
We match the eRASS1 cosmology sample at $z\ge0.25$ and with positive X-ray masses with 2 different radio source catalogs:  (1) mm-wave bright sources ($>2$ mJy at 95~GHz, but note this may also include a small number of the brightest DSFGs) from a dedicated source analysis of the SPT-3G maps and, as the mm-wave source selection may itself be impacted by SZ decrements from clusters, (2) the ``flat-spectrum'' ASKAP subsample \citep{duchesne24} we introduced in Sec. \ref{subsec:dsfg}. 
We use a 2\arcmin \ matching radius for sources and keep our original 3\arcmin \ matching radius for matching between clusters.  
In Table \ref{tab:erosita_rad_table} we provide a summary table of our matching results. 

With the above mass and redshift cuts on the eRASS1 sample, we have 285 eRASS1 clusters matched to an SPT-3G counterpart and 35 eRASS1 clusters without a counterpart.
Bright mm-wave sources were found nearby 8/285 (3\%) of eRASS1 clusters with SZ counterparts and 3/35 (8.5\%)  of eRASS1 clusters without.
We find 45/285 (16\%) of the SZ matched sample and 14/35 (40\%) of the unmatched sample have a flat-spectrum ASKAP counterpart. 
We highlight these matches in  Fig.~\ref{fig:erosita_comparison}. 
Over-plotted in orange squares are ASKAP matches and in blue diamonds are the SPT emissive source matches.
An identical fraction (14/35, 40\%) of the unmatched sample are flagged as having potentially higher X-ray AGN contamination ($P_\textrm{cont}^\textrm{AGN}>0.2$) in the unmatched eRASS1 sample; focusing solely on more secure eRASS1 systems, 11/21 eRASS1 clusters without SZ counterparts (52\% with $P_\textrm{cont}^\textrm{AGN}<0.2$) have nearby radio sources.
From this figure we can clearly see that for the X-ray systems not matched by SPT-3G, (1) the majority are at lower X-ray masses, (2) there are higher association rates with low-frequency radio sources than for the full sample, but also, (3) on the X-ray side, higher fractions of systems with potentially spurious AGN contamination for which we would not expect an SZ association. 

From Fig.~\ref{fig:erosita_comparison} we can also see that a number of the outliers in the eROSITA mass-SPT observable relation have nearby radio sources, but many clusters with similar sources fall cleanly within the typical scatter for the scaling between the two. 
Our scaling of radio source power from low frequencies using ASKAP measured spectral indices is a simplistic extrapolation and the contamination is highly sensitive to the spectral index model adopted \citep[e.g.,][]{dicker21}; if we instead adopt a typical spectral index of -0.7 found from an analysis of 1.4 and 30~GHz data in X-ray selected clusters \citep{coble07}, we find that 5\% of both the SZ matched and unmatched sample have ASKAP counterparts. 
This second cut with ASKAP also had a more significant impact in the internal frequency tests we conducted in Sec. \ref{subsec:outliers}. 

The above analysis was conducted for clusters in unmasked regions in the SPT-3G footprint, now we consider clusters in the eRASS1 cosmology sample falling in regions masked in SPT-3G data owing to bright mm-wave emissive sources. 
As discussed in Sec. \ref{sec:cluster_identification}, to reduce spurious detections we mask the SPT maps at the locations of bright radio sources; preferential correlation of the brightest mm-wave sources with clusters could lead to additional missed clusters over what might be expected from the lost area.  
We expect this effect to be too small to be well constrained by the eRASS1 sample \citep[see e.g., discussion in Sec. 3.6 in][]{bleem20} but report statistics for completeness here. 
We find 9  clusters at $z\ge0.25$, 8/9 at $P_\textrm{cont}^\textrm{AGN}<0.2$ (the one remainder, 1eRASS J013954.2-435507, at $P_\textrm{cont}^\textrm{AGN}=0.9$). 
Of these, 5/9 have SPT sources offset $>3\arcmin$ from the eRASS1 cluster (so likely unassociated), 3 (1eRASS J000557.9-562841, 1eRASS J013954.2-435507, 1eRASS J231306.7-550417) have mm- and ASKAP-bright sources within $0\farcm{25}$ of the eRASS1 cluster, and one is an ambiguous case (1eRASS J205208.7-691105) with a 4 mJy SPT source---but no low frequency ASKAP flat-spectrum detection---$1\farcm{2}$ away; this test suggests a small number of clusters missed at $z\ge0.25$ owing to very bright sources. 
If we remove the redshift restriction, the number of missed clusters increases to 17 in the masked footprint. 

Overall, our various explorations of the radio contamination of the SPT-3G sample---both with internal frequency checks (Sec. \ref{subsec:outliers}) and through cross-matching with the eRASS1 catalog---have shown that only a few percent of the SZ and X-ray samples have SZ signal strongly impacted by radio sources. 
This result is consistent with the previous works discussed above (Sec.~\ref{subsec:radio}) but questions remain both at lower mass where the majority of the unmatched eRASS1 clusters reside and for higher-redshift clusters beyond the reach of the current eROSITA sample. 
A more detailed exploration of radio source contamination for the SPT-3G sample will be reported in an upcoming work. 

\section{Conclusion and Future Outlook}
\label{sec:conclusion}

In this paper we introduced a new cluster sample from an analysis of five years of observations of the SPT-3G Main field. 
Here we summarize our main results and highlight several planned future studies with this sample. 

\begin{itemize}
\item We have identified \ncand \ cluster candidates detected above $\xi=4$ in \skyarea \ \sqdeg \ of sky mapped at 95, 150, and 220~GHz. Using optical (primarily from DES) and infrared (WISE, \textit{Spitzer}, and Magellan/FourStar) observations, we confirmed \nconfirm \ of these candidates as clusters, with an expected confirmed sample purity exceeding 96\%.  The clusters span $0.03 < z \lesssim 2$ in redshift and \mlow \ $< \mass < $ \mhigh \ in mass. The median redshift of the sample is $z_\textrm{med}=\medianredshift$ and 25\% of the sample is at $z>1$ ($\sim 4\%$ at $z>1.5$). The mass-completeness threshold decreases with redshift; at redshift $z=0.25$ (the lowest redshift for which filtering is not expected to strongly impact the SZ selection), the sample is expected to be $>90\%$ complete for masses $M_{500\mathrm c}>2.1\times10^{14} M_\odot/h_{70}$. 

\item We estimate the masses of the SZ clusters using scaling relations estimated by abundance-matching to our fiducial $\Lambda$CDM cosmological model. In this work we introduce an optical-richness-SZ scaling relation comparison to test our SZ candidate-counterpart assignments, finding a small (five) number of clusters for which foreground low-$z$ systems were mistakenly assigned to higher-redshift clusters. 

\item We flag \nstronglensing \ SPT clusters as strong gravitational lenses based on either visual evidence of strong lensing in optical survey data, or prior follow-up observations of a small number of bright DSFGs in close proximity to SPT cluster centers. 

\item Comparisons with previous generations of SZ surveys from ACT and SPT in the Main field footprint show good agreement in SZ (mass, centroid) properties, with the SPT-3G sample detection significance typically 2-4$\times$ higher for clusters in common. 

\item We perform several cross checks with clusters and point sources from eROSITA.  In the SPT-3G cluster analysis footprint we cross-match 62\% of the eRASS1 cluster sample from  \citet{bulbul24}, rising to 87\% for the subset in the eRASS1 cosmology sample at $z>0.25$. 
Examining in more detail this subset of 320 clusters (35 unmatched), we find 40\% (14/35) of the unmatched X-ray clusters have potentially flat-spectrum low-frequency radio sources nearby that could bias the SZ selection, but also 40\% of these clusters had higher likelihoods of X-ray AGN contamination (potentially biasing the X-ray selection).  

\item We also compare to three representative optical/IR wide-field cluster catalogs in the SPT-3G footprint: the red-sequence-based DES \textsc{redMaPPer} year 3 sample \citep{Abbott25clusters} and the optical/IR-photo-$z$-based cluster samples of \citet{WenAndHan24} and MadCOWS2 \citep{thongkham24}. While historically SZ catalogs have probed the high-richness end of optical/IR samples, the depth and redshift reach of the SPT-3G main sample (combined with scatter in the optical richness-to-mass relation) results in a more mixed view here, with the optical sample completeness for SPT-3G clusters ranging from $\sim52-72\%$. In general we find good agreement between MCMF redshifts and those from the external catalogs, with varying levels of scatter between the SZ mass and optical richness proxy. 

\item Finally, internal data splits using subsets of the SPT-3G bands show a $\sim17\times$ increase from redshifts 0.25 to 1.6 in the correlated 220~GHz emission at SPT cluster locations, but limited signs of correlated AGN contamination. 
A detailed study on a smaller set of clusters in  \citet{kornoelje26} showed that the three band SPT-3G selection is robust to the correlated dusty emission at the mass and redshift ranges probed by SPT-3G; the multi-band checks here support this conclusion.

\end{itemize}

Looking forward, work is underway to obtain cosmological constraints from this new cluster sample, both from internal mass calibration with SPT-3G data via CMB cluster lensing \citep[e.g.,][]{raghunathan19a} and cluster clustering \citep[e.g.,][]{lima05, cromer19,zhan26}, and with optical weak lensing from DES \citep{yamamoto25} and from the \textit{Euclid} mission \citep{mellier25}. 
This Main field release is the first wide-field release from SPT-3G, but the SPT-3G camera has been used to observe an additional $\sim 8,400$ \sqdeg \ of the southern sky \citep[for a total of $\sim 10,000$ \sqdeg,][]{prabhu24,vitrier25}.  Cluster catalogs from these data, the Simons Observatory \citep{simonsobservatorycollab19}, as well as future catalogs from the planned SPT-3G+ instrument---which will reach depths of 0.7 \muk-arcmin at 95 and 150 GHz, our most constraining SZ science bands---will continue to expand the capabilities of SZ cluster science. 

\facilities{Blanco (DECAM), Hubble Space Telescope (WFC3), Magellan:Baade (FourStar), Magellan:Clay (PISCO), NSF/US Department of Energy 10m South Pole Telescope (SPT-3G), Spitzer (IRAC)}

\section*{Acknowledgments} 

The South Pole Telescope program is supported by the National Science Foundation (NSF) through awards OPP-1852617 and OPP-2332483. Partial support is also provided by the Kavli Institute of Cosmological Physics at the University of Chicago. 
Argonne National Laboratory’s work was supported by the U.S. Department of Energy, Office of High Energy Physics, under contract DE-AC02-06CH11357. 
The UC Davis group acknowledges support from Michael and Ester Vaida. 
Work at the Fermi National Accelerator Laboratory (Fermilab), a U.S. Department of Energy, Office of Science, Office of High Energy Physics HEP User Facility, is managed by Fermi Forward Discovery Group, LLC, acting under Contract No. 89243024CSC000002.
The Melbourne authors acknowledge support from the  Australian Research Council’s Discovery Project scheme (No. DP260100705). 
The Paris group has received funding from the European Research Council (ERC) under the European Union’s Horizon 2020 research and innovation program (grant agreement No 101001897), and funding from the Centre National d’Etudes Spatiales. 
The SLAC group is supported in part by the Department of Energy at SLAC National Accelerator Laboratory, under contract DE-AC02-76SF00515.
The Innsbruck authors acknowledge support provided by the Austrian Research Promotion Agency (FFG) and the Federal Ministry of the
Republic of Austria for Innovation, Mobility, and Infrastructure (BMIMI) via the Austrian Space Applications Program with grant numbers 899537, 900565, 911971, and 928759, as well as the Austrian Science Fund (FWF) via project 10.55776/F101300. 

This work is based in
part on observations made with the Spitzer Space Telescope (PIDs 60099, 70053, 80012 and 10101; PI Brodwin), (PIDs 11096, 12073, 14096; PI Bleem) which was operated by the Jet Propulsion Laboratory, California Institute of Technology under a contract with NASA.
This research is based on observations made with the NASA/ESA Hubble Space Telescope obtained from the Space Telescope Science Institute, which is operated by the Association of Universities for Research in Astronomy, Inc., under NASA contract NAS 5-26555. These observations are associated with GO programs 12246, 12477, 13412, 14252, 14677, 15307, and 16017. 
This paper also uses data obtained with the 6.5-m Magellan Telescopes located at Las Campanas Observatory, Chile. Magellan observing time for this project was granted by the time allocation committees of the University of Chicago and the Center for Astrophysics at Harvard/Smithsonian. We would also like to express our gratitude to the staff of Las Campanas Observatory in Chile whose support was instrumental in facilitating data collection.

This work utilized data and resources provided by the Strong Lensing Database (SLED).
This research has made use of the VizieR catalogue access tool, CDS, Strasbourg Astronomical Observatory, France (DOI : 10.26093/cds/vizier).  Some results in this work were obtained using services provided by the OSG Consortium \citep{osg07, osg09, osg06, osg15}. which is supported by the National Science Foundation awards \#2030508 and \#2323298. 

Funding for the DES Projects has been provided by the U.S. Department of Energy, the U.S. National Science Foundation, the Ministry of Science and Education of Spain, 
the Science and Technology Facilities Council of the United Kingdom, the Higher Education Funding Council for England, the National Center for Supercomputing 
Applications at the University of Illinois at Urbana-Champaign, the Kavli Institute of Cosmological Physics at the University of Chicago, 
the Center for Cosmology and Astro-Particle Physics at the Ohio State University,
the Mitchell Institute for Fundamental Physics and Astronomy at Texas A\&M University, Financiadora de Estudos e Projetos, 
Funda{\c c}{\~a}o Carlos Chagas Filho de Amparo {\`a} Pesquisa do Estado do Rio de Janeiro, Conselho Nacional de Desenvolvimento Cient{\'i}fico e Tecnol{\'o}gico and 
the Minist{\'e}rio da Ci{\^e}ncia, Tecnologia e Inova{\c c}{\~a}o, the Deutsche Forschungsgemeinschaft and the Collaborating Institutions in the Dark Energy Survey. 

The Collaborating Institutions are Argonne National Laboratory, the University of California at Santa Cruz, the University of Cambridge, Centro de Investigaciones Energ{\'e}ticas, 
Medioambientales y Tecnol{\'o}gicas-Madrid, the University of Chicago, University College London, the DES-Brazil Consortium, the University of Edinburgh, 
the Eidgen{\"o}ssische Technische Hochschule (ETH) Z{\"u}rich, 
Fermi National Accelerator Laboratory, the University of Illinois at Urbana-Champaign, the Institut de Ci{\`e}ncies de l'Espai (IEEC/CSIC), 
the Institut de F{\'i}sica d'Altes Energies, Lawrence Berkeley National Laboratory, the Ludwig-Maximilians Universit{\"a}t M{\"u}nchen and the associated Excellence Cluster Universe, 
the University of Michigan, NSF NOIRLab, the University of Nottingham, The Ohio State University, the University of Pennsylvania, the University of Portsmouth, 
SLAC National Accelerator Laboratory, Stanford University, the University of Sussex, Texas A\&M University, and the OzDES Membership Consortium.

Based in part on observations at NSF Cerro Tololo Inter-American Observatory at NSF NOIRLab (NOIRLab Prop. ID 2012B-0001; PI: J. Frieman), which is managed by the Association of Universities for Research in Astronomy (AURA) under a cooperative agreement with the National Science Foundation.

The DES data management system is supported by the National Science Foundation under Grant Numbers AST-1138766 and AST-1536171.
Data access is enabled by Jetstream2 and OSN at Indiana University through allocation PHY240006: Dark Energy Survey from the Advanced Cyberinfrastructure Coordination Ecosystem: Services and Support (ACCESS) program, which is supported by U.S. National Science Foundation grants 2138259, 2138286, 2138307, 2137603, and 2138296.
The DES participants from Spanish institutions are partially supported by MICINN under grants PID2021-123012, PID2021-128989 PID2022-141079, SEV-2016-0588, CEX2020-001058-M and CEX2020-001007-S, some of which include ERDF funds from the European Union. IFAE is partially funded by the CERCA program of the Generalitat de Catalunya.
We  acknowledge support from the Brazilian Instituto Nacional de Ci\^encia
e Tecnologia (INCT) do e-Universo (CNPq grant 465376/2014-2).
This document was prepared by the DES Collaboration using the resources of the Fermi National Accelerator Laboratory (Fermilab), a U.S. Department of Energy, Office of Science, Office of High Energy Physics HEP User Facility. Fermilab is managed by Fermi Forward Discovery Group, LLC, acting under Contract No. 89243024CSC000002.

We thank Zoe, the SPT support cat for this analysis. Inquiries about SPT support cats shall be
directed to T.C.

\bibliographystyle{yahapj}
\bibliography{spt,second}


\appendix

\section{Matches to Other eROSITA Samples}
\begin{figure}
    \centering
    \includegraphics[width=\linewidth]{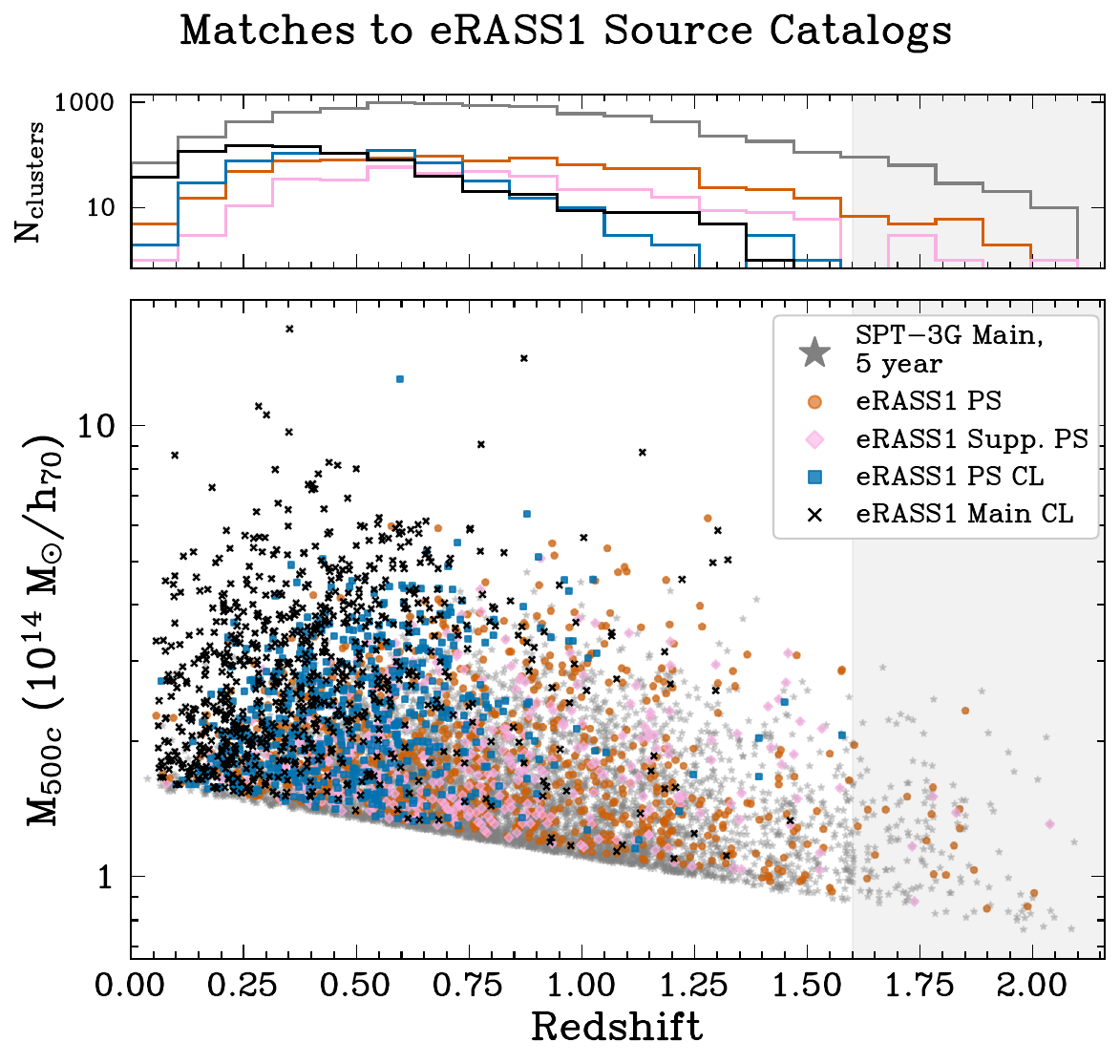}
    \caption{Redshift and mass distribution of matches between eRASS1 sources and the SPT-3G Main field  cluster sample. For the clusters reported in \citet{bulbul24} (Main CL) and \citet{balzer25} (PS CL) we adopt a matching radius of 3\arcmin. For sources from the Main (PS) and Supplemental  (Supp. PS) catalog of \citet{merloni24} we adopt a 1\arcmin \ matching radius to reduce the probability of random associations to under 4\%. We also exclude from these matches any SPT clusters also matched with X-ray clusters from the above catalogs. In total 1,348 eRASS1 clusters and 1,319 additional X-ray point sources have counterparts in the SPT-3G cluster sample. Masses and redshifts in this figure are from the SPT sample measurements.}
    \label{fig:erosita_spt_match2}
\end{figure}

\label{sec:xrayappend}
In Sec. \ref{subsec:erosita_comparison} we cross-matched the SPT-3G sample to the first cluster catalogs from the eRASS1 survey \citep{bulbul24}.  
In this appendix we extend this matching to other eRASS1 source samples, focusing on associations with (1) the expanded X-ray cluster sample of \citet{balzer25} and (2) the full eRASS1 point source catalog of \citet{merloni24}.
The large overlap in samples we show here briefly highlights the potential for future joint SPT-3G-eROSITA studies, particularly in understanding the evolution of X-ray AGN in clusters as well as in placing tighter constraints on biases in X-ray selection in moderate ($\gtrsim30\arcsec$) resolution data owing to the mis-identification of clusters as X-ray point sources.

\subsection{eROSITA Cluster Search in the eRASS1 Point Source Catalog}

\citet{balzer25} searched for clusters misclassified as X-ray point sources through a careful examination of the  
point-like subset of the full eRASS1 source catalog \citep{merloni24}. 
Such misidentification can arise from multiple sources including X-ray bright AGN near cluster centers \citep[e.g.,][]{green17}, difficulties in identifying faint sources as extended in the low SNR regime, or biased extent measurements from the presence of strong cool cores \citep[e.g.,][]{mcdonald12,donahue20,somboonpanyakul21}. 
After examination the authors found an additional 8,347 clusters in the eROSITA data.
Of this new sample, 1,003 clusters are in the SPT-3G footprint, and 58\% have matches within 3\arcmin \ in the SPT-3G cluster sample. 
This is slightly lower than the 62\% matched for the full eRASS1 cluster sample, and the median angular separation is higher (0\farcm{49} v.s. 0\farcm{36}), potentially reflecting more false associations or more uncertainty in the X-ray ICM centroids. 

\citet{balzer25} classify the new cluster detections based on potential for the X-ray signal to be from  Galactic, X-ray AGN, or ICM emission. 
Of the clusters matched to the SPT-3G sample, 60\% are in the most reliable classes (\texttt{CLASS 4}, potential cool core, or \texttt{CLASS 5}, most flux expected to come from ICM), compared to 40\% in the unmatched sample; this fraction further raises to 68\% with a stricter distance association of 1\arcmin \ to the SPT sample. The \texttt{CLASS 4} and \texttt{5} systems also have a smaller median offset from SPT clusters ($0\farcm{41} \pm 0.02$) than the lower classes ($0\farcm{74}^{+0.06}_{-0.04}$). 

\subsection{SPT-3G Cluster Matches to the eRASS1 Point Source Catalog}
The full eRASS1 point source sample  contains over 1.25 million sources  with detection likelihood greater than 5 \citep{merloni24}; the reanalysis of \citet{balzer25} focused on the 576,154 sources with likelihood greater than 8 to improve the overall X-ray sample purity.
Here we match SPT-3G clusters to both the main catalog as well as the more contaminated supplemental catalog which contains detections with likelihoods between 5 and 6. The eRASS1 survey essentially covers the full SPT-3G Main Field footprint; using a million random locations within the field we estimate 4\% (1.5\%) probability of chance associations within 1\arcmin \  for sources in the eRASS1 main point source (supplemental) catalog. 
Cross-matching these two lists using a 1\arcmin \ association radius to the SPT-3G cluster catalog (and excluding any SPT clusters with a more relaxed matching criterion of 3\arcmin \ to clusters in either \citealt{bulbul24} or \citealt{balzer25}), we find 921 matches from the main catalog (54\% at X-ray detection likelihood less than 8, the examination threshold in previous reported work) and an additional 398 unique matched SPT cluster candidates using the supplemental catalog (a total of 1,319 additional matched SZ candidates); 90.5\% of these candidates are confirmed as clusters through our optical analysis. 

We plot the distribution of SPT-3G clusters matched to the different eRASS1 catalogs in Fig \ref{fig:erosita_spt_match2}. 
The median redshift of these matched samples increases from the primary cluster sample of \citet{bulbul24} ($z_\textrm{med}=0.36$) to \citet{balzer25} ($z_\textrm{med}=0.49$) to purely point source matched clusters ($z_\textrm{med}=0.74$/0.73, for main/supplementary point source sample).
We expect a large number of these cross-matches between SPT-3G clusters and the eRASS1 point source catalog to be cluster X-ray AGN---whose abundance in clusters have been shown in previous studies to have a strong evolution with redshift \citep[see e.g.,][]{martini09,martini13,bufanda17}---but also expect the matches to contain a number of galaxy clusters misidentified in X-ray data.
As such, this large sample of cross-matched objects will provide a promising avenue for future explorations of both cluster AGN activity as well as the eROSITA selection function at high redshift.

\section{redMaPPer Richness-Mass Relation}
\label{sec:opticalappend}

We perform one additional characterization of the \textsc{redMaPPer} Y3  mass-richness relation, as the significant increase in sample matches and lowered mass threshold of the SPT-3G survey allows us to expand upon previous work presented in \citet{bleem20}. 
Here we follow the framework discussed in Sec.~\ref{sec:richnessmassrelation}, however when evaluating the likelihood given by Eq.~\eqref{eq:abundance_3_likelihood}, we do not account for a richness cut according to $\lambda_\mathrm{min}^\mathrm{MCMF}(z)$, but instead we model the optical selection as $\lambda_\mathrm{RM}>20$. As in the study of MCMF richness, we consider the subsample of SPT-3G clusters with $\xi>5$ and $0.25<z<1$.
\begin{figure}
  \includegraphics[width=\linewidth]{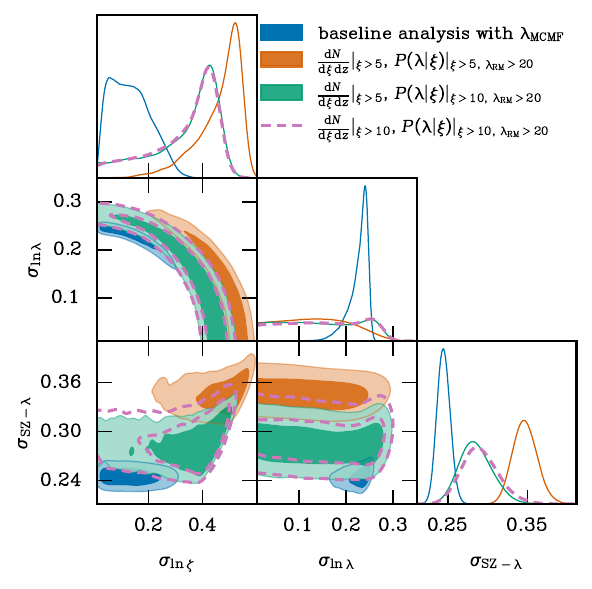}
  \caption{Inferred scatter in the SZ and optical observables. Our baseline analysis using MCMF richness is shown in blue (same as in Fig.~\ref{fig:GTC}). The quantity that is well measured is the combined, quadrature-summed scatter $\sigma_{\mathrm{SZ}-\lambda}$, whereas the individual scatter components exhibit a narrow and bent degeneracy. The baseline \textsc{redMaPPer}  analysis (orange) yields the largest amount of scatter. Limiting the \textsc{redMaPPer}  analysis to higher-significance clusters reduces the total scatter (green), whereas applying a higher-significance cut also to the abundance analysis does not further shift the constraints (purple dashed).}
  \label{fig:SZrichnessscatter}
\end{figure}

We constrain the parameters of the richness-mass relation (Eq.~\ref{eq:richness-mass}) as follows:
\begin{align}
  \ln A_{\lambda_\mathrm{RM}} = 3.04\pm0.04\\
  B_{\lambda_\mathrm{RM}} = 0.90\pm0.03\\
  C_{\lambda_\mathrm{RM}} = 0.37\pm0.10\\
  \sigma_{\ln\lambda_\mathrm{RM}} = 0.14\pm0.07
\end{align}
The amplitude (considering the different pivots), redshift evolution, and scatter are consistent with \cite{bleem20}, while the slope has decreased by $1.4\sigma$; the error bars on the parameters are reduced by 2$\times$.
Comparing to MCMF (Table \ref{tab:obs-mass}), the amplitude is lower than that of the MCMF richness-mass relation owing to the different definitions of richness (MCMF probes further down in the cluster luminosity function and with different radial cuts). Similarly, the evolution with mass and redshift differ. The marginalized scatter in \textsc{redMaPPer}  richness appears to be significantly lower than that in MCMF richness. We plot the posterior distributions of the SZ and richness scatter in Fig.~\ref{fig:SZrichnessscatter}.

The figure suggests that the lower scatter in \textsc{redMaPPer} richness is compensated by a larger scatter in $\zeta$. Note that our analysis is mostly sensitive to the total observable-to-observable scatter
\begin{equation}
  \sigma_{\mathrm{SZ}\,-\,\lambda} \equiv \sqrt{\left(\frac{\sigma_{\ln\zeta}}{B_\mathrm{SZ}}\right)^2 + \left(\frac{\sigma_{\ln\lambda}}{B_\lambda}\right)^2}.
\end{equation}
The posterior distributions in Fig.~\ref{fig:SZrichnessscatter} suggest that there is significantly more scatter between \textsc{redMaPPer} richness and the SZ significance than there is between the MCMF richness and the SZ significance (orange vs. blue distributions). To investigate further, we restrict the analysis of \textsc{redMaPPer} richness only to SPT-3G clusters with $\xi>10$. The constraint on $\sigma_{\mathrm{SZ}\,-\,\lambda}$ moves halfway to the value we obtain with MCMF (and is generally comparable to the simplified analysis of Sec. \ref{sec:comparingrichness} for higher SZ mass systems), while the constraints on all other parameters ($\ln A_\mathrm{SZ}$, $B_\mathrm{SZ}$, $C_\mathrm{SZ}$, $\ln A_\lambda$, $B_\lambda$, and $C_\lambda$) remain essentially unchanged. This finding might suggest that some element of the analysis of the $5<\xi<10$ sample is inadequate. We now restrict the entire SPT-3G sample to $\xi>10$ (instead of just discarding all \textsc{redMaPPer} measurements for $\xi<10$ clusters). The constraints on scatter remain essentially unchanged (green vs. purple dashed distributions in Fig.~\ref{fig:SZrichnessscatter}) as do the constraints on all other parameters except for $C_\mathrm{SZ}$ which shifts up by about $1\sigma$. The stability of the SZ-mass relation to these sample cuts suggests that there might be room for a more sophisticated modeling of the \textsc{redMaPPer} richness-mass relation \citep[c.f.,][]{grandis25}; further work here will be important in the context of upcoming optical cluster surveys with the Vera Rubin Observatory Legacy Survey of Space and Time \citep[LSST;][]{ivezic19, lsst18}. 

\newpage
\clearpage

\newpage
\onecolumngrid
\section{Column Descriptions}
\label{sec:tableoverview}
\begin{table}[H]
\centering
\begin{tabular}{ll p{120mm}}
\colhead{Column} &
\colhead{Symbol} &
\colhead{Description} \\
\hline
\texttt{SPT\_ID} & \nodata & Cluster name in the format SPT-CL~JHHMM.m$-$DDMM.m, except for clusters published previously with SPT designation, which are labeled as SPT-CL~JHHMM$-$DDMM to be consistent with previous IAU naming scheme \\
\texttt{RA} &\nodata & Right ascension (degrees)\\
\texttt{DEC} & \nodata & Declination (degrees)\\
\texttt{XI} & $\xi$ & SPT detection significance \\
\texttt{THETA\_CORE} & $\theta_\textrm{c}$  & Filter core size corresponding to detection \\
\texttt{YSZ} & \nodata & Integrated Comptonization measured within $0\farcm{75}$  \\
\texttt{YSZ\_UNC} & \nodata & 1 $\sigma$ Uncertainty on integrated Comptonization measured within $0\farcm{75}$ \\
\texttt{PS\_FLAG} & \nodata & (1) if cluster within 4\arcmin \ of a bright source $>6$ mJy at 150~GHz, (2) if $y_\textrm{SZ}$ measurement impacted by nearby source/massive cluster, (3) if both \\ 
\texttt{REDSHIFT} & $z$ & Redshift \\
\texttt{REDSHIFT\_UNC} & $\sigma_z$ & Uncertainty on redshift \\ 
\texttt{REDSHIFT\_SOURCE} & \nodata & Source of redshift (spectroscopic, MCMF, \textit{Spitzer}, etc.)\\
\texttt{SPECZ}& \nodata & Binary, 1 if redshift is spectroscopic, 0 if not \\
\texttt{SPECZ\_REF}& \nodata & Bibliographic reference for spectroscopic redshift \\
\texttt{M500C} & \mass & Mass, units of $10^{14}$ \mbox{$M_\odot/h_{70}$} \\
\texttt{M500C\_UERR} & \nodata & Upper 1 $\sigma$ uncertainty on the mass estimate \\
\texttt{M500C\_LERR} & \nodata & Lower 1 $\sigma$ uncertainty on the mass estimate \\
\texttt{LAMBDA} & $\lambda_\textrm{MCMF}$ & MCMF richness (where available) \\
\texttt{LAMBDA\_UNC} & \nodata & 1 $\sigma$ uncertainty on MCMF richness \\
\texttt{FCONT} & $f_\mathrm{cont}$ & Contamination estimation for structure associated with redshift assignment \\
\texttt{MCMF\_MASK1} & \nodata & Fraction of optical data within 1\arcmin \ masked in MCMF analysis \\ 
\texttt{MCMF\_MASK2} & \nodata & Fraction of optical data within 2\arcmin \ masked in MCMF analysis \\ 
\texttt{LOS} & \nodata & Binary flag identifying systems with multiple  significant optical/IR structures along the line of sight \\
\texttt{LAMBDA2 } & \nodata &  Richness estimation of secondary structure \\
\texttt{LAMBDA2\_UNC } & \nodata &  Uncertainty on richness estimation of secondary structure \\
\texttt{REDSHIFT2} & \nodata & Redshift estimate for secondary structure \\
\texttt{REDSHIFT2\_UNC} & \nodata & Redshift uncertainty for secondary structure \\
\texttt{FCONT2} & \nodata & Contamination estimate for secondary structure \\
\texttt{STRONG\_LENS} & \nodata & Gravitational strong lens flag,  (1) Optical strong lens flagged by our visual analysis, (2) Strong lens from the literature, (3) if both. \\
\texttt{STRONG\_LENS\_REF} & \nodata & Bibliographic reference for strong lens \\
\texttt{DSFG\_FLAG} & \nodata & (1) if DSFG within 30\arcsec \ of SZ centroid, (2) if $y_\textrm{SZ}$ potentially biased by dusty signal, (3) if both\\ 
\end{tabular}
\caption{Column descriptions in the SPT-3G \texttt{FITS}-formatted cluster catalog available online at \textnormal{\webaddress} \ and at LAMBDA at \textnormal{\url{https://lambda.gsfc.nasa.gov/product/spt/spt3g\_prod_table.html}}.}
\end{table}


\end{document}